\documentclass[11pt]{article}
\pdfoutput=1
\usepackage{jcapmod}
\usepackage{shorthand}
\usepackage[scaled]{helvet}
\usepackage[toc,page]{appendix}
\usepackage{mathtools}
\usepackage{booktabs}
\usepackage[english]{babel}
\usepackage{amsmath,amssymb,amsbsy,amstext, amsthm, simplewick, amsfonts}
\usepackage{hyperref}
\usepackage{graphicx}
\usepackage[small]{caption}
\usepackage{siunitx}
\usepackage{upgreek}
\usepackage{framed}
\usepackage{wrapfig}
\usepackage{multirow}
\usepackage{bbm}
\usepackage[svgnames,dvipsnames,x11names]{xcolor}
\usepackage[utf8x]{inputenc}
\usepackage{selinput}
\usepackage{bm}
\usepackage{float}
\usepackage{geometry}
\usepackage{yfonts}
\usepackage{caption}
\usepackage{subcaption}
\usepackage{sidecap}
\usepackage{longtable}
\usepackage{anyfontsize}
\usepackage{dsfont}
\usepackage{tikz}
\usepackage{relsize}
\usepackage{shuffle}
\usetikzlibrary{matrix}
\usepackage{bigstrut}
\usepackage{fnpct}
\usetikzlibrary{arrows.meta}
\usetikzlibrary{decorations.pathreplacing,calligraphy}
\usepackage{tensor}
\usepackage{mleftright}
\usepackage[most]{tcolorbox}
\usepackage{fancybox} 
\usepackage{tikz}
\usepackage{tikz-3dplot}
\usepackage{ytableau}
\usepackage{braket}
\usepackage{diagbox}
\usepackage{pifont}

\usetikzlibrary{decorations.markings, decorations.pathmorphing, decorations.pathreplacing, decorations.shapes}

\usepackage{xcolor}
\usepackage{xparse}
\definecolor{mplBlue}{rgb}{0,0,1}   %
\definecolor{mplRed}{rgb}{1,0,0}    %
\definecolor{mplGreen}{rgb}{0,0.5,0}   %

\usepackage{fontawesome5}

\makeatletter
\newcommand{\github}[1]{%
   \href{#1}{\faGithub}%
}
\makeatother

\NewDocumentCommand{\colornucleus}{omme{_^}}{%
  \begingroup\colorlet{currcolor}{.}%
  \IfValueTF{#1}
   {\textcolor[#1]{#2}}
   {\textcolor{#2}}
    {%
     #3%
     \IfValueT{#4}{_{\textcolor{currcolor}{#4}}}%
     \IfValueT{#5}{^{\textcolor{currcolor}{#5}}}%
    }%
  \endgroup
}

\usepackage{array}
\newcolumntype{L}[1]{>{\raggedright\let\newline\\\arraybackslash\hspace{0pt}}m{#1}}
\newcolumntype{C}[1]{>{\centering\let\newline\\\arraybackslash\hspace{0pt}}m{#1}}
\newcolumntype{R}[1]{>{\raggedleft\let\newline\\\arraybackslash\hspace{0pt}}m{#1}}

\usepackage[framemethod=default]{mdframed}
\newmdenv[skipabove=7pt,
skipbelow=7pt,
rightline=true,
leftline=true,
topline=true,
bottomline=true,
backgroundcolor=gray!10,
linecolor=black,
innerleftmargin=5pt,
innerrightmargin=5pt,
innertopmargin=5pt,
innerbottommargin=5pt,
leftmargin=0cm,
rightmargin=0cm,
linewidth=1pt]{eBox}

\definecolor{Red}{RGB}{214, 39, 40}
\definecolor{Blue}{RGB} {31, 119, 180}
\definecolor{Orange}{RGB}{255, 153, 51}
\definecolor{Purple}{RGB}{178, 102, 255}
\definecolor{Green}{RGB}{44, 160, 44}
\definecolor{regal}{RGB}{90,0,120}
\definecolor{darkblue}{rgb}{0.15,0.35,0.55}
\definecolor{reddish}{rgb}{0.65, 0.2, 0.2}
\definecolor{darkgreen}{RGB}{50,150,0}
\definecolor{greyish}{rgb}{.90,.90,.90}
\definecolor{greyish2}{rgb}{.96,.96,.96}
\definecolor{greyish3}{rgb}{.37,.37,.37}
\definecolor{darkblue2}{rgb}{0.3,0.4,0.9}
\definecolor{Blue3}{RGB}{31, 119, 180}

\definecolor{blue3}{RGB}{31, 119, 180}
\definecolor{red3}{RGB}{	214, 39, 40}
\definecolor{orange3}{RGB}{255, 127, 14}
\definecolor{green3}{RGB}{44, 160, 44}
\definecolor{repBlue}{RGB}{31, 119, 180}
\definecolor{repRed}{RGB}{	214, 39, 40}
\definecolor{repGreen}{RGB}{44, 160, 44}

\newcommand{\ydiagramalign}[1]{\raisebox{-5pt}{\ydiagram{#1}}}
\newcommand{\ytableaushortalign}[1]{\raisebox{-5pt}{\ytableaushort{#1}}}

\definecolor{vio}{RGB}{19, 130, 164}
\definecolor{vioo}{RGB}{89, 2, 155}
\newcommand{\Comment}[1]{{}}
\usepackage{empheq}

\usepackage[linktocpage=true]{hyperref}
\hypersetup{
colorlinks=true,
citecolor=darkblue,
linkcolor=reddish,
urlcolor=darkblue,
pdfauthor={},
pdftitle={},
pdfsubject={}
}

\usepackage{colortbl}
\definecolor{lightgreen}{cmyk}{0.2, 0, 0.2, 0.2}
\definecolor{lightgray2}{cmyk}{0.1,0.1,0,0.1}
\definecolor{Red2}{RGB}{214, 39, 40}
\definecolor{Blue2}{RGB} {31, 119, 180}
\definecolor{Orange2}{RGB}{255, 127, 14}
\definecolor{Green2}{RGB}{44, 160, 44}

\makeatletter
\newlength{\apb@width}
\newcommand{\autoparbox}[2][c]{\settowidth{\apb@width}{#2}\parbox[#1]{\apb@width}{#2}}

\makeatother


\def\beq{\begin{equation}}
\def\eeq{\end{equation}}
\def\be{\begin{equation}}
\def\ee{\end{equation}}
\allowdisplaybreaks[1]
\newcommand\sqmatrix[2][c]{%
  \fixTABwidth{T}%
  \setbox0=\hbox{$\tabbedCenterstack{#2}$}%
  \setstackgap{L}{\dimexpr\maxTAB@width+\tabbed@gap}%
  \tabbedCenterstack[#1]{#2}%
}

\usetikzlibrary{shapes.misc}

\tikzset{cross/.style={cross out, draw=black, minimum size=2*(#1-\pgflinewidth), inner sep=0pt, outer sep=0pt},
cross/.default={1pt}}

\usetikzlibrary{calc}

\tcbset{
	mybox/.style={
		colframe=black,
		colback=white,
		boxrule=0.5pt,
		arc=3pt,
		width=\textwidth-25pt,
		boxsep=8pt,
		left=10pt, right=5pt, top=-10pt, bottom=5pt
	}
}

\begin{document}

\newgeometry{top=2cm, bottom=2cm, left=2cm, right=2cm}

\begin{titlepage}
\setcounter{page}{1} \baselineskip=15.5pt 
\thispagestyle{empty}

\begin{center}
{\fontsize{18}{18} \bf Unraveling the Spectrum of the Open String}\vskip 4pt
\end{center}

\vskip 30pt
\begin{center}
\noindent
{\fontsize{12}{18}\selectfont 
Bruno Bucciotti$^{1,2}$,~Felipe Figueroa$^{3}$~and Guilherme L.~Pimentel$^1$}
\end{center}

\begin{center}
\vskip 4pt
$^{1}$\textit{Scuola Normale Superiore and INFN, Piazza dei Cavalieri 7, 56126, Pisa, Italy}\\
$^{2}$\textit{Department of Physics and Beyond: Center for Fundamental Concepts in Science,\\
Arizona State University, Tempe, AZ 85281, USA}\\
$^{3}$\textit{Service de Physique de l’Univers, Champs et Gravitation,}\\
\textit{Université de Mons, 20 place du Parc, 7000 Mons, Belgium}
\end{center}

\vspace{0.4cm}
\begin{center}{\bf Abstract}
\justify{We construct a large portion of the massive spectrum of the open bosonic string using light-cone quantization, providing explicit oscillator realizations for individual single-particle states as well as for full Regge trajectories. We show how combinations of transverse oscillators organize into irreducible $SO(25)$ representations, and provide an algorithm for constructing them level by level. We then develop a general method to “climb” the spectrum—adding oscillators in a controlled way that generates entire Regge trajectories from a finite set of seed states. Remarkably, the coefficients determining each state’s oscillator composition depend on the level in a simple way, allowing closed-form expressions for infinitely many states. Beyond individual trajectories, we explore internal regularities of the spectrum and establish relations among families of trajectories, extending the concept of a Regge trajectory to more general constructions. Our results expose a highly ordered and recursive structure underlying the open-string spectrum, suggesting that its massive excitations form an algorithmically constructible network. The framework presented here lays the groundwork for computing three-point amplitudes of arbitrary massive states, the essential building blocks of string interactions, which we tackle in upcoming work.}
 
\end{center}
\noindent 

\vskip 15 pt

\end{titlepage}
\restoregeometry

\newpage
\setcounter{tocdepth}{2}
\setcounter{page}{2}

\linespread{0.75}
\tableofcontents
\linespread{1.}

\clearpage{}%
\newpage
\section{Introduction} \label{sec:intro}

The scattering amplitudes of higher-spin particles diverge at large energies~\cite{Weinberg:1964cn}. Yet, string amplitudes manage to soften the amplitudes of high-energy scattering in a way that cannot be mimicked by quantum field theory~\cite{Amati:1987wq,Gross:1987ar,Gross:1987kza}. This remarkable property is key to the ability of string theory to provide an ultraviolet (UV) completion of theories of long-range forces, notably gravity~\cite{Scherk:1974ca,Amati:1988tn,Sagnotti:2010at}. How the plethora of string resonances manages to do this job in a fully consistent manner is still unclear from the target space point-of-view, and a subject of much exploration. We will revisit string perturbation theory and describe in detail its simplest building block: the tree-level, three-point scattering amplitude {\it of single particle states}.\footnote{There are other ways of preparing a scattering state, notably by bombarding the tachyon with a sequence of photons, first proposed by Di Vecchia, Di Giudice and Fubini (DDF)~\cite{DelGiudice:1971yjh}. For these states---which are a complicated superposition of single particle states, the three-point amplitude has long been known~\cite{Ademollo:1974kz}. For recent papers in this direction, see~\cite{Bianchi:2019ywd,Firrotta:2022cku,Firrotta:2024fvi,Firrotta:2024qel,Biswas:2024mdu,Biswas:2024unn}.} In this first paper we show how to explicitly construct a large portion of the spectrum of the bosonic open string, while we will study the interactions among these states in an upcoming paper.

We focus on this problem, as the three-point scattering amplitude is the simplest building block of theories of long-range forces. It is very reasonable to expect, in the spirit of the ``on-shell" approach to perturbation theory, that with a full grasp of the three-point amplitudes, we can learn much about the whole theory, by seeing how these building blocks stitch together, etc~\cite{Elvang:2015rqa}. Moreover, being such a tightly constrained object from kinematics, three-point amplitudes are ultimately a collection of numbers --- how triplets of single-particle states couple to each other according to a finite menu of options~\cite{Arkani-Hamed:2017jhn}.~\footnote{Strictly speaking, many three-point amplitudes (notably of massless particles) are zero in Lorentzian kinematics. As usual, we can complexify the momenta to assign a nonzero value to the amplitude.} 

At finite string coupling, the spectrum of the string should reorganize itself in an interesting, currently unknown way. However we do expect many of these states to ``look like" black holes from far away. Therefore it's interesting to speculate that this property persists even at zero coupling for the very massive states of the string. A potential diagnosis of this is to look at the three-point amplitude of a highly massive and spinning particle to a massless probe. In the case of the graviton there is a ``minimal coupling" that a black hole has, providing a diagnosis of whether a massive spinning particle behaves similarly to a black hole from far away~\cite{Arkani-Hamed:2017jhn,Guevara:2019fsj,Arkani-Hamed:2019ymq,Chung:2018kqs,Cangemi:2022bew,Cangemi:2023bpe,Cangemi:2022abk,Pichini:2023cqn}. There is a ``single copy" version of this phenomenon, which can also be investigated within the open string, by looking at couplings of the massless gauge boson in the spectrum to massive higher-spin states~\cite{Cangemi:2023ysz}. These analyses also raised the question of whether chaos can be detected in string amplitudes~\cite{Gross:2021gsj,Bianchi:2022mhs,Bianchi:2023uby,Firrotta:2023wem,Pesando:2025ztr}.

Why is such a basic problem not already solved, despite the enormous body of work in string perturbation theory? It becomes quickly clear that, though in principle what needs to be done is obvious, doing it in practice is a different matter. In order to build the spectrum, we need two ingredients: first, a single particle state is fully specified from its 26-momentum but also from the spin representation of its little group, $SO(25)$ (for massive particles). While in four dimensions the bosonic representations of the little group $SO(3)$ are labelled by their integer spin, a single number is insufficient for representations of $SO(25)$. These representations are instead neatly encoded by Young tableaux. The challenge then lies in writing all possible realizations of a single-particle state using the oscillators of the string. As we explain in Section~\ref{sec:review}, where we introduce the general setup and provide a short review of the quantization of the open string, in covariant form physical states must solve the Virasoro constraints as well as ensuring that the collection of oscillators realizes the spacetime symmetries of the Young tableau. It is the increasing complexity of the Virasoro constraints as one climbs up the spectrum what has mostly hindered progress in exploring the massive spectrum, limiting most approaches to level-by-level probes of the low-lying states~\cite{Koh:1987hm,Feng:2010yx,Bianchi:2010es,Feng:2011qc,Benakli:2021jxs,Lust:2023sfk,Pesando:2024lqa,Manes:1988gz}, or to the analysis of the leading Regge trajectory~\cite{Sagnotti:2010at,Taronna:2010qq,Rahman:2015pzl,Schlotterer:2010kk}. A remarkable exception is the recent set of articles~\cite{Markou:2023ffh,Basile:2024uxn}, where the authors develop a powerful method to solve the Virasoro constraints and access the deep string spectrum in an efficient way, finding structures and patterns reminiscent of the ones we observe.

Nonetheless, our strategy is to bypass the Virasoro constraint by working in light-cone gauge. This sacrifices covariance, but as our main interest is in three-particle amplitudes, which are fixed by kinematics, it is straightforward to identify the covariant amplitude. Moreover, we gain simplicity in working with physical oscillators and states with manifestly positive norm. The price to pay is that only a $SO(24)$ subgroup of the little group is manifest. We then need to apply Lorentz transformations to see how various $SO(24)$ representations with the same mass correspond to branches of a single $SO(25)$ irreducible representation (irrep). A collection of oscillator representations with fixed mass and rotated into each other then furnishes a single particle state of the open bosonic string. Implementing this in practice requires some work, which we explain in Section \ref{sec:singleps}.

The string has a very orderly spectrum, so it is natural to try and exploit its regularities.
For example, symmetrized products of light-cone gauge oscillators naturally organize in trajectories, where climbing a step corresponds to the insertion of an additional oscillator raising the mass level by one unit and contributing a free index.
Level by level, these structures can be used to obtain single particle $SO(25)$ irreps, which can be grouped into ``Regge trajectories".
Intuitively, to climb up a Regge trajectory we should look at the particle that simultaneously has one more box in the top row of the Young tableau and has mass level increased by one. The catch is that such a Regge trajectory can bifurcate at some level, giving rise to degeneracies and making the notion seemingly ambiguous.
We will show that bifurcations can only occur at finitely many levels as we climb up, so there are always finitely many Regge trajectories whose particles share the same Young tableau symmetry at the same mass level. We will call these trajectories ``degenerate," but we will nevertheless be able to unambiguously label them by their oscillator content. S

How do we then ``climb up" the spectrum, building such a Regge trajectory in terms of oscillators? In section \ref{sec:regget} we show a constructive algorithm to spin up a single particle state by ``adding a box" to its $SO(25)$ Young tableau, while increasing its mass one unit of string tension at a time. This shows that we do not have to work level by level, but we can rather access an entire Regge trajectory at once. Since every state in the spectrum is either the seed of a new trajectory or belongs to an existing one, this gives a systematic way of determining the full spectrum by focusing on all the -infinitely-many- seeds.

Our method also shows that the spectrum can be sliced according to ``generation."\footnote{We will define generation more carefully later in the text, but it amounts to the number of times the same Young tableau appeared before in the spectrum, when ordered by mass level. It is also referred to as ``depth" in~\cite{Markou:2023ffh,Basile:2024uxn}} In section \ref{sec:exploring} we uncover patterns linking different particles within the same generation, providing access to infinite {\it families} of Regge trajectories in closed form. A complete treatment, potentially encompassing all trajectories at fixed generation, is left for future work.
While our method can be carried out for any generation in principle, relating particles at different generations remains an important challenge. For an interesting recent development in this direction, we mention the construction in~\cite{Markou:2023ffh,Basile:2024uxn} which allows to connect states of different generation. It would be interesting to translate it to our light-cone language and try to exploit it.

We state our conclusions and an outlook in \ref{sec:Outlook}, and provide appendices with some details on string theory and light-cone quantization, the required background on Young tableaux, as well as additional results on the spectrum of single-particle states at high levels and on Regge trajectories.

{\bf Conventions:} 

Round brackets denote symmetrization between a set of indices, without any overall factor. For example, $t^{(ij)} \equiv t^{ij}+t^{ji}$. Square brackets denote antisymmetrization, $t^{[ij]}\equiv t^{ij}-t^{ji}$. 

The dot product of two light-cone gauge oscillators is to be understood as
\begin{equation}
    \alpha_{n}\cdot\alpha_{m} \equiv \sum_{k=1}^{24} \alpha_{n}^k\cdot\alpha_{m}^k
\end{equation}

When discussing symmetric irreducible tensors, we will often not explicitly subtract the trace part, but this operation is completely non-ambiguous so it is left implicit.
\smallskip

When ambiguities might arise, we specify the symmetry group associated to a given Young tableau as a subscript, for example $\ydiagram{2}_{24}$ or $\ydiagram{2}_{25}$.
\clearpage{}%
\clearpage{}%
\newpage
\section{Setup and Review} \label{sec:review}

\begin{figure}
    \centering
    \includegraphics[width=1\linewidth]{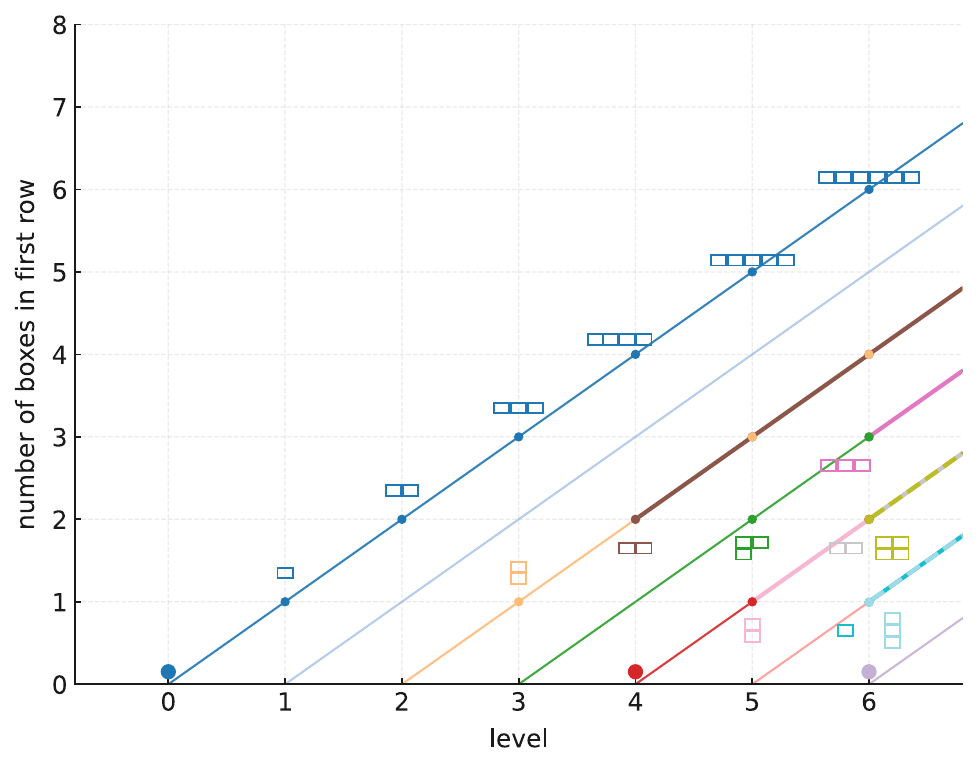}
    \caption{The whole spectrum, divided into Regge trajectories. The top blue one is the leading Regge trajectory. The one right below is the empty trajectory which does not contain any particles. Except for the leading Regge trajectory, we only draw the \emph{seed} of each trajectory, and we leave it implicit that, climbing the trajectory, new particles are found with greater mass and number of boxes in the first row of the Young tableau. For example, at level 4 there is a $\ydiagram{2,1}$ (implicit) originating from the $\ydiagram{1,1}$ (explicit) at level 3. Lastly, notice that at level 6 there is not only the (red) $\ydiagram{2}$ coming from the scalar at level 4, but also a \emph{new} (gray) $\ydiagram{2}$. This is an example of a degeneracy, which we will see is ubiquitous in the string spectrum.}
    \label{fig:regge_trajectories}
\end{figure}

In this section we setup the problem, explain the various necessary steps, and also discuss how we organize the spectrum, also showing how much of it we manage to construct. Most of the review consists of textbook material, and is developed in more detail in any standard string theory textbook, such as~\cite{Polchinski:1998rq,Green:1987sp,Blumenhagen:2013fgp}.

\subsection{Single-Particle States and Regge Trajectories}
We consider the open bosonic string. Consistency of the theory requires the string to be in 26-dimensional Minkowski space, and therefore the single-particle states are classified using the isometries of this spacetime. They carry a $26-$momentum which satisfies a constraint, fixing the mass of the particle, and they carry spin labels showing how they transform under Wigner's little group.\footnote{The little group is defined by fixing a reference momentum satisfying the mass constraint, and looking at the subgroup of Lorentz transformations that do not change the reference momentum. See~\cite{Wigner:1939cj,Bargmann:1948ck,Weinberg:1995mt,Bekaert:2006py} for more details.}
For the bosonic string, the little group is $SO(25)$ for massive particles and $SO(24)$ for massless particles, and their unitary representations are organized using Young tableaux, which in a sense provide a higher dimensional generalization of the spin label typical in 4-dimensional kinematics. A given Young tableau is a collection of ordered boxes specified by the lengths of its rows and columns, each row being of arbitrary length, provided it does not exceed the length of the one above it. The length of the columns is constrained to be smaller than half the number of spatial dimensions, and thus for our case of interest we are limited to Young tableaux with up to 12 rows. We provide further details on the Young tableaux in appendix~\ref{app:Young_Tableaux}.%

In string theory, furnishing a single-particle state requires building it from string excitations. These are constructed with oscillators corresponding to the string's quantized vibrational modes. The oscillators at the same time determine the mass of the state, forcing it to be integer multiples of the string tension, and carry Lorentz indices. Therefore, the mass and spin labels are not independent: at a given mass level, only some Young tableau symmetries will be possible, those that can be realized with the available oscillators.

On the other hand, one might expect extra labels to be necessary in order to fully determine a string state.
In fact, we find evidence of degeneracies in the spectrum -i.e. independent single-particle states at the same level with identical Young tableau symmetry, but realized with different oscillator structures- which suggests the possibility of extra quantum numbers required to distinguish them. These additional quantum numbers might correspond to the eigenvalue of the string states under the action of some hypothetical conserved charges. Yet, the nature of these charges is not a priori known, nor is it clear whether they generate a hitherto unidentified symmetry of the theory at zero string coupling.

In summary, to fully specify the string spectrum, one must determine, at every mass level, the list of Young tableaux that describe the spin labels of the single-particle states of that mass, as well as the multiplicity of each Young tableau, namely the number of particles in the spectrum with the same mass and transforming in the same representation of the Lorentz group. Furthermore, in order to distinguish among them, one should provide a labeling for the states sharing the same mass and spin quantum numbers. Assuming one needs $N_Q$ additional quantum numbers to label a state, and given that a Young tableau is fully specified by giving the length $s_1,\dots,s_{12}$ of each of its rows, one would need $N_Q + 12+1$ numbers to fully characterize a single-particle state, where the last term corresponds to the mass of the particle. This can be represented in a plot as the one in Fig.~\ref{fig:regge_trajectories}, which corresponds to a 2-dimensional projection of the full $N_Q+13$-dimensional space describing the spectrum.

There are different ways of quantizing the string; each way makes some of its properties manifest, typically at a price. In covariant quantization, one introduces oscillators exciting the string along all 26 space-time directions. This approach, while keeping Lorentz symmetry manifest after quantization, introduces negative-norm states that must be projected out by requiring that physical states satisfy a set of infinitely-many conditions known as the {\it Virasoro constraints}. Imposing these constraints systematically at high levels can be a daunting task, and this is the main reason why the excited string spectrum remained largely unexplored.

Our strategy to avoid the introduction of operators that create states of negative norm is to work in light-cone gauge. In that case, the time direction and one of the spatial directions define a light-front, with the other $24$ spatial directions being treated democratically. The advantage of such a method is that all states created by the action of string creation operators in the vacuum are physical. The disadvantage is that the oscillators transform covariantly only under an $SO(24)$ subgroup of the little group, and therefore various structures of $SO(24)$ oscillators must be combined in order to build a complete $SO(25)$ representation, causing the single-particle states built in this formalism to be made of many parts. We will find it nonetheless useful to work in this quantization, as the number of unknown coefficients to fix is much smaller. The main problem to solve is to identify, among the various structures that are $SO(24)$ covariant and carry the same $26-$momentum, which ones are different polarizations of {\it the same} particle, or---mathematically---branchings of the same $SO(25)$ irrep. To give a concrete example where the reader can contrast covariant and light-cone gauge expressions,~\cite{Sasaki:1985py} works out a representative of the scalar particle at level 4 in covariant gauge, finding
\begin{equation}
    \begin{split}
        &\Bigg\{ \left(\eta_{\mu\nu}+\frac{13}{3}k_\mu k_\nu\right)\alpha_{-2}^\mu \alpha_{-2}^\nu 
    +\left(\frac{20}{9}k_\mu k_\nu k_\rho + \frac{2}{3} k_\mu \eta_{\nu\rho}+\frac{13}{3} k_\rho \eta_{\mu\nu}\right)\alpha_{-1}^\mu\alpha_{-1}^\nu \alpha_{-2}^\rho \\
    &+\left(\frac{23}{81} k_\mu k_\nu k_\rho k_\sigma+\frac{32}{27} k_\mu k_\nu \eta_{\rho\sigma}+ \frac{19}{18}\eta_{\mu\nu}\eta_{\rho\sigma}\right)\alpha_{-1}^\mu\alpha_{-1}^\nu\alpha_{-1}^\rho\alpha_{-1}^\sigma \Bigg\} \ket{0} \; . 
    \end{split}
\end{equation}

The same particle in light-cone gauge quantization can be written more economically as
\begin{equation}
    \{(\alpha_{-1}\cdot\alpha_{-1}))^2+7(\alpha_{-2}\cdot\alpha_{-2})-
    10(\alpha_{-1}\cdot\alpha_{-3}) \} \ket{0}.
\end{equation}

We provide a systematic method to find all single-particle states at a given level in terms of $SO(24)$ oscillators, which corresponds to filling the spectrum by moving vertically in Fig.~\ref{fig:regge_trajectories}, finding all single-particles at constant mass at each given step. This not only gives the particle content at a given level, but also provides an explicit realization in terms of oscillators that can be used to compute their scattering amplitudes. However, while this level-by-level procedure allows us to go pretty far and excavate all single-particle states and their oscillator representations---we went explicitly up to level 15---, the exponential increase in the density of states with the energy calls for other ways of exploring the spectrum, as at large enough levels the problem becomes intractable even with a computer.

In this direction, we also find a way to obtain families of infinitely many states in the spectrum belonging to the same ``Regge trajectory," or rather a generalization of this concept. A natural extension of the idea of a Regge trajectory when one deals with the spectrum is to group states by starting from a seed state (characterized by a given mass, Young tableau and potentially other quantum numbers) that corresponds to the bottom of the trajectory, and then to climb up the trajectory by adding one unit of mass and one unit of spin at each step. Implementing this idea, we define ``Regge trajectories" to be given by states belonging to contiguous levels corresponding to Young tableaux that differ by one box in the first row. For instance, we could consider a trajectory generated by an antisymmetric rank 2 state at the bottom:
$$\underbrace{\ydiagram{1,1}}_{\text{level } N} \to \underbrace{\ydiagram{2,1}}_{\text{level } N+1} \to \underbrace{\ydiagram{3,1}}_{\text{level } N+2}\to \dots  $$

In terms of oscillators, the operation of adding a box to the top row of a diagram when going up one level corresponds to adding an $\alpha_{-1}^i$ oscillator and symmetrizing $i$ with the Lorentz indices corresponding to the top row of the original Young tableau. Note that the operation of adding a box to a given Young tableau makes sense both when considering $SO(24)$ and $SO(25)$ indices, so apart from defining Regge trajectories this way for physical $SO(25)$ particles, we can think of the trajectory generated by a given $SO(24)$ oscillator structure, even if it does not correspond to a single-particle state. However, it was mentioned that a single-particle state is built by a linear combination of many $SO(24)$ oscillator structures, and simply adding an $\alpha_{-1}$ to all of them will generically not yield a single-particle state one level higher. An important result presented here is that we find a way to, for a given seed state, obtain the level dependence of the coefficients determining all single-particle states on its Regge trajectory, thus providing at once the oscillator representation of infinitely many states in the spectrum. In terms of the plot in Fig.~\ref{fig:regge_trajectories}, this allows to explore the spectrum along infinite lines of constant $d\equiv N-s_1$, where we define $d$ as the {\it depth} of the trajectory.

In fact, we can do more than this. When analyzing the string spectrum in terms of light-cone gauge oscillators, we found numerous patterns relating states connected by the addition of oscillators not only on the first row, but also when adding oscillators in various rows simultaneously. Possibly the simplest example of this phenomenon is when one considers the family of states generated by starting from a seed single-particle state and adding simultaneously an $\alpha_{-1}$ oscillator to the first row and an $\alpha_{-2}$ to the second. This would correspond to a ``tilted Regge trajectory,"\footnote{This specific family of states was studied in~\cite{Alessio:2025nzd} in connection to black hole scattering.}  and the particular example in which the seed state is a scalar would look like
$$\bullet \rightarrow \ydiagram{1,1}\rightarrow \dots \rightarrow \ydiagram{3,3}\dots \ydiagram{1,1} \, \dots,$$

We show that the techniques that allowed us to obtain all the single-particle states in a Regge trajectory as a function of the level can be generalized to deal with these types of families of states forming tilted trajectories, and therefore allow us to work out the spectrum along lines of constant $N-\sum_{j=1}^k j s_j$, with the previous case corresponding to the line $N-s_1-2 s_2=0$.

One last relevant concept to chart the spectrum is what we call the {\it generation} of a given state at level $N$, and it refers to the number of levels lower than $N$ that contain a state with the same Young tableau as the state in consideration. This concept gives a measure of the complexity of the state. States at generation zero\footnote{These states are sometimes called Weinberg states, as Weinberg studied them in~\cite{Weinberg:1985tv}, or principally embedded states in~\cite{Markou:2023ffh}.}  correspond to Young tableaux filled with oscillators in a way that the $k$-th row contains $\alpha_{-k}$ oscillators only, and therefore correspond to the lightest state with that given Young tableau in the spectrum. These states are the easiest to deal with: In covariant language they satisfy the Virasoro constraint almost for free, and in light-cone language the matching between $SO(24)$ structures and the single-particle state is automatic, without the need of considering linear combinations of various structures with the same $SO(24)$ Young tableau symmetry. As generation rises, one goes deeper in the spectrum and the oscillator content of single-particle states becomes increasingly complicated. In terms of light-cone technology, the amount of different $SO(24)$ structures building a single state becomes bigger and bigger, and their analysis becomes hard, even if in principle doable at any finite generation. 

At present we do not have a systematic way of moving in generation while keeping the shape of the Young tableau defining the states fixed. What we can do is work out the expression of a given state at non-zero generation in terms of $SO(24)$ oscillators, and then use it as a seed to move along generalized Regge trajectories with finite generation.\footnote{In~\cite{Markou:2023ffh,Basile:2024uxn} the authors work in covariant quantization and propose a way of climbing up the generation for fixed Young tableau by the action of suitably defined raising operators and the posterior projection into the physical Hilbert space. While in this approach one needs to solve a system of equations that becomes increasingly complex as the generation grows, it provides a systematic way of exploring an independent direction in the spectrum and it would be very interesting to investigate if these ideas can be exploited in our language.}

With the setup in mind, in the rest of this section we will recall some basics of quantization of the open string, to establish notation and also highlight some important formulas, and to make the paper self-contained. A more detailed discussion is done in appendix \ref{app:lightcone}. 

\subsection{Review of Light-cone Quantization}

At the classical level, dynamics in string theory are dictated by the Polyakov action
\begin{equation}
\label{eq:PolyakovAction}
S_P = -\frac{1}{4\pi \alpha'} \int d^2 \sigma \, \sqrt{-h} \, h^{\alpha \beta} \partial_\alpha X^\mu \partial_\beta X_\mu,
\end{equation}
with $h_{\alpha\beta}$ being the worldsheet metric, 
and subject to the additional constraint of having a vanishing stress tensor:
\begin{equation}
\label{eq:Tvanishing}
T_{\alpha\beta}=\frac{1}{\sqrt{-h}}\frac{\delta S}{\delta h^{\alpha \beta}}=0.
\end{equation}

The quantization of constrained systems is a delicate issue, as one must deal with the unphysical degrees of freedom corresponding to configurations that do not satisfy the constraints. 

One possibility to do so is to quantize the system ignoring the constraints, and then imposing them at the quantum level as conditions that states in the Hilbert space must satisfy in order to be physical. This is the approach taken in covariant quantization mentioned in the previous subsection, in which the classical requirement~\eqref{eq:Tvanishing} becomes the Virasoro constraints that project negative norm states out of the spectrum. 

Another approach is to solve for the constraints at the classical level to find the physical phase space from the start, and then proceed to quantize these degrees of freedom only. This is what is achieved in light-cone quantization, which has the advantage of allowing to deal directly with the physical Hilbert space, without the need of imposing any supplementary constraints. This comes however with the drawback of spoiling the manifest Lorentz invariance of the theory, as the method requires the choice of a specific reference frame. 

The basic idea of light-cone quantization is to use the gauge freedom in~\eqref{eq:PolyakovAction} to choose a gauge in which the constraints $T_{\alpha \beta}=0$ are easily solved\footnote{The Polyakov action is invariant under reparametrizations of the worldsheet $\sigma^\alpha \to \tilde{\sigma}^\alpha(\sigma)$ and Weyl transformations $h_{\alpha \beta }(\sigma) \to \Omega^2(\sigma)h_{\alpha \beta}(\sigma)$. }. Introducing the spacetime light-cone coordinates as
\begin{equation}
\label{eq:lightconecoords}
X^{\pm}=\frac{X^0 \pm X^{25}}{\sqrt{2}},
\end{equation}
and the open string mode expansion
\begin{equation}
\label{eq:openstring-modeexp}
X^\mu (\sigma,\tau)=x^\mu+2\alpha'p^\mu \tau + i \sqrt{2\alpha'}\sum_{n\neq 0}\frac{1}{n}\alpha^\mu_n \, \cos n\sigma \, e^{-i n \tau},
\end{equation}
the convenient gauge choice consists of setting the worldsheet metric $h_{\alpha \beta}$ to be flat and fixing $X^+(\sigma, \tau)$ to be
\begin{equation}
\label{eq:light-conegauge}
X^+ = x^+ + 2 \alpha' p^+ \tau,
\end{equation}
which defines the {\it light-cone gauge}, trivializing the string excitations along the $+$ direction.

To see the value of this gauge choice, note that in  $\sigma^{\pm}=\tau \pm \sigma$ worldsheet coordinates, the constraints~\eqref{eq:Tvanishing} become
\begin{equation}
\label{eq:VirasoroConstraints}
\partial_+X^\mu \partial_+X_\mu=\partial_- X^\mu  \partial_-X_\mu=0,
\end{equation}
which in light-cone coordinates implies
\begin{equation}
    2 \partial_+ X^- \partial_+ X^+ = \sum_{i=1}^{24}\partial_+ X^i \partial_+ X^i, \quad \\,
    2 \partial_- X^- \partial_- X^+ = \sum_{i=1}^{24}\partial_- X^i \partial_- X^i.
\end{equation}

Since from the gauge choice~\eqref{eq:light-conegauge} we have $\partial_+ X^+ =\partial_- X^+ =\alpha ' p^+$, these become
\begin{equation}
\partial_+ X^-=\frac{1}{2 \alpha' p^+}\sum_{i=1}^{24}\partial_+ X^i \partial_+ X^i, \quad \partial_- X^-=\frac{1}{2 \alpha' p^+}\sum_{i=1}^{24}\partial_- X^i \partial_- X^i,
\end{equation}
showing that $X^-(\sigma^+,\sigma^-)$ is completely fixed by the other fields $X^i$, $X^+$, up to an integration constant. 

In terms of the mode expansion for $X^-$
\begin{equation}
X^-=x^-+2 \alpha' p^- \tau + i \sqrt{2 \alpha '}\sum_{n \neq 0}\frac{1}{n}\alpha_n^- \cos n \sigma \, e^{- i n \tau},
\end{equation}
$x^-$ corresponds to the integration constant, and $p^-$ and $\alpha_n^-$ are fixed by the constraints to be
\begin{equation}
\label{eq:alphaminus}
\alpha_n^-=\frac{1}{2\sqrt{2 \alpha '}p^+}\sum_{m=-\infty}^\infty \sum_{i=1}^{24}\alpha_{n-m}^i \alpha_m^i, \quad p^- = \frac{1}{\sqrt{2\alpha'}}\alpha_0^-,
\end{equation}
and are thus expressed in terms of oscillators made up of indices pointing along the $i=1,\dots,24,$ directions {\it only}. These {\it transverse} directions determine the physical excitations of the string and the oscillators along them correspond to the degrees of freedom that must be quantized in order to build the physical Hilbert space.

As standard, canonically quantizing the system amounts to promoting the Poisson brackets between the Fourier modes of the classical mode expansion by commutators between the quantum oscillators. For the transverse oscillators this gives
\begin{equation}
\label{eq:alphaminus}
[\alpha_n^i,\alpha_m^j]=[\tilde{\alpha}_n^i,\tilde{\alpha}_m^j]=n \delta^{ij}\delta_{n+m,0},
\end{equation}
and for the zero-modes 
\begin{equation}
[x^i,p_j]=i \delta^{ij}, \quad [x^-,p^+]=-i, \quad [x^+, p^-]=-i,
\end{equation}

The Fock space is built by acting on the string vacuum with transverse oscillators\footnote{Oscillators $\alpha_n^i$ with $n<0$ correspond to creation operators, while those with $n>0$ to annihilation operators.}, with generic states being of the form
\begin{equation}
\label{eq:lightconestate}
\alpha_{n_1}^{i_1} \alpha_{n_2}^{i_2}\dots |0;p \rangle, \quad n_1, \, n_2, \dots <0, \quad i_1,i_2,\dots=1, \dots , 24.
\end{equation}
Any such state is physical by construction, but transforms naturally only under an $SO(24)$ subgroup of the full little group: those rotations pointing along the transverse directions. This is the main drawback of working in light-cone quantization: as will be explained in detail in the next section, in order to build single-particle states corresponding to irreducible representations of the full little group $SO(25)$ we will need to combine various states of the form~\eqref{eq:lightconestate}.

To finish this review we provide a table of relevant commutation relations in light-cone gauge which shall be useful along the text. The commutators are written in the form $[r,c]$ with $r$ being the row entry, $c$ being the column entry:

\begin{center}
\label{tab:commutators}
\begin{tabular}{||c| c c c c c c c c||} 
 \hline
 $\left[ r\, , c\, \right]$ & $p^+$ & $p^-$ & $p^j$ & $x^+$ & $x^-$ & $x^j$ & $\alpha_m^j$ & $\alpha_m^-$ \\ [0.5ex] 
 \hline\hline
 $p^+$ & 0 & 0 & 0 & 0 & $i$ & 0 & 0 & 0\\ 
 \hline
 $p^-$ & 0 & 0 & 0 & $i$ & $-i \frac{p^-}{p^+}$ & $-i \frac{p^j}{p^+}$ & $-\frac{1}{2 \alpha' p^+}m \alpha_m^j$ & $-\frac{1}{\alpha' p^+}m \alpha_m^-$\\
 \hline
 $p^i$ & 0 & 0 & 0 & 0 & 0 & $-i \delta_{ij}$ & 0 & 0\\
 \hline
 $x^+$ & 0 & $-i$ & 0 & 0 & 0 & 0 & 0 & 0\\
 \hline
  $x^-$ & $-i$ & $i \frac{p^-}{p^+}$ & 0 & 0 & 0 & 0 & 0 & $i\frac{\alpha_m^-}{p^+} $\\
 \hline
  $x^i$ & 0 & $i \frac{p^i}{p^+}$ & $i \delta_{ij}$ & 0 & 0 & 0 & $i\sqrt{2 \alpha'} \delta_{ij}\delta_m$ & $i\frac{\alpha_m^i}{p^+} $\\
 \hline
 $\alpha_n^i$ & 0 & $\frac{1}{2 \alpha' p^+}n \alpha_n^i$ & 0 & 0 & 0 & $-i\sqrt{2 \alpha'} \delta_{ij}\delta_n$ & $ n \delta_{ij}\delta_{n+m}$ & $\frac{1}{\sqrt{2 \alpha'}p^+} n \alpha_{n+m}^i$\\
 \hline
 $\alpha_n^-$ & 0 & $\frac{1}{\alpha' p^+}n \alpha_n^-$ & 0 & 0 & $-i\frac{\alpha_n^-}{p^+}$ & $-i\frac{\alpha_n^j}{p^+}$ & $-\frac{1}{\sqrt{2 \alpha'}p^+} m \alpha_{n+m}^i$ & $\left[\alpha_n^-,\alpha_m^- \right] $\\ [1ex] 
 \hline
\end{tabular}
\end{center}
with $$\left[\alpha_n^-,\alpha_m^- \right]=\frac{2}{\sqrt{2\alpha'}p^+}(m-n)\alpha_{m+n}^- + \frac{1}{\alpha' (p^+)^2}m(m^2-1)\delta_{m+n}.$$
Note that the modes along the $-$ direction do not commute with the modes along the other directions nor with the center of mass degrees of freedom. This is another price to pay for working in a non-covariant framework.

\clearpage{}%
\clearpage{}%
\newpage

\section{Building a Single Particle State} \label{sec:singleps}

In the previous section we reviewed the method of light-cone quantization and showed how it can be used to remove all the unphysical degrees of freedom from the Hilbert space. However, even if every state created by the resulting transverse oscillators is physical by construction, they generically do not correspond to the single-particle states that we are interested in, but rather different states created by light-cone oscillators must be combined to build a single-particle state. In this section we present a systematic way of obtaining single-particle states from light-cone oscillators at a given level.

Single-particle states are irreducible, unitary representations of the $26-D$ Poincaré group, and are labeled by their mass and spin. The spin labels are organized as irreps of the little group, which for the massive spectrum is %
$SO(25)$\footnote{In arbitrary dimensions, the corresponding little groups are $SO(d-2,1)$ for tachyons, $ISO(d-2)$ for massless particles and $SO(d-1)$ for massive particles, where $SO(d-2,1)$ is the homogeneous Lorentz group with one spatial dimension less, $ISO(d-2)$ corresponds to the isometry group of $(d-2)$-dimensional Euclidean space, and $SO(d-1)$ is simply the rotation group in $(d-1)$ spatial dimensions. In the massless case, the group $ISO(d-2)$ allows representations labeled by a continuous parameter, sometimes referred to as {\it continuous spin representations}. If one demands that the action of the little group generators that correspond to this continuous parameter is trivial, the little group becomes $SO(d-2)$.}. However, the states obtained by exciting the vacuum with light-cone gauge transverse oscillators as in~\eqref{eq:lightconestate} are labeled by indices $i$ that run from $i=1$ to $i=24$, and therefore furnish representations of the group $SO(24)$. This means that a string of light-cone oscillators on their own, such as~\eqref{eq:lightconestate}, does not furnish a complete single-particle state. Instead, various combinations of states created by transverse oscillators must be grouped together to create a given single-particle state\footnote{The only exceptions to this are the tachyon and the massless vector at level 1. Since for massless particles the little group is $SO(24)$, the state created by $\alpha_{-1}^i$ acting on the vacuum has the right number of degrees of freedom to furnish a single-particle on its own.}.

\subsection{Warm-Up Examples}

Let us illustrate the problem with simple examples, which will then pave the way for a general method to build a single-particle state. 

The first example is of a massive spin-2 particle at level 2. It should be described by a rank~2, traceless, symmetric tensor $\mathcal{E}_{\mu \nu}$ that satisfies the transversality condition $\mathcal{E}_{\mu \nu}p^\mu=0$. Choosing the particle to be at rest in the frame that we used to define the light-cone gauge in eq.~\eqref{eq:lightconecoords}, transversality is satisfied by setting to zero all components involving the time direction in this frame. Denoting $X^{\ell}$ the spatial coordinate used to define $X^\pm$\footnote{In eq.~\eqref{eq:lightconecoords} we had $\ell=25$, but here we keep $\ell$ generic, as any spatial direction is equally good for our purposes. }, we see that the non-trivial components of the tensor are $\mathcal{E}_{i j}$, $\mathcal{E}_{i \ell}$, $\mathcal{E}_{\ell \ell}$, where $i,j=1,\dots,24$ correspond to transverse directions.

Now let us consider the states created from light-cone oscillators at level two.  In terms of their transformation properties with respect to $SO(24)$, they are a symmetric-traceless rank 2 tensor, a vector and a scalar, given by
\begin{equation}
\left(\alpha_{-1}^{ i}\alpha_{-1}^{j}-\frac{\delta^{ij}}{24}(\alpha_{-1}\cdot\alpha_{-1}) \right)| 0 \rangle, \quad \alpha_{-2}^{i}| 0 \rangle, \quad (\alpha_{-1}\cdot\alpha_{-1})| 0 \rangle.
\end{equation}
Throughout the remainder of the paper, traces will not be subtracted explicitly from tensors; this operation is unambiguous and understood implicitly.
As anticipated, these different structures combine to furnish the degrees of freedom of a single massive spin-2 $SO(25)$ irreducible representation, respectively $\mathcal{E}_{i j}$, $\mathcal{E}_{i \ell}$, and $\mathcal{E}_{\ell \ell}$. The three structures provide the $324$ polarization degrees of freedom of a massive spin-2 particle as $324=299+24+1$. Therefore, we see that the single-particle state is made up of a combination of different light-cone states. 

This branching rule is efficiently expressed diagrammatically, using Young tableaux. Each representation is encoded by an arrangement of boxes encoding the symmetries of the representation. Each box corresponds to an index of the tensor characterizing the representation, and each (row/column) of boxes corresponds to (symmetrization/antisymmetrization) of the corresponding indices. We review this in detail in appendix~\ref{app:Young_Tableaux}. For the case at hand, the decomposition of the spin-2 particle at level 2 into light-cone states reads
\begin{center}
		\begin{tabular}{lcl}
			$\ydiagramalign{2}_{{}_{25}}$  & $\longrightarrow$ & $\ydiagramalign{2}_{{}_{24}}$\quad$\ydiagramalign{1}_{{}_{24}}$\quad${}^\bullet_{{}_{24}}$
		\end{tabular}
	\end{center}
    
In mathematical terms, this corresponds to the restriction of a representation of the larger group $SO(25)$ into representations of its subgroup $SO(24)$, with the spin-2 representation of $SO(25)$ being decomposed into the direct sum of the spin-2, spin-1 and spin-0 $SO(24)$ representations. In general, the decomposition of the irreducible representations of a group $G$ into irreducible representations of one of its subgroups $H$ are described by the {\it branching rules} from $G$ to $H$. See appendix~\ref{app:Young_Tableaux} and references therein of more details.

Let us now work out the spectrum at level 3. From the light-cone oscillator point of view we have three possible structures acting on the vacuum which can be decomposed into the following $SO(24)$ irreducible representations:
\begin{align}
\alpha^i_{-1}\alpha^j_{-1}\alpha^k_{-1} &\longrightarrow  \, \ydiagramalign{3}_{{}_{24}} \quad \ydiagramalign{1}_{{}_{24}}\\ \alpha^i_{-2}\alpha^j_{-1} &\longrightarrow  \, \ydiagramalign{2}_{{}_{24}} \quad \ydiagramalign{1,1}_{{}_{24}} \quad {}^\bullet_{{}_{24}} \\
		\alpha^i_{-3}&\longrightarrow \ydiagramalign{1}_{{}_{24}},
\end{align}
where in the first row the product of three $\alpha_{-1}$ $SO(24)$ oscillators is decomposed into a symmetric and traceless part  $(\alpha_{-1}^{i}\alpha_{-1}^{j}\alpha_{-1}^{k }-{\rm Trace)}$ and a vector $(\alpha_{-1}\cdot \alpha_{-1}) \alpha_{-1}^{i}$, and in the second row the different combinations are obtained by symmetrizing, antisymmetrizing and tracing over the indices respectively. 

To find out the $SO(25)$ irreducible representations at this level we need the simple but crucial observation that the $SO(24)$ representations with the largest number of indices and a given symmetry pattern must trivially translate to the corresponding $SO(25)$ representation, as there is no larger representation of which they could be a part of. In this case, the rank 3 tensor $\alpha_{-1}^{i}\alpha_{-1}^{j}\alpha_{-1}^{k}$ necessarily implies the existence of a rank 3 symmetric single-particle state of which this oscillator combination gives the fully transverse part. Similarly, the antisymmetric rank 2 tensor $\alpha_{-2}^{[i}\alpha_{-1}^{j]}$ implies the existence of the corresponding $SO(25)$ representation, and we see that the single-particle states
\begin{equation}
\ydiagramalign{3}_{{}_{25}}, \quad \ydiagramalign{1,1}_{{}_{25}}
\end{equation}
are part of the spectrum at level 3. However, we must still determine whether all three remaining $SO(24)$ structures also belong to these two $SO(25)$ representations, or if some of them create a new single-particle state of its own.
For this it is enough to look at the branching rules of $SO(25) \to SO(24)$ for the two particles we already found:

\begin{center}
		\begin{tabular}{lcl}
			$\ydiagramalign{3}_{{}_{25}}$  & $\longrightarrow$ & $\ydiagramalign{3}_{{}_{24}}\quad\ydiagramalign{2}_{{}_{24}}\quad\ydiagramalign{1}_{{}_{24}}\quad \bullet_{{}_{24}}$\\
			\rule{0pt}{7ex}
			$\ydiagramalign{1,1}_{{}_{25}}$& $\longrightarrow$ & $\ydiagramalign{1,1}_{{}_{24}}\quad\ydiagramalign{1}_{{}_{24}}$
		\end{tabular}
	\end{center}
This exhausts all the $SO(24)$ structures, and therefore the single-particle states at level 3 are just the rank 3 fully symmetric and the rank 2 antisymmetric tensors.

This characterizes the spectrum at level 3, but raises another question. If we want to find the combinations of oscillators that give the polarizations of the single-particle states along non-transverse directions as we did in the level 2 example, we need to understand which $SO(24)$ vector is inside each $SO(25)$ irreducible representation.

The way to proceed, which will be at the basis of our general method to obtain the
spectrum at any given level, is to use the fact that $SO(24)$ states belonging to the same
$SO(25)$ irreducible representation must mix exclusively among themselves under little group
transformations\footnote{This observation is also at the basis of Pesando's approach in \cite{Pesando:2024lqa}.}. Therefore, we can start with the $SO(24)$ structures that can be trivially
located inside an $SO(25)$ irrep and rotate them by acting with the
combinations of Lorentz generators spanning the little group to obtain the remaining $SO(24)$
structures. We will refer to these special $SO(24)$ states that map directly to single-particle
states as highest spinning states or highest spinning structures, and they will play an essential role in what follows.~\footnote{The name highest spinning structures makes reference to the fact that they are the $SO(24)$ structures that have the same number of indices as the $SO(25)$ irreducible representation where they are embedded, while all the other components have fewer. It does not necessarily mean that these are the objects with the highest number of indices at a specific level.}\footnote{An alternative approach that we explored consisted in considering a generic state obtained by linearly combining all oscillator structures that share the same $SO(24)$ symmetry, and to compute its $SO(25)$ symmetry by evaluating all $12$ Casimir operators of the $SO(25)$ group on the state. States belonging to irreducible representations must be eigenstates of these Casimirs, and the set of eigenvalues completely determines the $SO(25)$ representation. The question is then turned into an eigenvalue problem. While attempting to develop this approach, we noticed that combining the quadratic Casimir with arguments on the size of representations would often be enough to completely classify oscillator structures at each level, but computations are nevertheless more complex than in the previous approach because the operator employed is quadratic in the Lorentz generators, instead of being linear.}

Using the mode expansion~\eqref{eq:openstring-modeexp} we can express the Lorentz generators in terms of oscillators as
\begin{equation}
\label{eq:J_covariant}
J^{\mu\nu} = x^\mu p^\nu - p^\nu x^\mu +\frac{i}{2}\sum_{n\neq0}\frac{\alpha^\mu_n\alpha^\nu_{-n}-\alpha^\nu_n\alpha^\mu_{-n}}{n},
\end{equation}

In this expression the relevant part is the sum over oscillators, as this is the one that will mix the different $SO(24)$ structures, while the contribution from the center of mass variables must vanish by definition if the transformation belongs to the little group. This point is subtle in light-cone gauge, as the center-of-mass degrees of freedom do not commute with $\alpha^-_n$; nonetheless, one can check that the contribution from the center-of-mass terms drops.

Let us show that the center-of-mass part of the angular momentum operators does not contribute. We will focus on the example of the light-cone state $\alpha_{-3}^i |0 \rangle$, but the computation is analogous in every case.

In the rest frame, $p^+=p^-$, which together with the mass shell condition fixes both of them to be
\begin{equation}
p^+=p^-=\sqrt{\frac{N-1}{2\alpha'}},\quad N = 3.
\end{equation}

Since $p^+$ commutes with the oscillators, the value of $p^+$ of the excited state is the same as for the tachyonic vacuum:
\begin{equation}
		\hat{p}^+\alpha_{-3}^i\ket{0} = p^+\alpha_{-3}^i\ket{0} =
		\alpha_{-3}^i\hat{p}^+\ket{0} = \alpha_{-3}^ip^+_{tac}\ket{0} =
		p^+_{tac}\alpha_{-3}^i\ket{0},
\end{equation}
where we exceptionally use the hat notation to distinguish the operators from the eigenvalues.

Naively one would expect a decoupling between the momentum and the internal degrees of freedom; this is not the case for $p^-$, as exciting internal degrees of freedom must turn the spacelike momentum of the tachyon into a timelike momentum. Indeed, with essentially the same logic,
	\begin{equation}
		\hat{p}^-\alpha_{-3}^i\ket{0} = p^-\alpha_{-3}^i\ket{0} =
		\left(3\frac{\alpha_{-3}^i}{2\alpha'p^+}+\alpha_{-3}^i\hat{p}^-\right)\ket{0}
	\end{equation}
	thus
	\begin{equation}
		\hat{p}^-\ket{0} = p^- - \frac{3}{2\alpha'p^+}\ket{0} = \frac{1}{\sqrt{\alpha'}}(1-\frac{3}{2})\ket{0} = -\frac{1}{2\sqrt{\alpha'}}\ket{0}.
	\end{equation}

Since the particle is at rest, the little group is simply the $SO(25)$ corresponding to spatial rotations generated by the $J^{ij}$ and $J^{\ell i}$, where the $J^{ij}$ are rotations involving purely transverse directions, and $J^{\ell i}$ includes the light-cone spatial direction, denoted $X^{\ell}$. As the light-cone states are by construction representations of $SO(24)$ already, acting with the rotation generators $J^{ij}$  along the transverse directions will not yield new information, and we must instead rotate them using the remaining little group element $J^{ \ell i}$. To express this operator in terms of transverse oscillators, note that the light-cone coordinates are defined by
\begin{equation}
X^{\pm}=\frac{X^0 \pm X^\ell}{\sqrt{2}},
\end{equation}
therefore 
\begin{equation}
		J^{+i} = \frac{1}{\sqrt{2}} (J^{0i}+J^{\ell i}),\qquad
		J^{-i} = \frac{1}{\sqrt{2}} (J^{0i}-J^{\ell i}),
\end{equation}
and therefore we can express $J^{\ell i}$ as
\begin{equation}
J^{\ell i}=\frac{J^{+i}-J^{-i}}{\sqrt2}.
\end{equation}

Now we write these generators in light-cone gauge, where we have effectively set to zero all the oscillators along the $+$ direction, as can be seen from eq.~\eqref{eq:light-conegauge}. A subtlety is that since $p^-$ does not commute with $x^{i}$, to promote the classical expression for the rotation generators in eq.~\eqref{eq:J_covariant} to a hermitian operator we must symmetrize the term $x^i p^- \to (x^i p^- + p^- x^i)/2$. This yields
\begin{equation}
J^{+i} = x^+ p^i - x^i p^+,\qquad
J^{-i} = x^-p^i - \frac{x^ip^-+p^-x^i}{2}+J_{int}^{-i}
\end{equation}
where $J_{int}$ is the part of~(\ref{eq:J_covariant}) involving the oscillator internal degrees of freedom.
	
We are now ready act with $J^{\ell i}$ on $\alpha_{-3}^j\ket{0}$. In the rest frame, the action of the center-of-mass part $J_{CoM}^{\ell i}$ is
	\begin{equation}
	J_{CoM}^{\ell i}\,	\alpha_{-3}^j\ket{0}=\frac{1}{\sqrt{2}}\left(-x^ip^++\frac{x^ip^-+p^-x^i}{2}\right)\alpha_{-3}^j\ket{0},
	\end{equation}
Now $\dfrac{x^ip^-+p^-x^i}{2}\alpha_{-3}^j\ket{0} = \dfrac{x^i(\hat{N}-1)}{2\alpha'p^+}\alpha_{-3}^j\ket{0}$, and since $p^+=\alpha'^{-1/2}$ the contribution from $J_{CoM}$ vanishes as claimed.

We see that the relevant part of the generator $J^{\ell i}$ that corresponds to the excitations of the string is proportional to the corresponding part of $J^{-i}$. Splitting the center-of-mass and the oscillator contributions as $J=J_{CoM}+J_{int}$, we have
\begin{equation}
J^{\ell i}_{int}=-\frac{1}{\sqrt{2}}J^{- i}_{int},
\end{equation}
and therefore we can equivalently act with $J^{- i}_{int}$ on the $SO(24)$ structures to group them into $SO(25)$ representations. Explicitly, $J^{- i}_{int}$ is given by
\begin{equation}
J^{- i}_{int}=\frac{i}{2}\sum_{n\neq0}\frac{\alpha^{-}_n\alpha^i_{-n}-\alpha^{i}_n\alpha^{-}_{-n}}{n},
\end{equation}
whose commutator with a transverse oscillator is
\begin{equation}
		\label{eq:commutator_J-ialphaj}
		[J_{int}^{-i},\alpha^j_l] = 
		-i\delta^{ij}\alpha^-_l-i\frac{l}{\sqrt{2\alpha'}p^+} \sum_{n\neq0} \frac{1}{n}\left(\alpha^i_{-n}\alpha^j_{n+l}-\alpha^i_{n}\alpha^j_{l-n} \right),
	\end{equation}
as can be worked out using the commutators in table~\eqref{tab:commutators}. 

Armed with these commutators, we can simply act with $J^{-i}_{int}$ on the states $\alpha_{-1}^{ i}\alpha_{-1}^{j}\alpha_{-1}^{k } |0 \rangle$\footnote{Strictly speaking we should also account for the term that makes this state traceless. However, one can just choose all the indices to be different and use $\alpha_{-1}^i\alpha_{-1}^j\alpha_{-1}^k$ to determine uniquely the composition of the single-particle states. The terms proportional to Kronecker deltas that ensure we have traceless states can be reconstructed a posteriori if desired, but in practice this is never needed. } and $(\alpha_{-2}^{ i}\alpha_{-1}^{j}-\alpha_{-2}^{ j}\alpha_{-1}^{i} )|0 \rangle$ to find the $SO(24)$ vector belonging to each single-particle state. It turns out that it is a linear combinations of both structures that go into each representation, namely
\begin{align}
		(-2\alpha_{-3}^i& + \alpha_{-1}^i(\alpha_{-1}\cdot\alpha_{-1})) |0 \rangle \in \ydiagramalign{1,1}_{{}_{25}},\\
		(26\alpha_{-3}^i &+ 3 \alpha_{-1}^i(\alpha_{-1}\cdot\alpha_{-1}))|0 \rangle \in \ydiagramalign{3}_{{}_{25}}.
\end{align}
Note that the states are not normalized, as the relevant information is simply the relative weight of the two structures.

Instead of acting on the highest-spinning states $\alpha_{-1}^{ i}\alpha_{-1}^{j}\alpha_{-1}^{k } |0 \rangle$ and $\alpha_{-2}^{[ i}\alpha_{-1}^{j]} |0 \rangle$ with $J^{-i}_{int}$, we could have used a little group element $J^{-m}_{int}$ pointing along a direction $m\neq\{i,j,k\}$. In this case we would have got a vanishing result, as otherwise the resulting $SO(24)$ structure would have had an extra index with respect to the highest-spinning ones, which is impossible by construction. 

The previous example at level 3 is especially simple, because for each $SO(25)$ particle the highest-spinning states can be accessed by symmetrizing some strings of $SO(24)$ oscillators, and non-trivial linear combinations of $SO(24)$ oscillators are only required to describe the remaining polarizations. More generally, even highest-spinning states could be given by linear combinations of such symmetrized products. However, it remains true that combinations of $SO(24)$ oscillators that yield the highest-spinning component of an $SO(25)$ single particle are annihilated by little group elements that involve a new transverse $m$ direction.

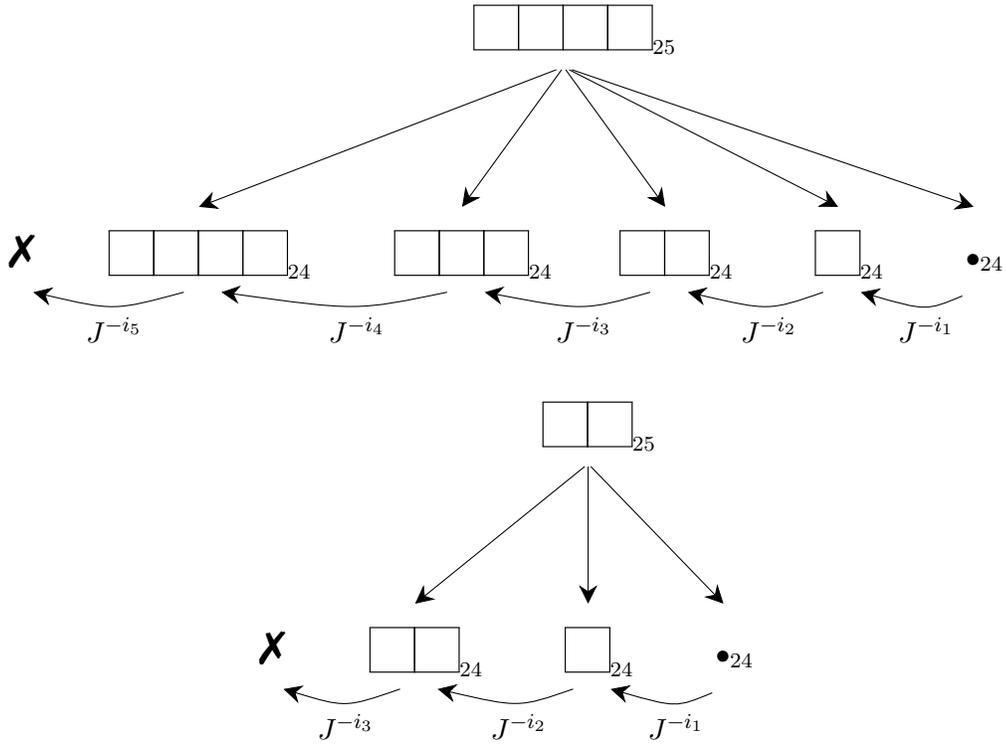
\begin{figure}
\centering

\begin{tikzpicture}[>={Stealth[length=7pt,width=7pt]},baseline=(T25.base)]

\node (T25) at (1.5,0) {$\ydiagram{4}_{25}$};
\node (branchingpoint) at ($(T25)+(-0.15,-0.5)$) {};

\node (Vzero) at (-6,-2.5) {}; %
\node (V4) at ($(Vzero)+(2.5,0.0)$) {};
\node (V3) at ($(V4)+(3.5,0)$) {};
\node (V2) at ($(V3)+(2.7,0)$) {};
\node (V1) at ($(V2)+(2.3,0)$) {};
\node (V0) at ($(V1)+(1.8,0)$) {};

\node (Tzero) at ($(Vzero)+(0.15,-0.45)$) {{\LARGE \ding{55}}}; %
\node (T4) at ($(V4)+(0.15,-0.5)$) {$\ydiagram{4}_{24}$};
\node (T3) at ($(V3)+(0.15,-0.5)$) {$\ydiagram{3}_{24}$};
\node (T2) at ($(V2)+(0.15,-0.5)$) {$\ydiagram{2}_{24}$};
\node (T1) at ($(V1)+(0.15,-0.5)$) {$\ydiagram{1}_{24}$};
\node (T0) at ($(V0)+(0.15,-0.6)$) {$\bullet_{24}$};

\draw[->] ($(branchingpoint)!0.02!(V4)$) -- (V4.north);
\draw[->] ($(branchingpoint)!0.02!(V3)$) -- (V3.north);
\draw[->] ($(branchingpoint)!0.02!(V2)$) -- (V2.north);
\draw[->] ($(branchingpoint)!0.02!(V1)$) -- (V1.north);
\draw[->] ($(branchingpoint)!0.02!(V0)$) -- (V0.north);

\node (JZ4) at ($(Vzero)!0.55!(V4)+(0.0,-1.5)$) {$J^{-{i_5}}$};
\node (J43) at ($(V4)!0.6!(V3)+(0.0,-1.5)$) {$J^{-{i_4}}$};
\node (J32) at ($(V3)!0.6!(V2)+(0.0,-1.5)$) {$J^{-{i_3}}$};
\node (J21) at ($(V2)!0.6!(V1)+(0.0,-1.5)$) {$J^{-{i_2}}$};
\node (J10) at ($(V1)!0.65!(V0)+(0.0,-1.5)$) {$J^{-{i_1}}$};

\draw[->] ($(V4)+(-0.2,-1)$) .. controls ($(JZ4)+(0.0,0.25)$) .. ($(Vzero)+(0.3,-1)$);
\draw[->] ($(V3)+(-0.2,-1)$) .. controls ($(J43)+(0.0,0.25)$) .. ($(V4)+(0.3,-1)$);
\draw[->] ($(V2)+(-0.2,-1)$) .. controls ($(J32)+(0.0,0.25)$) .. ($(V3)+(0.3,-1)$);
\draw[->] ($(V1)+(-0.2,-1)$) .. controls ($(J21)+(0.0,0.25)$) .. ($(V2)+(0.3,-1)$);
\draw[->] ($(V0)+(-0.15,-1.05)$) .. controls ($(J10)+(0.0,0.25)$) .. ($(V1)+(0.3,-1)$);

\end{tikzpicture}

\vspace{0.5cm} %

\begin{tikzpicture}[>={Stealth[length=7pt,width=7pt]},baseline=(T25.base)]

\node (T25) at (1.5,0) {$\ydiagram{2}_{25}$};
\node (branchingpoint) at ($(T25)+(-0.15,-0.5)$) {};

\node (Vzero) at (-3,-2.5) {}; %
\node (V2) at ($(Vzero)+(2.05,0)$) {};
\node (V1) at ($(V2)+(2.3,0)$) {};
\node (V0) at ($(V1)+(1.8,0)$) {};

\node (Tzero) at ($(Vzero)+(0.15,-0.45)$) {{\LARGE \ding{55}}}; %
\node (T2) at ($(V2)+(0.15,-0.5)$) {$\ydiagram{2}_{24}$};
\node (T1) at ($(V1)+(0.15,-0.5)$) {$\ydiagram{1}_{24}$};
\node (T0) at ($(V0)+(0.15,-0.6)$) {$\bullet_{24}$};

\draw[->] ($(branchingpoint)!0.02!(V2)$) -- (V2.north);
\draw[->] ($(branchingpoint)!0.02!(V1)$) -- (V1.north);
\draw[->] ($(branchingpoint)!0.02!(V0)$) -- (V0.north);

\node (JZ2) at ($(Vzero)!0.55!(V2)+(0.0,-1.5)$) {$J^{-{i_3}}$};
\node (J21) at ($(V2)!0.6!(V1)+(0.0,-1.5)$) {$J^{-{i_2}}$};
\node (J10) at ($(V1)!0.65!(V0)+(0.0,-1.5)$) {$J^{-{i_1}}$};

\draw[->] ($(V2)+(-0.2,-1)$) .. controls ($(JZ2)+(0.0,0.25)$) .. ($(Vzero)+(0.3,-1)$);
\draw[->] ($(V1)+(-0.2,-1)$) .. controls ($(J21)+(0.0,0.25)$) .. ($(V2)+(0.3,-1)$);
\draw[->] ($(V0)+(-0.15,-1.05)$) .. controls ($(J10)+(0.0,0.25)$) .. ($(V1)+(0.3,-1)$);

\end{tikzpicture}

\caption{Action of the Lorentz generator $J^{-i}$ on fully symmetric $SO(24)$ structures at level 4. When acting on lower-spinning components of a single-particle state it increases the number of free indices by one, while it annihilates maximally spinning components.}
\label{fig:Jaction}
\end{figure}

For example, at level 4 there is a rank 2, fully symmetric particle in the spectrum, while there are three rank 2 fully symmetric $SO(24)$ structures. As is displayed in Fig.~\ref{fig:Jaction}, acting with $J^{-k}$ on the rank 2 structure sitting inside the rank 4 particle simply moves one polarization direction from the light-cone direction to the transverse direction $k$, producing a rank 3 structure. In contrast, the combination of $SO(24)$ structures that describes the rank 2 particle at rest with polarizations $j,k$ fully in the transverse directions is annihilated when acted upon by $J^{-i}$.

The diagram in Fig.~\ref{fig:Jaction} shows that the oscillator structures that form a given single-particle state behave in a way reminiscent of a highest weight representation, with $J^{-i}$ acting like a raising operator and the highest spinning states being the highest weights annihilated by it. Following this analogy, at a given level we can find the highest spinning $SO(24)$ states that \emph{map directly} to single-particle states by looking at the kernel of $J^{-i}_{int}$. Since these \emph{completely} characterize the spectrum at any given level, this information would be enough if we were only interested in knowing the single-particle content.
If in addition we want to understand how the remaining $SO(24)$ structures accommodate themselves inside the different particles, we proceed as in the example and act on the highest spinning states with $J^{-i}_{int}$, with $i$ coinciding with one of the indices already appearing in the state, which in this analogy plays the role of the lowering operator\footnote{Actually the result is a combination of the $SO(24)$ structure without the $i$ index, and a structure where the index $i$ appears twice. As the latter is already known, it can be disentangled from the former.}.

\subsection{General Method}

Having seen how the $SO(24)$ states organize into single-particle states from a few examples, we proceed to generalize the approach to find the single-particle content at an arbitrary level $N$: 

\begin{itemize}
\item The first step is to find all the possible light-cone oscillator structures that contribute at level $N$. Since the sum of all the Virasoro indices of the oscillators at level $N$ must be $N$, this is simply given by all the partitions of $N$, namely all the different combinations of integers that give $N$ when summed. For example, at level 4 we would have $\{4,\, 3+1,\, 2+2, \, 2+1+1, \, 1+1+1+1 \}$, corresponding to structures with oscillator content $\alpha_{-4}$, $\alpha_{-3}\, \alpha_{-1}$, $(\alpha_{-2})^2$, $\alpha_{-2}\,(\alpha_{-1})^2 $, $(\alpha_{-1})^{4}$.

\item For each structure, consider all possible non-equivalent pairings of oscillators, where two pairings are equivalent if they group oscillators with the same Virasoro index. These pairings will correspond to $SO(24)$ traces in the light-cone structures. For example, the $\alpha_{-2}\,(\alpha_{-1})^2$ structure at level 4 can be paired as $\alpha_{-2}^{i} \, \alpha_{-1} \cdot \alpha_{-1}$, in which corresponds to tracing over the $-1$ oscillators, as $\alpha_{-2}\cdot \alpha_{-1} \, \alpha_{-1}^{i}$, where the trace is taken between a $-1$ and the $-2$ oscillators, and $\alpha_{-2}^{i} \alpha_{-1}^j \alpha_{-1}^{k}$, where no traces are taken. Note that the second trace is counted only once, even if there are two possible $-1$ oscillators. The transverse indices are just dummy variables at this stage, so either option corresponds to the same state.

\item For each pairing, write all possible symmetry structures that can be built up with the remaining free transverse indices, characterized by Young tableaux. Each group of oscillators with the same Virasoro index will form a fully symmetric substructure, and therefore we can obtain all the possible Young tableaux by taking tensor products of fully symmetric, one-row Young tableaux as $$\ydiagramalign{2}\dots \ydiagramalign{1} \otimes \ydiagramalign{2}\dots \ydiagramalign{1} \otimes \dots \ydiagramalign{2}\dots \ydiagramalign{1} . $$
In the previous example, the trivial pairing without traces will produce the Young tableaux given by the tensor product $$\ytableaushortalign{11} \otimes \ytableaushortalign{2}=\ytableaushortalign{112} \quad \ytableaushortalign{11,2} \,,$$
where the numbers inside the Young tableaux show the Virasoro index of the corresponding oscillator.\footnote{Note that while the relevant group for us is $SO(24)$, we can use the simpler tensor product rules of $SU(24)$, as the (anti)symmetrization patterns are the same. The difference is that in orthogonal groups we need to remove the traces from the irreducible representations, but his can be dealt with after the oscillator structures with their given symmetry are obtained.} The set of all Young tableaux obtained this way yields all the $SO(24)$ irreducible representations at level $N$

\item After obtaining all possible Young tableaux with Virasoro indices placed, we must turn these blueprints to actual symmetrized structures of oscillators with transverse Lorentz indices. The algorithm to go from a Young tableau to oscillator structures with the corresponding symmetry consists first in placing Lorentz indices in each box of the given Young tableau. Then, the indices in all rows are symmetrized, and the result is antisymmetrized over the indices in all columns. In the previous example, the $\ytableaushortalign{112}$ tableau corresponds to the oscillator structure $\alpha_{-2}^{i} \alpha_{-1}^j \alpha_{-1}^{k}+\alpha_{-1}^{i} \alpha_{-1}^j \alpha_{-2}^{k}+\alpha_{-1}^{i} \alpha_{-1}^j \alpha_{-2}^{k}$, while $\ytableaushortalign{11,2}$ corresponds to $(\alpha_{-2}^i\alpha_{-1}^j -\alpha_{-1}^i\alpha_{-2}^j) \alpha_{-1}^k$.

\item To find the single-particle content at the required level, we must obtain the highest spinning $SO(24)$ structures that map to single-particle states. This is done by writing an arbitrary linear combination of all the $SO(24)$ structures corresponding to a given Young tableau, and demanding that it is annihilated by the action of $J^{-i}$, where $i$ is different from all the indices of the $SO(24)$ structures. This yields a linear system where the coefficients of the linear combination are the unknowns, and non-trivial solutions correspond to single-particle states furnishing $SO(25)$ representations with the same Young tableau as the $SO(24)$ structures. If no solutions are found, then this specific Young tableau is not present in the spectrum, and all the $SO(24)$ structures with this symmetry must be embedded in a larger $SO(25)$ representation. Performing this procedure for all the $SO(24)$ Young tableaux determined in the previous step, we completely characterize the single-particle content at the required level.

\end{itemize}

If apart from the spectrum one is interested in understanding how the different $SO(24)$ structures that do not correspond to highest spinning structures fill the remaining polarizations in the rest frame, one can proceed to act on the highest spinning structures with $J^{-i}$, with $i$ repeated in the structures, to climb down the ladder and find the different components of the representation. 

The procedure above can be automated; we implemented this algorithm in both \texttt{Mathematica} and Python to obtain the single-particle content and its oscillator representations. 
The method is very efficient and allows to excavate the single-particle states from the spectrum up to levels that were not accessible before.

As an example we present the data at level 4, and we include higher levels in appendix~\ref{app:levelbylevelspectrum}.

The single-particle content at level 4 in terms of Young tableaux together with its decomposition into $SO(24)$ states is the following:
	\begin{center}
		\begin{tabular}{lcl}
			$\ydiagramalign{4}_{{}_{25}}$  & $\longrightarrow$ & $\ydiagramalign{4}_{{}_{24}}\quad\ydiagramalign{3}_{{}_{24}}\quad\ydiagramalign{2}_{{}_{24}}\quad\ydiagramalign{1}_{{}_{24}}\quad\bullet_{{}_{24}}$\\
			\rule{0pt}{7ex}
			$\ydiagramalign{2}_{{}_{25}}$& $\longrightarrow$ & $\ydiagramalign{2}_{{}_{24}}\quad\ydiagramalign{1}_{{}_{24}}\quad\bullet_{{}_{24}}$\\
			\rule{0pt}{7ex}
			$\ydiagramalign{2,1}_{{}_{25}}$& $\longrightarrow$ & $\ydiagramalign{2,1}_{{}_{24}}\quad\ydiagramalign{2}_{{}_{24}}\quad\ydiagramalign{1,1}_{{}_{24}}\quad\ydiagramalign{1}_{{}_{24}}$\\
			\rule{0pt}{7ex}
			$\bullet_{{}_{25}}$ & $\longrightarrow$ & $\bullet_{{}_{24}}$
		\end{tabular}
	\end{center}
It is given by a spin 4 state, namely a fully symmetric, traceless rank 4 tensor, a rank 3 tensor with mixed symmetry, a spin 2 state and a scalar. While the highest spinning $SO(24)$ structures that yield the spin 4 and the mixed symmetry tensor are trivially $\alpha_{-1}^{i}\alpha_{-1}^{j}\alpha_{-1}^{k}\alpha_{-1}^{\ell}$ and $(\alpha_{-2}^i\alpha_{-1}^j -\alpha_{-1}^i\alpha_{-2}^j) \alpha_{-1}^k$ respectively, as they are the only structures having this $SO(24)$ Young tableaux, the spin 2 and the scalar are not obvious. 

It turns out that the linear combination of light-cone oscillators that gives the scalar is
\begin{equation}
		{}^\bullet_{{}_{25}} = \left(((\alpha_{-1}\cdot\alpha_{-1}))^2+7(\alpha_{-2}\cdot\alpha_{-2})-10(\alpha_{-1}\cdot\alpha_{-3})\right) \ket{0},
	\end{equation}
while the $SO(24)$ structures associated to the spin 2 state are
\begin{align}
\label{eq:massivespin2}
		\ydiagramalign{2}_{{}_{24}} &= \left(\alpha_{-1}^i\alpha_{-1}^j(\alpha_{-1}\cdot\alpha_{-1})+7\alpha_{-2}^i\alpha_{-2}^j-4(\alpha_{-1}^i\alpha_{-3}^j+\alpha_{-1}^j\alpha_{-3}^i)\right) \ket{0},\\
		\ydiagramalign{1}_{{}_{24}} &= \left(4\alpha_{-4}^i+4\alpha_{-1}^i(\alpha_{-1}\cdot\alpha_{-2})-9\alpha_{-2}^i(\alpha_{-1}\cdot\alpha_{-1})\right)\ket{0},\\
		{}^\bullet_{{}_{24}} &= \left(8((\alpha_{-1}\cdot\alpha_{-1}))^2-19(\alpha_{-2}\cdot\alpha_{-2})+20(\alpha_{-1}\cdot\alpha_{-3})\right)\ket{0},
	\end{align}
with $\ydiagramalign{2}_{{}_{24}}$ corresponding to the massive spin 2 state at rest and polarized along the $i,j$ directions, $\ydiagramalign{1}_{{}_{24}}$ to the state polarized along the $i, \ell$ directions, with $\ell$ still corresponding to the spatial light-cone component, and ${}^\bullet_{{}_{24}}$ to the state with both indices polarized along $\ell$.

\subsection{Validity of the results in arbitrary frames} 
All our computations were carried in the reference frame in which the massive particles are at rest. Yet, one may wonder if the results therefore are only valid in this specific frame.
The spectrum cannot depend on the reference frame, and therefore the set of Young tableaux characterizing the $SO(25)$ irreducible representations at a given level that we found in the previous section is on solid ground. However, it could be that the specific oscillator structures furnishing the states are frame-dependent, since we computed them in the rest frame.
In this section we bridge this gap and show that the same oscillator structures that define single-particle states in the rest frame also define single-particle states in boosted reference frames. 

To generalize the analysis performed in the rest frame of the massive particle to arbitrary reference frames, we  must determine single-particle states by rotating the oscillator structures using the elements of the little group of the massive particle in this frame. Since the particle is not at rest, its little group will not be given by pure spatial rotations anymore, but it will still correspond to an $SO(25)$ subgroup of the full Lorentz group. Thus, a basis for its algebra can be written as linear combinations of the Lorentz generators $J^{\mu \nu}$, and must have the required dimension $\mathcal{N}=\dfrac{25 \times 24}{2}$.

We can determine the coefficients in this linear combination by requiring that the algebra elements annihilate the momentum $k^\mu$ of the massive particle, and a convenient choice is the following:
\begin{equation}
\label{eq:newlorentzbasis}
\begin{split}
   \tilde{J}^{ij}(k)=  &J^{ij}-\frac{k^j}{k^+} J^{i+}+\frac{k^i}{k^+} J^{j+}, \quad i,j\,\text{ transverse}, \\
   \tilde{J}^{i-}(k)= &J^{i-}-\frac{N}{2 \alpha' (k^+)^2} J^{i+}+\frac{k^i}{k^+} J^{-+},
\end{split}
\end{equation}
where $N$ is the level of the state, $N=\alpha' k^2+1$.

This set of generators has the right dimensionality to span the little group and maps to $\{J^{ij},J^{i-}\}$ in the rest frame\footnote{Strictly speaking, it maps to $\{J^{ij},J^{i\ell}\}$, but since $J^{i\ell}=\frac{J^{i+}-J^{i-}}{\sqrt{2}}$ and $J^{i+}$ acts trivially on the oscillators, $J^{i\ell} \propto J^{i-}$. }. It has the advantage that when used to classify particles, the terms containing $+$ directions do not contribute (as all $\alpha^+_n$ oscillators are required to vanish and the zero modes encoding the center of mass degrees of freedom act trivially, as necessary to annihilate $k^\mu$). Thus, these operators act on the oscillators in exactly the same way as in the rest frame. In particular, the only little group element that has a non-trivial action is $J^{i-}$. Therefore the single-particle states are still classified by it, showing that the oscillator structures that we identified with $SO(25)$ representations in the rest frame also constitute single-particle states when acting on a vacuum state with non-trivial spatial momentum. 

It is important to note that the same oscillator structure, when viewed in the rest frame and in a generic frame, generally corresponds to different physical polarizations of a given single-particle state. In other words, identical oscillator configurations can acquire distinct physical interpretations depending on the frame. This subtlety does not arise for maximally spinning representations for a broad class of center-of-mass momenta—specifically, for momenta aligned along the $\ell$ direction or lying in the transverse plane orthogonal to all $SO(24)$ indices of the oscillator structure. In these cases, one can reach the desired frame by boosting with generators $J^{t\ell},\,J^{ti}\propto J^{+-},\,J^{+i},\,J^{-i}$, all of which leave the oscillator structure acting on the vacuum invariant. (For $J_{+-}$ and $J_{+i}$ this is immediate, and for $J_{-i}$ the action would add an extra index—but this vanishes for maximal spin.) Hence, whenever $J_{int}$ acts trivially, the oscillator structure remains unchanged.

\clearpage{}%
\clearpage{}%
\newpage
\section{Building a Regge Trajectory}\label{sec:regget}

In this section we provide a method to obtain the oscillator representation of infinite families of states at arbitrary level, all of them belonging to a suitable ``Regge trajectory," which we define below. While this approach still does not suffice to provide the full spectrum, it allows to probe interesting infinite subsectors of the particle content at any mass. Therefore, it is a useful complement of the method outlined in the previous section, which exhausts all single-particle states at fixed level, but becomes inefficient as the level grows.

In section~\ref{sec:review} we presented our working definition of a Regge trajectory: it is specified by a seed state. The seed is generated by acting on the vacuum with a set of oscillators carrying a given symmetry that determines the bottom of the trajectory. From the seed, one climbs up the trajectory by adding $\alpha_{-1}$ oscillators corresponding to boxes in the first row of its Young tableau. This adds a unit of mass and a free index per level, which naturally connects with the standard notion of Regge trajectory.

Naively adding a string of $\alpha_{-1}$'s to a seed state is not enough to determine a Regge trajectory. As in the previous section, we must ensure that we obtain single-particle states at every rung of the ladder. Technically, a single-particle state is associated to a linear combination of $SO(24)$ structures, and the coefficients in the linear combination determining different particles on the same trajectory will depend on the level. 

Since the number of possible oscillator combinations grows with the level, it is plausible that the number of terms in the linear combination of a given particle grows as one climbs up its Regge trajectory, hindering the possibility of accessing the high energy states. 

We show that this is not the case: the number of terms making up a single particle is constant along a given trajectory, and most importantly we provide a mechanism to obtain the level dependence of the coefficients in the linear combination, allowing us to write in a single formula all the infinitely many single-particle states in the trajectory.

Each trajectory, comprising infinitely many particles, can then be determined by a finite amount of work. The existence of infinitely many trajectories prevents us from solving for the entire spectrum. In section~\ref{sec:exploring} we will show how infinite families of Regge trajectories, though not spanning the whole spectrum, can similarly be uncovered all at once.

To solve for full Regge trajectories we adapt the method of the previous section. The requirement of an oscillator string being a single-particle state can be expressed as finding the null space of the matrix representing the action of $J^{i-}$ on a suitable basis. The null vectors correspond to the coefficients in the linear combination of $SO(24)$ structures yielding single particles. It turns out that the level dependence of the entries in this matrix are very simple: they depend linearly on the level. This linearity fixes the matrix completely by computing it at two different levels, and in this way the full Regge trajectory can be obtained without much effort.

As in the previous section, we illustrate our approach in examples, before outlining the general strategy. 

\subsection{Some Examples of Trajectories}

In this section we show a couple of examples of construction of Regge trajectories. Several additional examples are shown in appendix~\ref{app:trajectories}.
\vskip 10pt
\noindent {\bf Fully symmetric subleading trajectory}:
to begin, let us revisit the method presented in section~\ref{sec:singleps} and reformulate it to obtain the oscillator content of full Regge trajectories. 

Consider the fully symmetric rank 2 single-particle state at level $N=4$, whose expression in terms of $SO(24)$ oscillators was given in eq.~\eqref{eq:massivespin2}. In particular, its highest-spinning component is 
\begin{equation}
\label{eq:spin2level4}
    \left(7\alpha_{-2}^i\alpha_{-2}^j-4(\alpha_{-1}^i\alpha_{-3}^j+\alpha_{-1}^j\alpha_{-3}^i)+\alpha_{-1}^i\alpha_{-1}^j(\alpha_{-1}\cdot\alpha_{-1})\right) \ket{0}.
\end{equation}

Recall that we obtained the coefficients in~\eqref{eq:spin2level4} by writing the most general linear combination of $SO(24)$ oscillators at level 4, symmetric in 2 indices, and demanding that it was annihilated by the action of the Lorentz generator $J^{-k}$.

We will now rederive this result in a slightly more abstract way by representing the action of the Lorentz generator $J^{-k}$ as a matrix in a suitable basis, and we will solve for the coefficients by finding the null space of this matrix. While at this stage this might sound as a trivial reformulation of the problem, its power will become clear soon.

Since acting with $J^{-k}$ on any of the terms in eq.~\eqref{eq:spin2level4} will produce an oscillator structure with an extra index without changing the level, we can think of $J^{-k}$ as a map between the set of oscillator structures with 2 indices at level 4 to the set of structures with 3 indices at the same level. Focusing on the structures in eq.~\eqref{eq:spin2level4}, the action of $J^{-k}$ is 
\begin{equation}
    \begin{split}
        J^{-k}\left(\alpha_{-2}^i\alpha_{-2}^j\right)&=\frac{2i}{\sqrt{2\alpha'}p^+}2\alpha_{-2}^{(i}\alpha_{-1}^{j)}\alpha_{-1}^k,\\
         J^{-k}\left(\alpha_{-3}^{(i}\alpha_{-1}^{j)}\right)&=\frac{2i}{\sqrt{2\alpha'}p^+}3\alpha_{-2}^{(i}\alpha_{-1}^j\alpha_{-1}^{k)},\\
         J^{-k}\left((\alpha_{-1}\cdot\alpha_{-1})\alpha_{-1}^i\alpha_{-1}^j \right)&=\frac{2i}{\sqrt{2\alpha'}p^+}\left(-14\alpha_{-2}^{(i}\alpha_{-1}^{j)}\alpha_{-1}^k+
    12\alpha_{-2}^{(i}\alpha_{-1}^j\alpha_{-1}^{k)}\right),
    \end{split}
\end{equation}
and thus is a map between the sets
\begin{equation}
    \left\{ \alpha_{-2}^i\alpha_{-2}^j,\quad\alpha_{-3}^{(i}\alpha_{-1}^{j)},\quad(\alpha_{-1}\cdot\alpha_{-1})\alpha_{-1}^i\alpha_{-1}^j\right\} \to \left\{\alpha_{-2}^{(i}\alpha_{-1}^{j)}\alpha_{-1}^k, \quad \alpha_{-2}^{(i}\alpha_{-1}^j\alpha_{-1}^{k)} \right\},
\end{equation}
which can be represented by the matrix
\begin{equation}
\label{eq:Jmatrix_level4}
    J^{-k}\left(N=4,\,\ydiagramalign{2}\right) \equiv \frac{2i}{\sqrt{2\alpha'}p^+}\begin{pmatrix}
        2 & 0 & -14\\
        0 & 3 & 12
    \end{pmatrix}.
\end{equation}

The null space of this matrix will thus encode all the single-particle spin 2 states at level 4. It is easy to check that it is one dimensional and spanned by $(7~~-4~~1)^T$, %
the already known coefficients in eq.~\eqref{eq:spin2level4}.\footnote{The reader might be surprised that we chose as a basis for the image of $J^{ik}$ structures that do not correspond to irreducible $SO(24)$ representations. The reason for this is that the null space of the map $J^{-k}$ is independent of the choice of basis, and in practice it is easier to use the set of structures that appears directly in the image of $J^{-k}$, rather than decomposing them into irreducible parts.}

The reason why this reformulation is useful is the crucial observation that the coefficients of the matrix representing the action of $J^{-k}$ on the oscillator structures associated to the single-particle states on a given trajectory depend {\it linearly} on the level. Thus, it suffices to compute this matrix at two different levels to determine it for the full trajectory, allowing to find its null space---and therefore the linear combination of oscillators building single-particle states---for the whole trajectory at once.

While we postpone the proof of this fact to subsections~\ref{sec:finite_number_structures} and~\ref{sec:level-dependence}, we finish this example by performing the previous analysis at level 5, fixing the matrix as a function of the level and determining the oscillator representation of the full Regge trajectory.

The next state in the Regge trajectory having as a seed the rank 2 fully symmetric state at level 4 is a fully symmetric rank 3 state at level 5. In this case, the action of $J^{-k}$ provides a map between the sets
\begin{equation}
    \left\{ \alpha_{-2}^{(i}\alpha_{-2}^j\alpha_{-1}^{\ell)},\quad\alpha_{-3}^{(i}\alpha_{-1}^{j}\alpha_{-1}^{\ell)},\quad(\alpha_{-1}\cdot\alpha_{-1})\alpha_{-1}^i\alpha_{-1}^j\alpha_{-1}^\ell\right\} \to \left\{\alpha_{-2}^{(i}\alpha_{-1}^j\alpha_{-1}^{\ell)}\alpha_{-1}^k, \quad \alpha_{-2}^{(i}\alpha_{-1}^j\alpha_{-1}^\ell \alpha_{-1}^{k)} \right\},
\end{equation}
which can be represented by the matrix
\begin{equation}
\label{eq:Jmatrix_level4}
    J^{-k}\left(N=5,\,\ydiagramalign{3}\right) \equiv \frac{2i}{\sqrt{2\alpha'}p^+}\begin{pmatrix}
        4 & -3/2 & -15\\
        0 & 9/2 & 13
    \end{pmatrix}.
\end{equation}

Admitting that these matrices depend linearly on $N$, we can use the values at $N=4$ and $N=5$ to solve for the action of $J^{-k}$ at an arbitrary level. The solution is 
\begin{equation}
\label{eq:Jmatrix_level4}
        J^{-k}\left(N,\,\underbrace{\ydiagramalign{2}\dots}_{N-2}\right) \equiv \frac{2i}{\sqrt{2\alpha'}p^+}\begin{pmatrix}
        2N-6 & 6-3N/2 & -10-N\\
        0 & -3+3N/2 & 8+N
    \end{pmatrix},
\end{equation}
and its null space as a function of $N$ is spanned by 
\begin{equation}\frac{1}{6}
    \begin{pmatrix}
    6N+18\\ -2N^2-10N+48\\ 3N^2-15N+18
\end{pmatrix}.
\end{equation}

This yields the oscillator structure associated to any state in the Regge trajectory originated by the spin 2 particle at level 4. At level $N$ it corresponds to a fully symmetric rank $N-2$ state whose highest spinning component is given by
\begin{equation}
\label{eq:subleadsymtrajectory}
\begin{split}
    \frac{1}{6}\bigg((6N & +18)\alpha_{-2}^{(i_1}\alpha_{-2}^{i_2}\dots\alpha_{-1}^{i_{N-2})}+(-2N^2-10N+48)\alpha_{-3}^{(i_1}\alpha_{-1}^{i_2}\dots\alpha_{-1}^{i_{N-2})} \\ & +(3N^2-15N+18)\alpha_{-1}^{(i_1}\alpha_{-1}^{i_2}\dots\alpha_{-1}^{i_{N-2})}(\alpha_{-1}\cdot\alpha_{-1})\bigg) \ket{0}.
\end{split}
\end{equation}

We see now that reformulating the problem of finding the spectrum at a fixed level paid off: Eq.~\eqref{eq:subleadsymtrajectory} condenses the information of infinitely-many particles in the spectrum of arbitrarily high mass, something that a priori was unthinkable in a level-by-level approach.

An important remark about this particular example is that all types of oscillator structures appearing in the linear combination forming the single-particle states were already present in the seed state generating the trajectory. This will generically not be the case, and for a given seed state we might need to go higher in the trajectory before the number of oscillator structures stabilizes. Since the number of oscillator structures determines the size of the matrix associated to $J^{-k}$, we can only carry out this strategy for the full trajectory when we reach the final number of oscillator structures. 

The fact that the number of oscillator structures always reaches a value independent of the level at high enough levels in any Regge trajectory is the other key ingredient---together with the matrix's linear dependence in the level---to apply this technique. Before demonstrating this claim, let us look at the simplest example where the seed state does not possess all the relevant oscillator structures for determining its trajectory.
\vskip 10pt
\noindent{\bf Mixed symmetry subleading trajectory}:
\noindent the simplest example of the phenomenon mentioned above is the Regge trajectory having the antisymmetric rank 2 state at level $N=4$ as a seed. At this level, the only possible oscillator structure with this symmetry is $\alpha_{-3}^{[i}\alpha_{-1}^{j]}$, and therefore the highest spinning component of the associated state is simply
\begin{equation}
\alpha_{-3}^{[i}\alpha_{-1}^{j]}\ket{0}.
\end{equation}

However, at level 5 there exist two possible oscillator structures with the Young symmetry required for the next state in the Regge trajectory, namely $\ydiagramalign{2,1}$, these being $\alpha_{-3}^{[i}\alpha_{-1}^{j]} \alpha_{-1}^k$ and $\alpha_{-2}^{[i}\alpha_{-1}^{j]}\alpha_{-2}^k$, and therefore the single-particle state at level 5 will in principle be composed of a linear combination of both structures.

In this example, these two type of structures are the only ones that will contribute to the rest of the Regge trajectory at any level. Applying the same logic as before, we have that at generic level $N$ the relevant structures are
\begin{equation}
    \underbrace{\alpha_{-3}^{[i}\alpha_{-1}^{j]}\alpha_{-1}^l\dots}_{N-2},\quad
    \underbrace{\alpha_{-2}^{[i}\alpha_{-1}^{j]}(\alpha_{-2}^l\alpha_{-1}^m\dots}_{N-2}+\alpha_{-1}^l\alpha_{-2}^m\dots+\dots)
\end{equation}
which upon the action of $J^{-k}$ get respectively mapped to
\begin{equation}
\label{eq:J_example_startingbasis}
    \frac{6i}{\sqrt{2\alpha'}p^+}\alpha_{-2}^{[i}\alpha_{-1}^{j]}\alpha_{-1}^k\alpha_{-1}^\ell \dots,\quad
    \frac{4(N-4)i}{\sqrt{2\alpha'}p^+}\alpha_{-2}^{[i}\alpha_{-1}^{j]}\alpha_{-1}^k\alpha_{-1}^\ell\dots,
\end{equation}
which shows that in this case the image of the structures forming the single-particle state under $J^{-k}$ is one-dimensional, and thus $J^{-k}$ can be represented as the row vector
\begin{equation}
\label{eq:J_example_trajectory_matrix}
     J^{-k}\left(N,\,\overbrace{\ydiagramalign{2,1}\dots}^{N-3}\right) \equiv  \frac{2i}{\sqrt{2\alpha'}p^+}\begin{pmatrix}
        3, & 2(N-4)
    \end{pmatrix},
\end{equation}
whose null space is spanned by 
\begin{equation}
    \begin{pmatrix} 2N-8\\ -3 \end{pmatrix},
\end{equation}
and therefore the single-particle states in the full Regge trajectory are given by
\begin{equation}
\left((2N-8)(\alpha_{-3}^{[i_1}\alpha_{-1}^{i_2]}\alpha_{-1}^{i_3}\dots\alpha_{-1}^{i_{N-2}})-3(\alpha_{-2}^{[i_1}\alpha_{-1}^{i_2]}(\alpha_{-2}^{i_3}\alpha_{-1}^{i_4}\dots+\alpha_{-1}^{i_3}\alpha_{-2}^{i_4}\dots) \right)\ket{0}, \quad N \geq 5.
\end{equation}

The previous example shows that while the method introduced in this section allows to solve for the coefficients of any Regge trajectory, care must be taken when considering the first levels of the trajectory, as the structure allowing to apply this construction might not have emerged fully yet.\footnote{Note that while for this example we only needed to climb up one step in the trajectory to uncover all types of structures, for more complicated Young tableaux the number of structures in a trajectory can take various levels to stabilize.} 

Having presented these examples, we now demonstrate the two results that we mentioned without proof: the number of structures contributing to the single-particle states in a given trajectory becomes constant at high enough levels; and the matrix associated to $J^{-k}$ on a given trajectory depends only linearly on the level.

\subsection{Finite number of structures}
\label{sec:finite_number_structures}

We want to argue that the number of oscillator structures in any $SO(24)$ Regge trajectory can increase at first, but eventually saturates to a finite value. The construction of tensor structures is done in two steps: first we list the oscillator monomials with the appropriate mass and number of free Lorentz indices, and then we symmetrize them. The outline of the argument is that, to stick to a given Regge trajectory, the number of monomials saturates to a constant, and the number of ways of symmetrizing also saturates.

The first step has a simple proof. We look for all monomials at level $N$ and with number of free Lorentz indices $S$. In a given Regge trajectory, these are linearly related as $d = N-S$ where $d$ is a constant we call \textquoteleft depth\textquoteright, characterizing the trajectory. Let $n_j$ be the number of oscillators $\alpha_j$ that appear in the monomial. Then
\begin{equation}
\label{eq:finite_structures_eqsystem}
    N = \sum_{j=1}^\infty j\, n_j,\quad
    S+2k = \sum_{j=1}^\infty n_j,\quad
    d=N-S
\end{equation}
where $k\ge0$ is the number of contractions of the oscillators. Solving for $n_1$ from the second equation we get
\begin{equation}
\label{eq:finite_structures_finalequation}
    N = S+2k + \sum_{j=2}^\infty (j-1)\, n_j\Rightarrow
    \sum_{j=2}^\infty (j-1)\, n_j = d-2k
\end{equation}
Thus for each $k=0,\dots,\lfloor \frac{d}{2}\rfloor$ we get an integer partition problem with weights $(j-1)$. These equations admit finitely many solutions. $n_1$ remains unconstrained, in agreement with the idea of a Regge trajectory being obtained by allowing the number of $\alpha_1$ oscillators to vary.

Now we select one of these monomials and symmetrize it. As we did when constructing the spectrum level-by-level, neglecting traces we can view the product of $n$ oscillators with the same Virasoro index as a single row Young tableau with $n$ boxes. The full monomial then translates to the tensor product of single row Young tableaux, each with $n_j$ boxes. For example,
\begin{equation}
    \prod_{j=1}^\infty (\alpha_j)^{n_j} \longrightarrow
    \ydiagramalign{4}\,\otimes\, \ydiagramalign{2}\,\otimes\,\ydiagramalign{3}\dots
\end{equation}
having $n_1=4,\,n_2=2,\,n_3=3$ etc. as solutions to the system of equations in \eqref{eq:finite_structures_eqsystem}.
We then have to decompose this product and find the irreducible representations with the correct symmetry for our $SO(24)$ trajectory. To account for traces, for each pair of dotted oscillators, say $\alpha_p\cdot\alpha_q$, we must erase one box from the $p$-th and $q$-th single line Young tableaux before taking the tensor product.

It is convenient to first take the tensor product of all Young tableaux corresponding to Virasoro indices $j\ge2$, because their $n_j$ are level-independent. This step clearly gives finitely many Young tableaux. Considering one of them, the final step is to take the tensor product with the remaining single line Young tableau with $n_1$ indices, whose length depends on the level. This step also generates finitely many tableaux, actually a fixed number for large enough level, because boxes from the $\alpha_1$ tableau can eventually only be added to the top row, else they end up on the same column giving zero because of antisymmetry.

This concludes the proof, which also provides a constructive method for obtaining the desired oscillator structures. As a final remark, let us mention that the reasons why this regularity is not manifest at low levels is that oscillators with large Virasoro index might not be allowed, or that more $\alpha_1$ oscillators might be needed to properly symmetrize the structure.

\subsection{Level dependence of the matrices}
\label{sec:level-dependence}

We now prove that the coefficients in the matrix associated to the action $J^{-i}$ on the oscillator structures in a given Regge trajectory are either constant in the level or scale linearly with it. Throughout, we assume that the oscillator structures are normalized so that the coefficient of each monomial is $\pm 1$. 

The first step is to generalize the choice of bases in which we represent the action of $J^{-i}$ viewed as a linear transformation between the sets of $SO(24)$ oscillator structures with $k$ Lorentz indices to the set of oscillator structures with $k+1$ indices, at fixed level. 

As in the examples from the previous sections, the natural choice for a basis of the domain is the set of oscillator structures carrying the symmetry of the Young tableau defining the trajectory. For the image, however, we saw that it was more convenient to use the sets of structures that do not correspond to $SO(24)$ irreps. 

For any tensor resulting from the application of $J^{-i}$ the following decomposition is always possible:
\begin{equation}
\label{eq:JbasisImage}
    \sum_{k=1}^{N} \alpha^i_{-k} T_{(k)}^{i_1\dots},
\end{equation}
where $T_{(k)}$ are oscillator structures that depend on the original structure we transform and carry its indices. The $T$'s can then be further decomposed into irreducible representations. Thus, our choice of basis for the image if $J^{-i}$ is the set $\{\alpha_{-k} t_{(k)}\}$, with $k$ generic and $t_{(k)}$ any $SO(24)$ trajectory sharing the symmetry of the original one, but at depth decreased by $k$.\footnote{This basis spans a bigger space than what we obtain by applying $J^{-i}$, but this is not a problem.} 

To proceed, we need to characterize the states relevant for climbing along a Regge trajectory, keeping the level as a free parameter. They are built from a fixed product of oscillators defining the trajectory's seed, multiplied by a variable number of $\alpha_{-1}$ oscillators that extend the first row of the Young tableau, with their number scaling proportionally to the level. This product is then to be symmetrized and then antisymmetrized according to the Young tableau under consideration. In particular, symmetrizing the first row of the Young tableau gives rise to a \emph{symmetric tail}, a sum of terms whose number can depend on the level, multiplying the oscillators sitting in the other rows. After antisymmetrizing, one is left with a sum of such terms, each possessing a symmetric tail. As an example, the product $\alpha_{-3}^i\alpha_{-2}^j\alpha_{-1}^k$ can be given the symmetry of the Young tableau $\ydiagramalign{2,1}$ by first symmetrizing $j,k$, and then antisymmetrizing $i,j$, resulting in
\begin{equation}
    \alpha_{-3}^i\alpha_{-2}^j\alpha_{-1}^k\to
    \alpha_{-3}^i\alpha_{-2}^{(j}\alpha_{-1}^{k)}\to
    \alpha_{-3}^i\alpha_{-2}^{(j}\alpha_{-1}^{k)}-\alpha_{-3}^j\alpha_{-2}^{(i}\alpha_{-1}^{k)}
\end{equation}
where the symmetric tails $\alpha_{-2}^{(j}\alpha_{-1}^{k)},\,\alpha_{-2}^{(i}\alpha_{-1}^{k)}$ are clearly displayed, each being the sum of two terms. By increasing the level and the number of boxes in the first row of the Young tableau, we increase the number of $\alpha_{-1}$ oscillators and the length of the symmetric tails.

We will show that the image of $J^{-i}$ acting on this class of states can be decomposed in the basis introduced in eq.~\eqref{eq:JbasisImage}, and that the coefficients in the resulting state are either independent of the level $N$ or scale linearly with it. Choosing the bases of oscillator structures for the domain and image of $J^{-i}$ as explained, this implies that the matrix associated to $J^{-i}$ in a given Regge trajectory depends only linearly in $N$.

Individual oscillators carry no information about the level of the full structure, hence the only level dependence comes from the length of the symmetric tail corresponding to the top row of the Young tableau, specifically those boxes extending beyond the second row that are never antisymmetrized. More precisely, level dependence arises because the symmetric tail is in general a sum of many terms, whose number increases with the level, many of which can collapse to identical expressions when $J^{-i}$ is applied. As a simple example, symmetrizing $\alpha_{-2}$ with $n$ oscillators $\alpha_{-1}$ gives a sum of $n+1$ terms, each of which rotates to a product of $\alpha_{-1}$. These terms sum to give an overall coefficient proportional to $4(n+1)$
\begin{equation}
\label{eq:linear_scaling_example}
\begin{split}
    J^{-i}\alpha_{-2}^j\ket{0} = 4\frac{i}{\sqrt{2\alpha'}p^+}\alpha_{-1}^i\alpha_{-1}^j\ket{0}\,,\\
    J^{-i}\alpha_{-2}^{(j}\alpha_{-1}^{k)}\ket{0} = 8\frac{i}{\sqrt{2\alpha'}p^+}\alpha_{-1}^i\alpha_{-1}^j\alpha_{-1}^k\ket{0}\,,\\
    J^{-i}\alpha_{-2}^{(j}\alpha_{-1}^{k}\alpha_{-1}^{l)}\ket{0} = 12\frac{i}{\sqrt{2\alpha'}p^+}\alpha_{-1}^i\alpha_{-1}^j\alpha_{-1}^k\alpha_{-1}^l\ket{0}\,.
\end{split}
\end{equation}

To derive this level dependence in full generality, we write the original oscillator structure as some level-independent structure $\mathcal{O}$ times the fully symmetric tail $t$ (whose length will depend on the level) which we assume is obtained from the full symmetrization of a set of oscillators containing $n_j$ oscillators $\alpha_{-j}$, for each value of the Virasoro index $j$. Because Young tableaux involve antisymmetrization, the most general case is actually a linear combination of such products $\sum_\lambda \mathcal{O}_\lambda t_\lambda$, but our argument straightforwardly generalizes to this case as the number of terms in this linear combination is determined by the number of antisymmetric indices, which is finite and fixed by the seed state, and in particular cannot scale with $N$. Out of all the $n_j$, only $n_1$ scales linearly with $N$, while the rest are constant and define the type of structure appearing in the Regge trajectory. 

Restricting our attention to a single $\mathcal{O}_\lambda t_\lambda$, we see that $t$ upon symmetrization is the sum of
\begin{equation}
\label{eq:J_theorem_Ni}
    N_i = \frac{(\sum_j n_j)!}{\prod_j n_j!}
\end{equation}
different terms.

The action of $J^{-i}$ can be decomposed as
\begin{equation}
\label{eq:linear_scaling_decomposition_of_J_on_O_and_t}
    J^{-i}(\mathcal{O} \cdot t)\ket{0} = [J^{-i},\mathcal{O}]t\ket{0} + \mathcal{O} [J^{-i},t]\ket{0},
\end{equation}
Let us start analyzing second term. Assuming $\alpha^j_{-m}$ appears in $t$, recall that
\begin{equation}
\label{eq:J_theorem_schematicJ}
    [J^{-i}, \alpha_{-m}^j] \sim \sum_k \alpha^i_{-k} \alpha^j_{-m+k}
\end{equation}
for $i\neq j$ and up to numerical coefficients that will be irrelevant for our purposes. We assume as before that we can choose all Lorentz indices to be different, so only creators are allowed in the sum.
From $[J^{-i},t]\ket{0}$ we can extract the tensor structure multiplying $\alpha^i_{-k}$ for any fixed $k>0$. We see from \eqref{eq:J_theorem_schematicJ} that we are trading one $\alpha^j_{-m}$ for a $\alpha^j_{-m+k}$ ($m$ generic), but the overall sum is still fully symmetric, and actually described by $n'_m = n_m-1$, $n'_{m-k}=n_{m-k}+1$, while the number of the other oscillators is unchanged $n_j'=n_j$, $j\neq m,m-k$. The number of terms in this new sum is
\begin{equation}
\label{eq:J_theorem_Nf}
    N_f = \frac{(\sum_j n'_j)!}{\prod_j n'_j!}
\end{equation}
so that the coefficient in front of the sum representing $[J^{-i},t]\ket{0}$ is, up to a level-independent constant coming from the precise expression for \eqref{eq:J_theorem_schematicJ}, given by the ratio $\frac{N_i}{N_f}$, given by \eqref{eq:J_theorem_Ni},\eqref{eq:J_theorem_Nf}. This ratio represents the number of terms that collapsed onto the same expression, as in \eqref{eq:linear_scaling_example}, and evaluates to
\begin{equation}
    \frac{N_i}{N_f} = \frac{n_{m-k}+1}{n_m}
\end{equation}

The only cases in which this ratio can depend on the level are $m-k=1$ or $m=1$, else it will be level-independent. Actually our argument does not quite apply to $m=1$ because we see from \eqref{eq:J_theorem_schematicJ} that the sum only contains annihilators (or terms proportional to the transverse momentum, here zero), so no terms are actually produced. The other case $m-k=1$ gives a ratio that scales with $n_1$ and therefore with the level $N$, as we wanted to show.

A similar argument applies to the first term $[J^{-i},\mathcal{O}]t\ket{0}$ in \eqref{eq:linear_scaling_decomposition_of_J_on_O_and_t}. The only way this term can contribute something that scales with the level is if it produces annihilators $\alpha_{>0}$ that collapse many terms in $t$ into a single one, as $J^{-i}$ was doing in the second term of \eqref{eq:linear_scaling_decomposition_of_J_on_O_and_t}. Actually, if $\mathcal{O}$ does not contain any dotted Lorentz indices then the annihilators will have no Lorentz indices in common with $t$, and therefore will commute with it giving no contribution. The only important case is if $\mathcal{O}$ contains a dot product $\alpha_{-n}\cdot\alpha_{-m}$, in which case schematically
\begin{equation}
    [J^{-i},\alpha_{-n}\cdot\alpha_{-m}] \supset \alpha^i_{-k}(\alpha_{k-n}\cdot\alpha_{-m}),\,\alpha^i_{-k}(\alpha_{k-m}\cdot\alpha_{-n})
\end{equation}
where $\alpha_{k-n},\,\alpha_{k-m}$ can be annihilators and not commute with $t$. Their action on $t$ is then trading an oscillator $\alpha^j_{n-k},\,\alpha^j_{m-k}$ for an oscillator $\alpha^j_{-m},\,\alpha^j_{-n}$. The counting arguments provided before can then be repeated, showing once more linear scaling.
\clearpage{}%
\clearpage{}%
\section{Exploring the Spectrum}\label{sec:exploring}

In this section we present additional results which allow for further exploration of the string spectrum in an efficient and systematic way. 

First we describe a method to obtain the single-particle states at an arbitrary level and fixed depth based solely on group-theoretical considerations. We demonstrate the procedure with the single-particle content at depth 3. 
This method does {\it not} provide the oscillator realizations of the single-particle states and therefore is not suitable for amplitude computations. Nonetheless, it provides the most direct way of obtaining the single-particle content of the theory depth by depth. While carrying out this analysis, we observe surprising relations that would be interesting to prove.

Next, we generalize the framework of the previous section to obtain the oscillator representation of other types of Regge trajectory. They are parametrized not just by the length of the first row of the corresponding Young tableau, but rather by the length of arbitrarily many rows. This method hints at a connection among families of states beyond the notion of Regge trajectories; it is very reminiscent of the observations in~\cite{Markou:2023ffh,Basile:2024uxn} when solving for the Virasoro constraints in the covariant formalism. 

We have not studied these {\it generalized Regge trajectories} in full generality, as in section \eqref{sec:regget}. However we observed that the construction works in various examples. We show in detail the case where the lengths $s_1$ and $s_2$ of the first and second rows in the Young tableau associated to the states are variable, obtaining a closed form formula for all the states in the generalized trajectory as functions of $s_1$ and $s_2$. Similarly, we study $L-$shaped trajectories with first column of variable length.

\subsection{Listing single-particle states depth by depth}

Group theory can take us a long way toward understanding the physical spectrum, at least at the level of counting particles, if we do not require explicit expressions in terms of oscillators for the single-particle states. In this subsection we present a method based on simple Young tableau manipulations to recursively compute the single-particle content at a given level in the first place, and then to the single-particle content of entire $SO(24)$ and $SO(25)$ (physical) Regge trajectories.\footnote{While there exist partition functions that allow to count the number of states at a given mass (see for instance~\cite{Hanany:2010da,Lust:2012zv}, with our approach it is possible to specialize to the counting of states with specific symmetry properties only.} 

For the version of the method at fixed level $N$, our strategy will be to first list all oscillator structures at that level. This step does involve oscillators, but only to determine the $SO(24)$ group theoretic properties of the level $N$ states: Every other step will never involve oscillators explicitly. We will be able to immediately identify the $SO(24)$ structures that correspond to highest spinning representatives of $SO(25)$ representations, thus discovering part of the physical spectrum. We will then use branching rules $SO(25)\to SO(24)$ to recursively understand how many $SO(24)$ representations are particular polarizations of the known physical particles, declaring that the remaining ones must be highest spinning representatives of new particles.\footnote{To recall the notion of highest spinning representative, see the beginning of section~\eqref{sec:singleps}.}

This strategy extends in a straightforward way to all the states at arbitrary level and fixed depth, allowing to find the single-particle states at depth $d$ assuming that the ones at depth $d-1$ are already known. Notice that this allows us to recursively obtain all the states in full Regge trajectories at once, as all states in a given trajectory have the same depth.

Both procedures are best illustrated through explicit examples, which we provide below. We summarize our results in table~\ref{tab:trajectories_depth}, where we list all $SO(24)$ and $SO(25)$ Regge trajectories with their multiplicities for various values of $d$.

\subsubsection{The Spectrum at Fixed Level}

In this subsection we present the method to determine the single-particle content at a fixed level by working out explicitly the spectrum of physical particles at level $5$ as an example, only using group theory and without ever applying $J^{-i}$.

The basic building block for any state at a given level is the list of all possible monomials of oscillators with the right mass, which at level 5 can be written schematically as 
\begin{equation}
\label{eq:grouptheory_fixedlevel_monomials}
    (\alpha_{-1})^5,\;\;
    (\alpha_{-1})^3\alpha_{-2},\;\;
    \alpha_{-1}(\alpha_{-2})^2,\;\;
    (\alpha_{-1})^2\alpha_{-3},\;\;
    \alpha_{-2}\alpha_{-3},\;\;
    \alpha_{-1}\alpha_{-4},\;\;
    \alpha_{-5},
\end{equation}
where we omitted Lorentz indices in the oscillators, as they can be either free or contracted among pairs of oscillators depending on the Young symmetry of the required state. 

As before, the first step is to classify all possible $SO(24)$ irreps that can be built from each of these monomials. To do so in a systematic way note that each monomial is given by products of factors of the type $(\alpha_k)^n$, built of $n$ copies of the same oscillator. 

From a group theory perspective $(\alpha_k)^n$ is necessarily a fully symmetric $SO(24)$ tensor, so we can decompose it into irreducible components by removing the various possible traces among its indices, which is equivalent to specifying the number of contractions among pairs of oscillators in $(\alpha_k)^n$. In Young tableau notation this is represented by
\begin{equation}
    (\alpha_k)^n \longrightarrow \overbrace{\ydiagramalign{5}}^{n}\quad\overbrace{\ydiagramalign{3}}^{n-2}\quad\dots,
\end{equation}
where the process goes all the way down to either a vector or a scalar depending on the parity of the initial number of oscillators of type $k$. In our current example the structure $(\alpha_{-1})^5$ is of this kind, and it decomposes into the following $SO(24)$ irreps:
\begin{equation}
   (\alpha_{-1})^5 \longrightarrow \alpha_{-1}^{i_1}\alpha_{-1}^{i_2}\alpha_{-1}^{i_3}\alpha_{-1}^{i_4}\alpha_{-1}^{i_5}, \quad \left(\alpha_{-1}\cdot\alpha_{-1}\right)\alpha_{-1}^{i_1}\alpha_{-1}^{i_2}\alpha_{-1}^{i_3}, \quad \left(\alpha_{-1}\cdot\alpha_{-1}\right)^2\alpha_{-1}^{i_1}.
\end{equation}

To decompose generic structures into irreducible $SO(24)$ components we simply need to consider the tensor product of the fully symmetric objects coming from the factors corresponding to a single oscillator. For example, to decompose the structure $(\alpha_{-1})^3\alpha_{-2} $ we must consider the tensor product between the decomposition of $(\alpha_{-1})^3$ into irreducible $SO(24)$ components and a single vector corresponding to $\alpha_{-2}$:

\begin{equation}
\label{eq:YTtensorproduct}
\begin{split}
    &\hspace{1.5cm} (\alpha_{-1})^3\alpha_{-2} \longrightarrow \left(\ydiagramalign{3}\oplus \ydiagramalign{1}\right) \otimes \ydiagramalign{1} = \\
    =& \left(\ydiagramalign{4}\oplus \ydiagramalign{3,1}\oplus\ydiagramalign{2}\right) \oplus \left(\ydiagramalign{2}\oplus\ydiagramalign{1,1}\oplus \bullet \right).
\end{split}
\end{equation}
Note that in the second line we used the fact that the tensor product of $SO(N)$ tensors differs from the usual one for $GL(N)$ in that one must consider possibly index contractions.

Repeating this procedure for all products in \eqref{eq:grouptheory_fixedlevel_monomials}, we obtain the list of all $SO(24)$ irreps at level 5:
\begin{align}
\label{eq:grouptheory_fixedlevel_SO24tensors}
    \ydiagramalign{5}\quad\ydiagramalign{4}\quad\ydiagramalign{3}(\times 3)\quad\ydiagramalign{2}(\times 4)\\\
    \ydiagramalign{1}(\times 6) \quad\bullet(\times 3)\quad
    \ydiagramalign{3,1}\quad\ydiagramalign{2,1}(\times 2)\quad\ydiagramalign{1,1}(\times 3)
\end{align}

As was explained in section~\eqref{sec:singleps}, at each level there exist $SO(24)$ tensors which immediately imply the existence of their $SO(25)$ analog in the spectrum, being these the simplest examples of highest spinning structures. In the current example, the $SO(24)$ tensors $\ydiagramalign{5}$ and $\ydiagramalign{3,1}$ determine the presence of identical $SO(25)$ representations in the physical spectrum. 

To determine which of the remaining $SO(24)$ structures furnish new single-particle states we must subtract the ones that correspond to components of these $SO(25)$ irreps from the list~\eqref{eq:grouptheory_fixedlevel_SO24tensors}. The branching rules from $SO(25)$ to $SO(24)$ for these representations are
\begin{align}
    \ydiagramalign{5}  & \longrightarrow & \ydiagramalign{5}\quad\ydiagramalign{4}\quad\ydiagramalign{3}\quad\ydiagramalign{2}\quad\ydiagramalign{1}\quad {}^\bullet\\
    \ydiagramalign{3,1}& \longrightarrow & \ydiagramalign{3,1}\quad\ydiagramalign{2,1}\quad\ydiagramalign{3}\quad\ydiagramalign{2}\quad\ydiagramalign{1,1}\quad\ydiagramalign{1}
\end{align}
and therefore we can account for the presence of many $SO(24)$ representations in \eqref{eq:grouptheory_fixedlevel_SO24tensors}.

The remaining $SO(24)$ irreps
\begin{equation}
    \ydiagramalign{3}\quad\ydiagramalign{2}(\times 2)\quad \ydiagramalign{1}(\times 4) \quad\bullet(\times 2)\quad \ydiagramalign{2,1}\quad\ydiagramalign{1,1}(\times 2)
\end{equation}
Analogously to the $\ydiagramalign{5}$ and $\ydiagramalign{3,1}$ irreps, we can deduce the presence of the $SO(25)$ representations $\ydiagramalign{3}$ and $\ydiagramalign{2,1}$. Indeed, once the $SO(24)$ structures associated with components of the former representations are removed, no higher-rank objects remain to which the latter could belong as subcomponents. 

We can now iterate the procedure, removing from the list of $SO(24)$ irreps the ones corresponding to the branching of $\ydiagramalign{3}$ and $\ydiagramalign{2,1}$ from $SO(25)$ to $SO(24)$. As before, the highest spinning objects that remain necessarily correspond to their analogous $SO(25)$ irreps, whose $SO(24)$ components can then be subtracted. Repeating this process systematically exhausts the list, yielding the single-particle content at level 5:
\begin{equation}
    \ydiagramalign{5}\quad
    \ydiagramalign{3}\quad
    \ydiagramalign{3,1}\quad
    \ydiagramalign{2,1}\quad
    \ydiagramalign{1,1}\quad
    \ydiagramalign{1}
\end{equation}

To summarize, the procedure to obtain the spectrum at a given level consists of the following steps:
\begin{enumerate}
    \item List all possible $SO(24)$ oscillator structures with the right mass, without any symmetry.
    \item Obtain all $SO(24)$ irreps by considering the tensor product of the fully symmetric $SO(24)$ irreps corresponding to each kind of oscillator.
    \item From the list of $SO(24)$ irreps not yet identified with a given particle, identify the highest spinning ones that can be lifted directly to their $SO(25)$ analog. Add these states to the list of single-particle states, and use their $SO(25) \to SO(24)$ branching rules to remove all their components from the list.
    \item Iterate the last step until there are no more unclassified $SO(24)$ irreps. Once the process is finished, the obtained single-particle states yield the full spectrum at this given level. 
\end{enumerate}

\subsubsection{Working out the spectrum at fixed depth}

We now show how the method from the previous subsection can be generalized to go beyond working at fixed levels, and use it to obtain the single-particle spectrum at fixed depth $d$, namely all the particles at arbitrary level $N$ satisfying $N-s_1=d$. To better illustrate the procedure, we obtain as an example the contribution of depth $d=3$ single-particle states to the spectrum.

In section~\eqref{sec:finite_number_structures} we showed that the possible oscillator structures at fixed depth can be found by solving \eqref{eq:finite_structures_finalequation} for arbitrary $N$. In the case of depth $d=3$, the resulting structures are
\begin{equation}
\begin{split}
    \left(\alpha_{-4}^{i_1} \right)&\alpha_{-1}^{i_2}\dots\alpha_{-1}^{i_{N-3}},\;\;
    \left(\alpha_{-3}^{i_1}\alpha_{-2}^{i_2} \right)\alpha_{-1}^{i_3}\dots\alpha_{-1}^{i_{N-3}},\;\;
    \left(\alpha_{-2}^{i_1}\alpha_{-2}^{i_2}\alpha_{-2}^{i_3}\right)\alpha_{-1}^{i_4}\dots\alpha_{-1}^{i_{N-3}},\\
    &\left(\alpha_{-2} \cdot \alpha_{-1}\right) \alpha_{-1}^{i_1}\dots\alpha_{-1}^{i_{N-3}},\;\;
    \left((\alpha_{-1}\cdot\alpha_{-1})\alpha_{-2}^{i_1}\right)\alpha_{-1}^{i_2}\dots\alpha_{-1}^{i_{N-3}},
\end{split}
\end{equation}
where we used parentheses to separate the seed structures from the additional $\alpha_{-1}$ oscillators with free Lorentz indices that must dress them so that they have the required mass to belong to the level $N$. Note also that the first level for which all possible structures at depth 3 are present is $N=6$.

Using the techniques from the previous subsection we can decompose these products into irreducible $SO(24)$ tensor structures. The resulting tableaux with their total multiplicity are:
\begin{align}
\label{eq:grouptheory_fixeddepth_SO24trajectories}
    &(A): \overbrace{\ydiagramalign{3}\dots\ydiagramalign{1}}^{N-3}(\times 5),\;\;
    (B): \overbrace{\ydiagramalign{3,1}\dots\ydiagramalign{1}}^{N-4}(\times 5),\;\;
    (C): \overbrace{\ydiagramalign{3,2}\dots\ydiagramalign{1}}^{N-5}(\times 2),\\
    &\hspace{2cm}(D): \overbrace{\ydiagramalign{4,3}\dots\ydiagramalign{1}}^{N-6}(\times 1),\;\;
    (E): \overbrace{\ydiagramalign{3,1,1}\dots \ydiagramalign{1}}^{N-5}(\times 1)
\end{align}
where the overbrace indicates the number of boxes in the top row as a function of the level $N$. In particular, each $SO(24)$ structure produces the following $SO(24)$ irreps:
\begin{equation}
\begin{split}
    &\left(\alpha_{-4}^{i_1} \right)\alpha_{-1}^{i_2}\dots\alpha_{-1}^{i_{N-3}} \longrightarrow \left\{A,B \right\},\\
    &\left(\alpha_{-3}^{i_1}\alpha_{-2}^{i_2} \right)\alpha_{-1}^{i_3}\dots\alpha_{-1}^{i_{N-3}}\longrightarrow \left\{A,B\times2,C,E \right\},\\
    &\left(\alpha_{-2}^{i_1}\alpha_{-2}^{i_2}\alpha_{-2}^{i_3}\right)\alpha_{-1}^{i_4}\dots\alpha_{-1}^{i_{N-3}} \longrightarrow \left\{A,B,C,D \right\},\\
    &\left(\alpha_{-2} \cdot \alpha_{-1}\right) \alpha_{-1}^{i_1}\dots\alpha_{-1}^{i_{N-3}}\longrightarrow \left\{A \right\},\\
    &\left((\alpha_{-1}\cdot\alpha_{-1})\alpha_{-2}^{i_1}\right)\alpha_{-1}^{i_2}\dots\alpha_{-1}^{i_{N-3}}\longrightarrow \left\{A,B \right\}.
\end{split}
\end{equation}

As before we determine the spectrum iteratively, obtaining the spectrum at depth $d$ from the spectrum at depth $d-1$. For the example at hand, we require the spectrum of depth 2 states at level $N$. While we could have worked it out as the chosen example for the method, starting all the way from the depth 0 states consisting of the leading Regge trajectory, we choose to give it as an input and determine the spectrum at depth 3 instead, as the examples for physical $SO(25)$ irreps at $d=0,1,2$ are somewhat trivial.

The spectrum of physical single-particle states at depth $d=2$ is given by
\begin{equation}
\label{eq:grouptheory_fixeddepth_d2}
\begin{split}
    & \hspace{2cm}(a): \overbrace{\ydiagramalign{2}\dots \ydiagramalign{1}}^{N}\;\;
    (b): \overbrace{\ydiagramalign{2,1}\dots \ydiagramalign{1}}^{N-2}\;\;\\
    &(c): \overbrace{\ydiagramalign{2}\dots\ydiagramalign{1}}^{N-2}\;\;
    (d): \overbrace{\ydiagramalign{2,1}\dots \ydiagramalign{1}}^{N-3}\;\;
    (e): \overbrace{\ydiagramalign{3,2}\dots\ydiagramalign{1}}^{N-4}  
\end{split}
\end{equation}

Starting from the knowledge of the $d-2$ single-particle states, we proceed with the same logic as in the fixed level case from the previous subsection. For example, the first four $SO(25)$ tableaux $(a,b,c,d)$ in \eqref{eq:grouptheory_fixeddepth_d2} all give the $(A)$ tableau in \eqref{eq:grouptheory_fixeddepth_SO24trajectories}, leaving only one of the $(A)$ tableaux as a new $SO(25)$ representation. Similarly, $(b,d,e)$ all branch into $(B)$, leaving only two physical $(B)$. Applying the same logic repeatedly, we discover all $d=3$ single-particle states, with their given multiplicity. This procedure can be iterated to increase the depth and uncover deeper single-particle states in the spectrum. In table~\eqref{tab:trajectories_depth} we provide the full spectrum at arbitrary $N$ for depths $d=0,\dots,4$.

Notice that, when determining if a given $SO(24)$ trajectory is physical, we only need to check the part of the physical spectrum where tableaux have more boxes, with equal level. These, by definition, are particles at lower depth, meaning we only need to know the physical spectrum up to depth $d-1$ in order to determine it at depth $d$. This is one of the reasons why exploring the spectrum at a fixed depth is useful: It provides a systematic way of accessing states deeper in the spectrum.

\begin{table}
    \centering
    \begin{tabular}{cc}
        $d$ & Multiplicity of trajectories $(SO(24),\,SO(25))$ \\
        \noalign{\vskip 4pt}
        \hline
        \noalign{\vskip 4pt}
        0 & $\overbrace{\ydiagramalign{3}}^{N}~(1,\,1)$ \\
        \noalign{\vskip 4pt}
        \hline
        \noalign{\vskip 4pt}
        1 & $\overbrace{\ydiagramalign{3}}^{N-1}~(1,\,0)$, $\overbrace{\ydiagramalign{3,1}}^{N-1}~(1,\,1)$ \\
        \noalign{\vskip 4pt}
        \hline
        \noalign{\vskip 4pt}
        2 & $\overbrace{\ydiagramalign{3}}^{N-2}~(3,\,1)$, $\overbrace{\ydiagramalign{3,1}}^{N-3}~(2,\,1)$, $\overbrace{\ydiagramalign{4,2}}^{N-4}~(1,\,1)$ \\
        \noalign{\vskip 4pt}
        \hline
        \noalign{\vskip 4pt}
        3 & $\overbrace{\ydiagramalign{3}}^{N-3}~(5,\,1)$, $\overbrace{\ydiagramalign{3}}^{N-4}~(4,\,1)$, $\overbrace{\ydiagramalign{4,2}}^{N-5}~(2,\,1)$, $\overbrace{\ydiagramalign{4,3}}^{N-6}~(1,\,1)$, $\overbrace{\ydiagramalign{3,1,1}}^{N-5}~(1,\,1)$ \\
        \noalign{\vskip 4pt}
        \hline
        \noalign{\vskip 4pt}
        4 & $\overbrace{\ydiagramalign{3}}^{N-4}~(11,\,3)$, $\overbrace{\ydiagramalign{3,1}}^{N-5}~(10,\,3)$, $\overbrace{\ydiagramalign{4,2}}^{N-6}~(6,\,3)$, $\overbrace{\ydiagramalign{4,3}}^{N-7}~(2,\,1)$,\\
        \noalign{\vskip 6pt}
        &$\overbrace{\ydiagramalign{5,4}}^{N-8}~(1,\,1)$, $\overbrace{\ydiagramalign{2,1,1}}^{N-6}~(2,\,1)$, $\overbrace{\ydiagramalign{3,2,1}}^{N-7}~(1,\,1)$ \\
        \noalign{\vskip 4pt}
        \hline
    \end{tabular}
    \caption{For each depth $d$ up to $4$, we list all $SO(24)$ and $SO(25)$ trajectories indicating how many copies of each trajectory appear in brackets.}
    \label{tab:trajectories_depth}
\end{table}

\subsection{Relating different trajectories}

Looking at table \eqref{tab:trajectories_depth}, where we list $SO(24)$ and physical $SO(25)$ trajectories for small values of depth, we notice some patterns. The most obvious one is that if a $SO(24)$ Young tableau appears for the first time at some depth, it does so with multiplicity one, and since it has never appeared at lower depth that structure also corresponds to a physical particle. This corresponds to the fact that given a Young tableau we can construct an oscillator structure by replacing each box at row $k$ with an oscillator $\alpha_k$, all free Lorentz indices, and then antisymmetrizing the columns as prescribed by the Young tableau. This oscillator structure obviously respects the symmetry of the given Young tableau, but it also realizes the symmetry at the lowest possible level allowed in the string spectrum, implying that the trajectory cannot appear at lower depth. These states are sometimes called \textquoteleft Weinberg states\textquoteright, because Weinberg investigated them in~\cite{Weinberg:1985tv}. In our language, these trajectories belong to generation $G=0$, and are the simplest possible states to construct in our formalism.\footnote{These states are referred to as {\it principally embedded states} in~\cite{Markou:2023ffh}, and are also the simplest ones to deal with in their construction.} The generation of a particle is in general defined as $G=N-\sum\limits_{j} j\, s_j$, where $N$ is the level of the particle and $s_j$ is the length of the $j-$th row of the Young tableau. Intuitively, $G$ measures the number of times the same Young tableau has already appeared in the spectrum at lower levels, because $G$ scales linearly with $N$ once we fix all the $\{s_j\}$ and is zero for Weinberg states.

\subsubsection{Adding boxes to the second row}
\label{sec:adding_boxes_second_row}

One more easily noticeable pattern is that we can observe families of states at different depth that have to the same number of $SO(24)$ structures and correspond to a single $SO(25)$ trajectory. For example, the $L-$shaped representation at depth $d=2$ and the representations at depth $d=2+k$ obtained from the one at depth $d=2$ by removing $2k$ boxes from the top row of the Young tableau, and adding instead $k$ boxes to the second row (at fixed level $N$) all share the same $SO(24)$ multiplicities and correspond to a single $SO(25)$ trajectory, albeit different ones. We will now make some statements about this family of Regge trajectories.

A closed formula for the shape of the tableaux is $\overbrace{\underbrace{\ydiagramalign{5,3}\hspace{-31pt}}_{d-1}\hspace{31pt}}^{N+1-2d}$. The depth parametrizes which trajectory of the family we are discussing, but crucially all of them sit at generation $G=N-(s_1+2s_2)\equiv1$. This possibility of trading two boxes in the top row for one in the second is simple to understand at generation $G=0$ in the string spectrum, where we can easily see the relation as coming from trading two $\alpha_{-1}$ oscillators for one $\alpha_{-2}$, an operation that takes Weinberg states to Weinberg states at higher depth. However, at non-zero generation the explicit relation among the oscillators realizing these trajectories might be more involved than this simple replacement.

Even more strikingly, similar relations hold for more complicated trajectories. In order to test this idea of trading two boxes in the top row of the representation for a single box in the second row, we restrict our attention to representations consisting of only two rows, which can be generically parametrized as $\overbrace{\underbrace{\ydiagramalign{5,3}\hspace{-31pt}}_{d-G}\hspace{31pt}}^{N+G-2d}$, $d\ge G$, where $d$ is the depth and $G$ the generation. Looking at table~\eqref{tab:counting_addbox_secondrow_2rowstotal_SO24} for the $SO(24)$ trajectories and at table~\eqref{tab:counting_addbox_secondrow_2rowstotal_SO25} for the physical $SO(25)$ trajectories, we see that for all generations $G$ these families of trajectories have a multiplicity that, although initially increasing with depth, quickly saturates to some finite value for large enough depth. Once more, hidden structure is revealed when we look at states deep in the string spectrum.
The fact that the number of $SO(24)$ structures saturates to a constant for large depth is reminiscent of what we observed while climbing a given Regge trajectory at large level. There, we were able to exploit this feature to compare the matrices determining the highest spinning representatives at different levels, and we observed and proved a linear scaling for the coefficients. In the next subsection, we will again exploit the finite and constant number of structures, this time at high depth, to compare matrices determining entire Regge trajectories. Once more, we will observe a linear scaling of the coefficients of the matrices, this time in depth, will will allow us to extract the maximally spinning states for generic depth.

\begin{table}
    \centering
    \begin{tabular}{c|cccccccccccccccc}
        \diagbox{$G$}{$d$} & 0 & 1 & 2 & 3 & 4 & 5 & 6 & 7 & 8 & 9 & 10 & 11 & 12 & 13 & 14 & 15\\
        \hline
        0 & 1 & 1 & 1 & 1 & 1 & 1 & 1 & 1 & 1 & 1 & 1 & 1 & 1 & 1 & 1 & 1\\
        1 & 0 & 1 & 2 & 2 & 2 & 2 & 2 & 2 & 2 & 2 & 2 & 2 & 2 & 2 & 2 & 2\\
        2 & 0 & 0 & 3 & 5 & 6 & 6 & 6 & 6 & 6 & 6 & 6 & 6 & 6 & 6 & 6 & 6\\
        3 & 0 & 0 & 0 & 5 & 10 & 12 & 13 & 13 & 13 & 13 & 13 & 13 & 13 & 13 & 13 & 13\\
        4 & 0 & 0 & 0 & 0 & 11 & 21 & 27 & 29 & 30 & 30 & 30 & 30 & 30 & 30 & 30 & 30\\
        5 & 0 & 0 & 0 & 0 & 0 & 18 & 39 & 51 & 57 & 59 & 60 & 60 & 60 & 60 & 60 & 60\\
        6 & 0 & 0 & 0 & 0 & 0 & 0 & 35 & 74 & 101 & 114 & 120 & 122 & 123 & 123 & 123 & 123\\
        7 & 0 & 0 & 0 & 0 & 0 & 0 & 0 & 57 & 131 & 182 & 211 & 224 & 230 & 232 & 233 & 233
    \end{tabular}
    \caption{Number of $SO(24)$ trajectories at depth $d$ and generation $G$ whose Young tableaux have only two lines. These tableaux are all of the form $\ydiagramalign{5,3}$ with $N+G-2d$ boxes in the first row and $d-G$ boxes in the second row in general.
    }
    \label{tab:counting_addbox_secondrow_2rowstotal_SO24}
\end{table}

\begin{table}
    \centering
    \begin{tabular}{c|cccccccccccccccc}
        \diagbox{$G$}{$d$} & 0 & 1 & 2 & 3 & 4 & 5 & 6 & 7 & 8 & 9 & 10 & 11 & 12 & 13 & 14 & 15\\
        \hline
        0 & 1 & 1 & 1 & 1 & 1 & 1 & 1 & 1 & 1 & 1 & 1 & 1 & 1 & 1 & 1 & 1\\
        1 & 0 & 0 & 1 & 1 & 1 & 1 & 1 & 1 & 1 & 1 & 1 & 1 & 1 & 1 & 1 & 1\\
        2 & 0 & 0 & 1 & 2 & 3 & 3 & 3 & 3 & 3 & 3 & 3 & 3 & 3 & 3 & 3 & 3\\
        3 & 0 & 0 & 0 & 1 & 3 & 4 & 5 & 5 & 5 & 5 & 5 & 5 & 5 & 5 & 5 & 5\\
        4 & 0 & 0 & 0 & 0 & 3 & 6 & 9 & 10 & 11 & 11 & 11 & 11 & 11 & 11 & 11 & 11\\
        5 & 0 & 0 & 0 & 0 & 0 & 3 & 9 & 13 & 16 & 17 & 18 & 18 & 18 & 18 & 18 & 18\\
        6 & 0 & 0 & 0 & 0 & 0 & 0 & 7 & 16 & 25 & 30 & 33 & 34 & 35 & 35 & 35 & 35\\
        7 & 0 & 0 & 0 & 0 & 0 & 0 & 0 & 8 & 24 & 37 & 47 & 52 & 55 & 56 & 57 & 57
    \end{tabular}
    \caption{Number of $SO(25)$ physical trajectories at depth $d$ and generation $G$ whose Young tableaux have only two lines. These tableaux are all of the form $\ydiagramalign{5,3}$ with $N+G-2d$ boxes in the first row and $d-G$ boxes in the second row in general.
    }
    \label{tab:counting_addbox_secondrow_2rowstotal_SO25}
\end{table}

\subsubsection{Oscillator content}
\label{sec:boxes_second_row_oscillators}

The strategy to look for the highest spinning $SO(24)$ oscillator structures at fixed level was to write down a matrix representing $J^{-i}$ on the relevant structures, as we did in \eqref{eq:Jmatrix_level4}, and look for its null space. This strategy was extended to entire Regge trajectories in \eqref{eq:J_example_trajectory_matrix} where the matrix was made level-dependent, leveraging that the number of structures is constant (at high level) and the level-dependence is very simple.

In this subsection, we will further extend the strategy to two infinite families of Regge trajectories, leveraging the constant number of structures that was shown in the previous sections using group-theoretic means.
\smallskip

The first example we look at is the family of trajectories discussed at the beginning of section~\eqref{sec:adding_boxes_second_row}, specifically the generation $1$ trajectories parametrized by their depth $d$
\begin{equation}
    \overbrace{\underbrace{\ydiagramalign{5,3}\hspace{-31pt}}_{d-1}\hspace{31pt}}^{N+1-2d}
\end{equation}

These trajectories contain two types of oscillator structures for each $N$ and $d$, given by
\begin{equation}
\label{eq:trajectorystructures}
\begin{split}
    (&\alpha_{-3}^{(i_1}   \alpha_{-2}^{i_2}\dots \alpha_{-2}^{i_{d-1})}\alpha_{-1}^{i_d} \dots\alpha_{-1}^{i_{2d-2}}-\text{antisym.})\alpha_{-1}^{i_{2d-1}}\dots\alpha_{-1}^{i_{N-d}},\\
   & (\alpha_{-2}^{(i_1}  \alpha_{-2}^{i_2}\dots \alpha_{-2}^{i_{d-1})}\alpha_{-1}^{i_d} \dots\alpha_{-1}^{i_{2d-2}}-\text{antisym.})\alpha_{-2}^{(i_{2d-1}}\dots\alpha_{-1}^{i_{N-4})},
\end{split}
\end{equation}
where $\text{antisym.}$ means that one should antisymmetrize $i_1\leftrightarrow i_d,\,i_2\leftrightarrow i_{d+1},\dots,i_{d-1}\leftrightarrow i_{2d-2}$.
We compute the matrix, which turns out to be $2\times 1$, representing $J^{-i}$ precisely as we did in \eqref{eq:J_example_trajectory_matrix} for $d=2$ to $5$, and we notice a linear dependence on $d$ which can be summarized as
\begin{equation}
\label{eq:J_two_rows}
    J_{N,\,d} \equiv \begin{pmatrix}
        6-6d, & -8+12d-4N
    \end{pmatrix}
\end{equation}
The null space is then $\begin{pmatrix}
    -4+6d-2N \\ -3+3d
\end{pmatrix}$. These coefficients describe how to linearly combine the two oscillator structures in~\eqref{eq:trajectorystructures} in order to get a single particle state, for generic level and depth, aside from very small values of either. 
\smallskip

Similarly, we also explore the second generation of the same Young tableaux, meaning
\begin{equation}
    \overbrace{\underbrace{\ydiagramalign{5,3}\hspace{-31pt}}_{d-2}\hspace{31pt}}^{N+2-2d}
\end{equation}
In this case we find a $6\times 3$ matrix
\begin{equation}
    J_{N,\,d} \equiv \begin{pmatrix}
        8 & -18+6d & -4 & 20-12d+4N & 0 & -8\\
        8-4d & 0 & -6+6d-3N & -6+3d & 0 & -24+4d-2N\\
        0 & 0 & -6 & 12-6d & -12+12d-4N & 4
    \end{pmatrix}
\end{equation}
The null space is then spanned by three vectors. They are
\begin{equation}
    \begin{pmatrix}
        \frac{2 d-N-7}{d-2} \\
        \frac{4 (-2 d+3 N+13)}{9 \left(d^2-5 d+6\right)} \\
        \frac{2}{3} \\
        0 \\
        0 \\
        1
    \end{pmatrix}
    ,\;
    \begin{pmatrix}
        \frac{(2 d-N-2) (3 d-N-3)}{2 (d-2)} \\
        -\frac{2 (4 d-3 N-2) (3 d-N-3)}{9 \left(d^2-5 d+6\right)} \\
        \frac{2}{3} (3 d-N-3) \\
        0 \\
        1 \\
        0 \\
    \end{pmatrix}
    ,\;
    \begin{pmatrix}
        -\frac{3}{4}(2 d-N-3) \\
        \frac{5 (2 d-N-3)}{3 (d-3)} \\
        2-d \\
        1 \\
        0 \\
        0
    \end{pmatrix}
\end{equation}
The number of physical particles in these two examples turned out to be $1$ and $3$ respectively, in agreement with the group theoretic prediction seen for $G=1,\,2$ in \eqref{tab:counting_addbox_secondrow_2rowstotal_SO25}.

\subsection{Some further explorations}

In this section we further explore the string spectrum, focusing on counting the number of structures generalizing what we did in section~\eqref{sec:adding_boxes_second_row} to tableaux with more than two rows.

\subsubsection{Exploring Young tableaux with three rows}

We increase the number of rows of the Young tableaux to three, letting $s_3$ be the length of the third row. Fixing $s_3$, we can parametrize the number of particles and $SO(24)$ tensors again using the generation and the depth. In particular, we will look at Young tableaux of the form
\begin{equation}
\label{eq:tableau_shape_3rows}
    \overbrace{\underbrace{\ydiagramalign{5,3,2}\hspace{-31pt}}_{d-G-2s_3}\hspace{31pt}}^{N+G-2d+s_3}
\end{equation}
Our first motivation is to test if the idea, suggested in section~\eqref{sec:adding_boxes_second_row}, of trading boxes in the top row for boxes in the lower rows of a Young tableau can be applied in this case. In this case, we would expect to be able to exchange three boxes in the top row for one box in the third row, since this transformation would preserve the generation and clearly works for Weinberg states. Our second motivation is that we already established in section~\eqref{sec:adding_boxes_second_row} that the most interesting features of the string spectrum are revealed for large depth, where the number of $SO(24)$ structures and the number of particles both saturated to constant values for $s_3=0$, as seen from tables~\eqref{tab:counting_addbox_secondrow_2rowstotal_SO24} and \eqref{tab:counting_addbox_secondrow_2rowstotal_SO25}. We want to confirm that this is still the case even for $s_3>0$.

The number of $SO(24)$ structures when $s_3=1$ is reported in table~\eqref{tab:counting_addbox_secondrow_1boxthirdrow}, as a function of the generation and depth. This table mirrors table~\eqref{tab:counting_addbox_secondrow_2rowstotal_SO24}, which instead focused on $s_3=0$. In all examples, for every generation there is a value of depth above which these numbers remain constant. Our first observation when comparing tables~\eqref{tab:counting_addbox_secondrow_2rowstotal_SO24} and \eqref{tab:counting_addbox_secondrow_1boxthirdrow} is that our naive idea of trading three boxes in the top row for a single box in the third row fails already at $G=1$ for $s_3=0\to\, s_3=1$. The reason is that the number of $SO(24)$ structures increases from $2$ to $3$, even for large depth, so they cannot be matched.

Regarding our second question, we do see saturation in the number of $SO(24)$ structures at large depth even for $s_3=1$. We checked that this remains true in all examples with $s_3>1$ as well, for both $SO(24)$ structures and for the number of particles. We will not report the whole table (generation$\times$depth) for more complicated representations with $s_3>1$, but we will rather limit ourselves to summarizing the asymptotic large-depth limits in table~\eqref{tab:counting_addbox_thirdrow_threerowstotal}, where we counted the number of $SO(24)$ structures and particles for tableaux of the form \eqref{eq:tableau_shape_3rows}, letting $G$ and $s_3$ vary. In our next section, we will reveal very non-trivial properties one can notice in table~\eqref{tab:counting_addbox_thirdrow_threerowstotal}.

\begin{table}
    \centering
    \begin{tabular}{c|ccccccccccccccccccc}
        \diagbox{$G$}{$d$} & 0 & 1 & 2 & 3 & 4 & 5 & 6 & 7 & 8 & 9 & 10 & 11 & 12 & 13 & 14 & 15 & 16 & 17 & 18\\
        \hline
        0 & 0 & 0 & 0 & 1 & 1 & 1 & 1 & 1 & 1 & 1 & 1 & 1 & 1 & 1 & 1 & 1 & 1 & 1 & 1\\
        1 & 0 & 0 & 0 & 0 & 2 & 3 & 3 & 3 & 3 & 3 & 3 & 3 & 3 & 3 & 3 & 3 & 3 & 3 & 3\\
        2 & 0 & 0 & 0 & 0 & 0 & 6 & 8 & 9 & 9 & 9 & 9 & 9 & 9 & 9 & 9 & 9 & 9 & 9 & 9\\
        3 & 0 & 0 & 0 & 0 & 0 & 0 & 12 & 19 & 21 & 22 & 22 & 22 & 22 & 22 & 22 & 22 & 22 & 22 & 22\\
        4 & 0 & 0 & 0 & 0 & 0 & 0 & 0 & 27 & 42 & 49 & 51 & 52 & 52 & 52 & 52 & 52 & 52 & 52 & 52\\
        5 & 0 & 0 & 0 & 0 & 0 & 0 & 0 & 0 & 51 & 86 & 102 & 109 & 111 & 112 & 112 & 112 & 112 & 112 & 112\\
        6 & 0 & 0 & 0 & 0 & 0 & 0 & 0 & 0 & 0 & 101 & 171 & 209 & 225 & 232 & 234 & 235 & 235 & 235 & 235\\
        7 & 0 & 0 & 0 & 0 & 0 & 0 & 0 & 0 & 0 & 0 & 182 & 325 & 403 & 442 & 458 & 465 & 467 & 468 & 468
    \end{tabular}
    \caption{Number of $SO(24)$ trajectories at depth $d$ and generation $G$ whose Young tableaux have three lines, and the third has strictly one box. These tableaux are all of the form $\ydiagramalign{5,3,1}$ with $N+G-2d+1$ boxes in the first row and $d-G-2$ boxes in the second row in general.}
    \label{tab:counting_addbox_secondrow_1boxthirdrow}
\end{table}

\begin{table}
    \centering
    \begin{tabular}{c|cccc}
        \diagbox{$G$}{$s_3$} & 0 & 1 & 2 & 3\\
        \hline
        0 & 1,\,1 & 1,\,1 & 1,\,1 & 1,\,1 \\
        1 & 2,\,1 & 3,\,2 & 3,\,2 & 3,\,2 \\
        2 & 6,\,3 & 9,\,5 & 10,\,6 & 10,\,6 \\
        3 & 13,\,5 & 22,\,10 & 25,\,12 & 26,\,13 \\
        4 & 30,\,11 & 52,\,21 & 62,\,27 & 65,\,29 \\
        5 & 60,\,18 & 112,\,39 & 137,\,51 & 147,\,57 \\
        6 & 123,\,35 & 235,\,74 & 297,\,101 & 323,\,114 \\
        7 & 233,\,57 & 468,\,131 & 605,\,182 & 670,\,211
    \end{tabular}
    \caption{Each entry lists, separated by a comma, the number of $SO(24)$ and $SO(25)$ trajectories whose Young tableau has three rows, with the third row of length $s_3$. We present results for each generation $G$ but only in the large depth limit. These tableaux have row lengths $s_1=N+G-2d+s_3$, $s_2=d-G-2s_3$. The values for $s_3=0$ coincide with the large depth limit reported in tables~\eqref{tab:counting_addbox_secondrow_2rowstotal_SO24} and \eqref{tab:counting_addbox_secondrow_2rowstotal_SO25}, while the $SO(24)$ large-depth results for $s_3=1$ can also be found in table~\eqref{tab:counting_addbox_secondrow_1boxthirdrow}.}
    \label{tab:counting_addbox_thirdrow_threerowstotal}
\end{table}

\subsubsection{Regularities for large $s_3$}
The first observation we can draw from table~\eqref{tab:counting_addbox_thirdrow_threerowstotal} is that, while at $G=1$ the number of $SO(24)$ and $SO(25)$ structures increases when going from $s_3=0$ to $s_3=1$, constant values are reached for $s_3\ge1$. This appears to remain true even at generation $G=2$, although we could not explore $s_3$ large enough to perform the same check at higher generations.

Another less apparent feature can also be noticed in table~\eqref{tab:counting_addbox_thirdrow_threerowstotal}. For both $SO(24)$ and $SO(25)$ multiplicities $M_{SO(24)}(G,s_3)$ and $M_{SO(25)}(G,s_3)$ as functions of $G$ and $s_3$ in the large depth limit, the following relation holds
\begin{equation}
\label{eq:multiplicities_relation}
    M(G,s_3) = M(G,s_3-1) + M(G-s_3,s_3)
\end{equation}
where it is convenient to define $M\equiv 0$ when either argument is negative. This relation, while empirically found, can be postulated to reconstruct the whole table~\eqref{tab:counting_addbox_thirdrow_threerowstotal} only from the first column $s_3=0$. Actually, if the first column is known only up to a given generation $G$, then the whole table can be recovered up to that generation. To show this, it is important to observe that the first row $G=0$ is known to be identically $(1,1)$, because Weinberg states are unique. The result of the extrapolation is in table~\eqref{tab:counting_addbox_thirdrow_threerowstotal_extrapolation}.
The new table makes it apparent that saturation is indeed achieved for large $s_3$, and it can be readily shown from \eqref{eq:multiplicities_relation} that this should happen for $s_3\ge G$ because the $M(G-s_3,s_3)$ term becomes zero. This is significant because it means that there could be a closed formula for the coefficients of all particles described by Young tableaux with three rows at given generation, at least for $s_3$ large enough. We leave this important check for future work.
In a very surprising twist, the number of $SO(25)$ particles in the large $s_3$ limit precisely matches the number of $SO(24)$ structures at $s_3=0$, for every generation. 

\begin{table}
    \centering
    \begin{tabular}{c|cccccccccccccc}
        \diagbox{$G$}{$s_3$} & 0 & 1 & 2 & 3 & 4 & 5 & 6 & 7 & 8\\
        \hline
         0 & 1, 1 & 1, 1 & 1, 1 & 1, 1 & 1, 1 & 1, 1 & 1, 1 & 1, 1 & 1, 1 \\
         1 & 2, 1 & 3, 2 & 3, 2 & 3, 2 & 3, 2 & 3, 2 & 3, 2 & 3, 2 & 3, 2 \\
         2 & 6, 3 & 9, 5 & 10, 6 & 10, 6 & 10, 6 & 10, 6 & 10, 6 & 10, 6 & 10, 6 \\
         3 & 13, 5 & 22, 10 & 25, 12 & 26, 13 & 26, 13 & 26, 13 & 26, 13 & 26, 13 & 26, 13 \\
         4 & 30, 11 & 52, 21 & 62, 27 & 65, 29 & 66, 30 & 66, 30 & 66, 30 & 66, 30 & 66, 30 \\
         5 & 60, 18 & 112, 39 & 137, 51 & 147, 57 & 150, 59 & 151, 60 & 151, 60 & 151, 60 & 151, 60 \\
         6 & 123, 35 & 235, 74 & 297, 101 & 323, 114 & 333, 120 & 336, 122 & 337, 123 & 337, 123 & 337, 123 \\
         7 & 233, 57 & 468, 131 & 605, 182 & 670, 211 & 696, 224 & 706, 230 & 709, 232 & 710, 233 & 710, 233
    \end{tabular}
    \caption{Precisely the same as table~\eqref{tab:counting_addbox_thirdrow_threerowstotal}, but values extrapolated using \eqref{eq:multiplicities_relation}. Each entry lists the number of $SO(24)$ and $SO(25)$ trajectories, separated by a comma, whose Young tableau has three rows in the large depth limit.}
    \label{tab:counting_addbox_thirdrow_threerowstotal_extrapolation}
\end{table}

\subsubsection{Adding boxes to the first column}
Motivated by the study of Weinberg states, we first conjectured that two boxes in the top row of a Young tableau could be traded for a single box in the second row. We checked that the number of structures would match in tables~\eqref{tab:counting_addbox_secondrow_2rowstotal_SO24} and \eqref{tab:counting_addbox_secondrow_2rowstotal_SO24}, and in section~\eqref{sec:boxes_second_row_oscillators} we substantiated our claim by showing that a single closed formula could describe the coefficients of the $SO(24)$ structures for any length of the second row. In the previous section, we extended the counting of structures and particles to Young tableaux with three rows, showing that three boxes in the top row could be traded for a single box in the third row, at least for Young tableaux whose third row has enough boxes.
In this section we want to test this idea further, by trading $k$ boxes in the first row for a box in the $k-$th row. We will do this by recursively adding boxes to the first column of a given tableau. These transformations preserve the generation, as for all previous examples, but they increase the depth by $k-1$. We could start checking the leading Regge trajectory, but this would be trivially true, as all these checks are at generation $G=0$. We thus start from the generation $G=1$ clone of the Regge trajectory. Adding boxes to the first column we obtain representations with $r$ many rows shaped as follows
\begin{equation}
\label{eq:hookRT}
    r\text{ rows, }\overbrace{\ydiagram{5,1,1}}^{N-r(r+1)/2},\quad
    d = \frac{r^2-r+2}{2},\quad
    G\equiv 1
\end{equation}
Running our counting algorithms, we see that the number of $(SO(24),\,SO(25))$ trajectories begins as $(1,\,0)$ for the fully symmetric $r=1$ representation, and equals $(2,\,1)$ for all $r\ge 2$. In this case, the two structures are given by
\begin{equation}
\begin{split}
  &\left( \alpha_{-(r+1)}^{[i_1} \alpha_{-(r-1)}^{i_2}\alpha_{-(r-2)}^{i_3}\dots\alpha_{-1}^{i_r]}\right) \alpha_{-1}^{i_{r+1}}\dots \alpha_{-1}^{i_{N-d}},\\ 
  & \hspace{.3cm}\left( \alpha_{-r}^{[i_1} \alpha_{-(r-1)}^{i_2}\alpha_{-(r-2)}^{i_3}\dots\alpha_{-1}^{i_r]}\right) \alpha_{-2}^{(i_{r+1}}\dots \alpha_{-1}^{i_{N-d})}.
\end{split}
\end{equation}
This shows that saturation is reached even in this very peculiar way of slicing the spectrum, and that we can attempt to find a closed formula for all the states in the family defined by~\eqref{eq:hookRT} for $r \geq 2$, as for $r=1$ the number of structures in the family is not yet stabilized. The reader should notice that there is no large depth limit here, the depth is finite and a function of $r$.

When writing the matrices describing the action of the Lorentz generator on the two aforementioned $SO(24)$ structures, we notice that this family of matrices can be expressed in closed form as
\begin{equation}
\label{eq:J_L_shaped}
    J \equiv \begin{pmatrix}
        -2r-2, & -4s_1+4
    \end{pmatrix}
\end{equation}
so that the scaling is found to be linear in $r$ for the first structure, but actually quadratic in $r$ for the second (because $s_1=N-r(r+1)/2$). However this quadratic dependence in $r$ can actually be rephrased as a linear dependence in $s_1$, hinting at the fact that in more general settings we might expect linear scalings in the length of each row $s_i$ of the Young tableau. The null space of this matrix is
\begin{equation}
    \begin{pmatrix}
        2-2s_1\\
        1+r
    \end{pmatrix}
\end{equation}
When $r=2$ the L-shaped structures under consideration overlap with the structures with two rows scrutinized in section~\eqref{sec:adding_boxes_second_row}. In particular, at generation 1 the $d=2$ tableau there is the $r=2$ tableau here. Indeed, the matrix \eqref{eq:J_two_rows} can be shown to coincide with the matrix \eqref{eq:J_L_shaped} in this case, providing a good consistency check.
\clearpage{}%
\clearpage{}%
\newpage
\section{Outlook}\label{sec:Outlook}
Building the whole spectrum of the open string is in principle understood, but in practice still an outstanding problem. We made some progress in constructing it explicitly. Many outstanding questions remain, as well as future work that will utilize these explicit constructions. We list some interesting directions for future work below:
\begin{itemize}
    \item{\bf Other string theories:} One obvious direction is to study the spectrum of other string theories, close, with and without supersymmetry. In particular, the simplest example would be closed bosonic string theory, which contains gravitons at tree level and could be used to study possible relations between black holes and string microstates. It would also be interesting to understand how the spectrum gets reorganized if the target space has four large directions, with the others being compactified. 

    \item {\bf The spectrum level by level:} The method we propose can be implemented numerically in a simple and efficient way, allowing us to obtain oscillator representations of single-particle states to, as far as we know, higher-than-ever levels. These can be used to study different properties of the spectrum (couplings, symmetries, degeneracies, etc). If we forget about the oscillator representations and we only care about the particle content, we can go even further on a level-by-level basis using simple counting arguments, and we are able to obtain the full spectrum of single-particle states generation by generation.
    
   \item {\bf Towards the full spectrum:} Building the full spectrum is a herculean task: the number of states grows exponentially with the level, and simply working out all states level by level is an impossible task.  Moreover there exist degeneracies among states (many states with the same mass and spin quantum numbers) that should be understood, as they also obscure the choice of a “good basis” for the spectrum, as any linear combination of these states would also be a good single-particle state. 
While we do not manage to solve the problem fully, we are able to go beyond the level-by-level paradigm by exploiting the observation that we can relate the action of Lorentz transformations on states belonging to  infinite families that generalize the notion of Regge trajectories, allowing to obtain the oscillator representations for the single-particle states of a given family all at once. This permits to slice the spectrum depth by depth---or generation by generation---and while we still need infinitely many steps to count the full spectrum, at each step we have access to an infinite number of states. We show that this can be done rigorously for families of states differing by adding boxes on the first row -i.e. adding one unit of spin per mass, as in a typical Regge trajectory- but observe that we can generalize this to more complicated examples in which we add boxes to various rows/columns simultaneously. We conjecture that all states at fixed generation can be expressed in closed form using the techniques we developed, with the exception of those states lying at the bottom of generalized Regge trajectories for which the number of oscillator structures has not stabilized, and therefore do not fit the general formula for their corresponding family.

\item {\bf Amplitudes:} Our main goal in building the spectrum is to determine the basic building block of perturbation theory: the three-point amplitudes. As these amplitudes have a kinematic structure that is quite rigid, a computation of amplitudes essentially fixes the sizes of each such structure. With such amplitudes in hand we hope to uncover much of the dynamical structure of string theory, its relation to black holes at weak coupling, and potentially further applications. The use of the  machinery developed here to the exploration of the interaction among three single-particle states will be the focus of an upcoming paper.

\item {\bf Further questions about the spectrum:} Our results confirm the expectation that degeneracies in the string spectrum are generic. It would be interesting to find a way of understanding the multiplicities of particles with given mass and Young symmetry without explicitly going through the counting. Even more interesting would be to understand what distinguishes these particles. It is natural to imagine that that the degeneracies might be resolved by computing loop corrections, or that degenerate particles  could be distinguished by their couplings or their charge under some yet-unknown symmetry of the theory, which resonates with our findings revealing non-trivial relations among infinite families of states, as well as the mysterious relation between single-particle states and $SO(24)$ irreps at higher generation described in the last section. While these questions are intriguing, we leave their further investigation to future work.
\end{itemize}
String theory has advanced tremendously in the past decades. We believe that its intricate spectrum of excitations and interactions still hides much valuable information, which should explain its remarkable properties at low and high energies. 
\vskip 40 pt
\noindent{\bf Acknowledgments}
We are grateful to Nima Arkani-
Hamed, Thomas Basile, Nathan Berkovits, Augusto Sagnotti, Evgeny Skvortsov, Massimo Taronna, Emilio Trevisani, and especially to Chrysoula Markou for helpful discussions.

BB is supported by the U.S. Department of Energy under grant number DESC0019470 and by the Heising-Simons Foundation “Observational Signatures of Quantum Gravity” collaboration grants 2021-2818 and 2024-5305.

FF is supported by the European Research Council (ERC) under the European Union’s
Horizon 2020 research and innovation programme (grant agreement No 101002551).

GLP is supported by Scuola Normale, by
INFN (IS GSS-Pi), and by the ERC (NOTIMEFORCOSMO, 101126304). Views and opinions
expressed are, however, those of the author(s) only and do not necessarily reflect those of the
European Union or the European Research Council Executive Agency. Neither the European
Union nor the granting authority can be held responsible for them. GLP is further supported by
a Rita-Levi Montalcini Fellowship from the Italian Ministry of Universities and Research (MUR),
as well as under contract 20223ANFHR (PRIN2022).

\clearpage{}%

\appendix
\newpage
\section{String theory and light-cone quantization}\label{app:lightcone}

In this appendix we provide a review on string theory basics and light-cone gauge quantization, to make the treatment self-contained. This is standard material, and the interested reader is referred to any introductory book on the topic, such as~\cite{Polchinski:1998rq,Green:1987sp,Blumenhagen:2013fgp}.

\subsection{The classical string}

String Theory describes the quantum dynamics of fundamental one-dimensional objects, i.e. {\it strings}, rather than the zero-dimensional particles from Quantum Field Theory. When a classical particle propagates through spacetime, we can describe its dynamics by considering an action that extremizes the proper time along the worldline swept by the particle. Analogously, a string propagating through spacetime will swipe a two-dimensional {\it worldsheet}, and its dynamics are dictated by an action that extremizes the area swept by it. The action that performs this task is called the Nambu-Goto action. Denoting the string's spacetime coordinates as $X^{\mu}, \, \mu = 0, \dots , d-1$, and parametrizing the worldsheet with variables $\tau$ and $\sigma$, it reads
\begin{equation}
S_{\text{NG}} = -T \int d\tau d\sigma \, \sqrt{-\det(\gamma_{ab})},
\end{equation}
where 
\begin{equation}
\gamma_{\alpha \beta} = \frac{\partial X^\mu}{\partial \sigma^\alpha}\frac{\partial X^\nu}{\partial\sigma^\beta}\eta_{\mu \nu} ,
\end{equation}
is the induced metric on the worldsheet, $\sigma^{\alpha} =(\tau, \sigma)$, $\alpha=0,1$, and $T$ is the string tension related to $\alpha'$ by $T=\frac{1}{2\pi \alpha '}$. 

While the Nambu-Goto action already allows to describe the dynamics of the string moving in $d$ dimensional spacetime as a $2$-dimensional field theory living on the worldsheet, with string coordinates $X^\mu (\sigma, \tau)$ seen as scalar fields, its non-linear nature makes it hard to work with at the quantum level. The standard trick is to introduce an auxiliary dynamical metric $h_{\alpha \beta}$ on the worldsheet to obtain an action that eliminates the square-root and has the same equations of motion as the Nambu-Goto action. This is the Polyakov action, given by
\begin{equation}
S_P = -\frac{1}{4\pi \alpha'} \int d^2 \sigma \, \sqrt{-h} \, h^{\alpha \beta} \partial_\alpha X^\mu \partial_\beta X^\nu \eta_{\mu \nu},
\end{equation}
with $h = \det h_{\alpha \beta}$, and simply solving the equations of motion for $h_{\alpha \beta}$ and plugging the solution back into the Polyakov action yields the Nambu-Goto action. What is more, this action enjoys Weyl invariance which can be used to fix the worldsheet metric to be flat, in which case the Polyakov action becomes simply a 2-dimensional free theory of $d$ scalar fields. 

However, this simplification does not comes for free, as now we need to impose that the equations of motion for the auxiliary metric are satisfied. Since there is no kinetic term for the worldsheet metric, its equation of motion simply reads
\begin{equation}
\frac{1}{\sqrt{-h}}\frac{\delta S}{\delta h^{\alpha \beta}}=0 \, \iff T_{\alpha \beta}=0,
\end{equation}
Thus, we see that the equations of motion for $h_{\alpha \beta}$ imply that the worldsheet theory is constrained to have a vanishing stress tensor.

The best way to appreciate the consequences of the vanishing of the stress tensor is to look at the mode expansions for the $X^\mu$. 
Recall that the most general solution to the free wave equation $\partial^\alpha \partial_\alpha X^\mu=0$ is given by a superposition of left and right-moving waves:
\begin{equation}
X^\mu (\sigma,\tau) = X^\mu_L (\sigma^+)+X^\mu_R (\sigma^-), \quad \text{with }\sigma^{\pm}=\tau \pm \sigma,
\end{equation}
as in these coordinates the free wave equation becomes $\partial_{-} \partial_{+ }X^\mu =0$.

The most general solution can be expanded in Fourier modes. In this work we focus on open strings, which require the choice of boundary conditions to specify a solution. We will use Neumann boundary conditions,
\begin{equation}
\partial_\sigma X^\mu =0 \quad \text{at }\sigma=0, \pi,
\end{equation}
which correspond to a string moving freely through spacetime. The open string mode expansion reads
\begin{align}
\label{eq:OSmodeexp}
X^\mu_L (\sigma^+)&=\frac{1}{2}x^\mu+\alpha'p^\mu \sigma^+ + i \sqrt{\frac{\alpha'}{2}}\sum_{n\neq 0}\frac{1}{n}\tilde{\alpha}^\mu_n \, e^{-i n \sigma^+},\\
X^\mu_R (\sigma^-)&=\frac{1}{2}x^\mu+\alpha'p^\mu \sigma^- + i \sqrt{\frac{\alpha'}{2}}\sum_{n\neq 0}\frac{1}{n}\alpha^\mu_n \, e^{-i n \sigma^-},
\end{align}
where $x^\mu$ and $p^\mu$ correspond to the position and momentum of the string's center of mass. 

From the mode expansion we see that Neumann boundary conditions relate the left and right-moving modes of the open string by 
\begin{equation}
\alpha_n^\mu = \tilde{\alpha}_n^\mu,
\end{equation}
and therefore the open string depends on a single set of oscillators. Moreover, the reality of $X^\mu$ requires that the positive and negative frequency modes are related by
\begin{equation}
\alpha^\mu_n = (\alpha_{-n}^\mu)^*, \quad \tilde{\alpha}^\mu_n = (\tilde{\alpha}_{-n}^\mu)^*.
\end{equation}

Combining the left and right-moving waves in the case of the open string gives
\begin{equation}
\label{eq:openstringmodeexp}
X^\mu (\sigma,\tau)=x^\mu+2\alpha'p^\mu \tau + i \sqrt{2\alpha'}\sum_{n\neq 0}\frac{1}{n}\alpha^\mu_n \, \cos n\sigma \, e^{-i n \tau}.
\end{equation}

Using the $\sigma^{\pm}=\tau \pm \sigma$ worldsheet coordinates, the constraints from requiring a vanishing stress tensor become
\begin{equation}
\label{eq:VirasoroConstraints}
(\partial_+X)^2=(\partial_- X)^2=0,
\end{equation}
which can be expressed in terms of the oscillators $\alpha_n^\mu$ as
\begin{equation}
(\partial_- X)^2=\alpha' \sum_{n=-\infty}^\infty L_n \,e^{-i n \sigma^-},\quad L_n=\frac{1}{2}\sum_{m=\infty}^\infty \alpha_{n-m}\cdot \alpha_m,
\end{equation}
where we have defined the {\it Virasoro generators} $L_n$ corresponding to the Fourier modes of the $T_{--}$ component of the stress tensor, and we identified the zero mode $\alpha_0^\mu$ with the center of mass momentum as $\alpha_0^\mu  \equiv \sqrt{2\alpha'}p^\mu $.

The analogous formula for the $T_{++}$ component gives the corresponding relation for the left-moving modes:
\begin{equation}
(\partial_+ X)^2=\alpha' \sum_{n=-\infty}^\infty \tilde{L}_n \,e^{-i n \sigma^-},\quad \tilde{L}_n=\frac{1}{2}\sum_{m=\infty}^\infty \tilde{\alpha}_{n-m}\cdot \tilde{\alpha}_m,
\end{equation}
and $\tilde{\alpha}_0^\mu  \equiv \sqrt{2\alpha'}p^\mu $ for the open string.\footnote{Note that since left and right-moving oscillators are identified in the case of the open string, we have that $L_n = \tilde{L}_n$. This is would not be the case for the closed string.}

The vanishing stress tensor constraint expressed in terms of its Fourier modes says that any classical solution of the string must obey the infinitely many constraints
\begin{equation}
L_n = \tilde{L}_n=0, \quad \forall \, n \in \mathbb{Z}.
\end{equation}
Expressed this way these are called the {\it Virasoro constraints}, and are the ones that we will enforce upon quantization. 

\subsection{The quantum string}

To quantize the string we will use the method of canonical quantization, in which we promote the Fourier modes of the string to creation and annihilation operators. There exist basically two approaches to canonically quantize a constrained theory as the one we are facing: either one quantizes the systen first, and then imposes the constraints as operator equations to define the physical states of the system, or one solves the constraints at the classical level to determine the space of physical solutions and then quantizes these. An example of the second approach is the method of light-cone quantization that we use in this work. However, for illustrative purposes let us describe briefly what happens when using the first approach, usually called covariant quantization.

As standard, to canonically quantize a theory of $d$ free scalar fields $X^\mu$ we promote them and their conjugate momenta $\pi^\mu =\partial L / \partial \dot{X}_\mu$ to field operators satisfying canonical equal-time commutation relations. Translating these into commutation relations for the Fourier modes using the mode expansion we obtain
\begin{equation}
\label{eq:osc_commutationrels_cov}
[x^\mu,p_\nu]=i \delta^{\mu}_{\nu}, \quad [\alpha_n^\mu,\alpha_m^\nu]=[\tilde{\alpha}_n^\mu,\tilde{\alpha}_m^\nu]=n \eta^{\mu \nu}\delta_{n+m,0},
\end{equation} 
with all other commutators being zero.  

For eq.~\eqref{eq:osc_commutationrels_cov} we see that the Fourier modes $\alpha_n^\mu$ and $\tilde{\alpha}_n^\mu$ are nothing but harmonic oscillators up to a rescaling, and therefore the scalar fields give rise to infinite towers of creation and annihilation operators, with $\alpha_{n>0}$ corresponding to creation operators and $\alpha_{n<0}$ to annihilation operators. Naturally, the left-moving modes satisfy identical properties.

Armed with these relations we can easily build the Fock space of the theory by acting on the vacuum with creation operators. From now on we will focus for simplicity on the case of the open string, so we will only deal with the $\alpha_n^\mu$ oscillators.

The vacuum state of the theory corresponds to a string without excited modes and center of mass momentum $p^\mu$, and is defined to obey
\begin{equation}
\alpha_n^\mu \,|0;p \rangle =0 \quad \text{for }n>0, \quad \hat{p}^\mu\,|0;p \rangle=p^\mu \,|0;p \rangle,
\end{equation}
and a generic state corresponding to an excited state of the string is obtained by the action of oscillators with $n<0$ on the vacuum as
\begin{equation}
\alpha_{n_1}^{\mu_1} \alpha_{n_2}^{\mu_2}\dots |0;p \rangle, \quad n_1, \, n_2, \dots <0.
\end{equation}

We immediately see that there is a problem with our Fock space. The fact that the commutation relations for the oscillators shown in eq.~\eqref{eq:osc_commutationrels_cov} depend on the metric gives rise to negative norm states when we excite the timelike oscillators. For instance, the norm of the state created by acting with $\alpha_{-1}^0$ on the vacuum is
\begin{equation}
| \alpha_{-1}^0|0;p\rangle|^2=\langle 0;p| \alpha_1^0 \, \alpha_{-1}^0|0;p\rangle=\langle 0;p|[ \alpha_1^0 , \alpha_{-1}^0]|0;p\rangle= -1,
\end{equation}
explicitly showing the presence of ghosts in the Fock space generated by the string modes.

In covariant quantization, the way to get rid of the ghosts is to impose the Virasoro constraints, as demanding that the physical Hilbert space is the one annihilated by the Virasoro generators projects out the ghost excitations. More concretely, we demand that the matrix element of the Virasoro generators between physical states yield zero in the quantum theory. In fact, since $L^\dagger = L_n $, demanding that the constraints are satisfied inside matrix elements only requires that physical states are annihilated by half of the Virasoro generators, namely
\begin{equation}
L_n | \psi \rangle = 0, \quad \text{for }n>0\, \text{ and }| \psi \rangle \in \mathcal{H}_{\text{phys}}.
\end{equation} 

The constraint associated to $L_0$ is more delicate, as in this case each term in the sum is made of a pair of non-commuting oscillators, and therefore when going from the classical definition in terms of Fourier modes to the quantum mechanical one we face an ambiguity in the ordering of the oscillators. To deal with this ambiguity we simply define $L_0$ to be normal-ordered, and add an arbitrary constant $a$ to the Virasoro constraint, namely
\begin{equation}
L_0=\sum_{m=1}^\infty \alpha_{-m}\cdot \alpha_m +\frac{1}{2}\alpha_0^2, \quad (L_0 -a)| \psi \rangle=0, \quad \text{for } | \psi \rangle \in \mathcal{H}_{\text{phys}}.
\end{equation}

It turns out that the decoupling of the ghosts from the spectrum requires that the constant arising from the ordering ambiguity must be related to the spacetime dimension by $a = (d-2)/24$. While we refer the reader to the references mentioned earlier for details, let us say that this Virasoro constraint has a very important interpretation. Since $\alpha_0^\mu$ is proportional to the momentum of the string, we find that the $0^\text{th}$ Virasoro constraint corresponds to the on-shell condition determining the mass of the physical string states. For the open string $\alpha_0^\mu=\sqrt{2 \alpha'}p^\mu$, and therefore
\begin{equation}
M^2=\frac{1}{\alpha'}\left(\sum_{m=1}^\infty \alpha_{-m}\cdot \alpha_m-\frac{d-2}{24} \right),
\end{equation}
where the sum over oscillators sums the number of excited modes of the string weighted by their Virasoro index, defining the mass level $N$. In particular, by enforcing the $L_0$ constraint on the (level 0) vacuum we recover the known fact that the string's ground state is a tachyon, and since the vacuum is annihilated by all the Virasoro generators with $n>0$ by construction we find that as expected it belongs to the physical Hilbert space.

To work with states at higher levels as was our original goal one needs to enforce an increasingly large number of Virasoro constraints. To see how this would work, let us consider the first non-trivial example of a state at the first level, given by 
\begin{equation}
\mathcal{E} \cdot \alpha_{-1}| 0;p \rangle,
\end{equation}
where $\mathcal{E}$ is a $d$-dimensional polarization tensor. 

The non-trivial Virasoro generators in this case are $L_0$ and $L_1$. Acting on it with $L_0$ we find that the mass of the state is $M^2=1-(d-2)/24$, which already points to the fact that $d=26$ is relevant, as in this case we have a massless spin $1$ state in the spectrum. In fact, as is well known this is precisely the critical dimension of bosonic String Theory, which is the required one for the theory to be consistent when quantized in Minkowski space. 

Acting on the level 1 state with $L_1$ we obtain
\begin{equation}
L_1 \left( \mathcal{E} \cdot \alpha_{-1}| 0;p \rangle \right) = \sqrt{2\alpha'}\mathcal{E} \cdot p \, | 0;p \rangle \equiv 0,
\end{equation}
which implies that the polarization vector must be transverse to the momentum for the state to be physical.

In principle, we could proceed the same way for states at higher levels. Consider an arbitrary state at level $N$ of the form
\begin{equation}
\sum_{i}\mathcal{E}^{(i)}_{\mu_1^{(i)} \dots \mu_k^{(i)}} \, \alpha_{n_1^{(i)}}^{\mu_1^{(i)}} \dots \alpha_{n_k^{(i)}}^{\mu_k^{(i)}}| 0;p \rangle, \quad n_1^{(i)}+\dots+ n_k^{(i)}=N,
\end{equation} 
where $i$ sums over all possible combinations of oscillators consistent with the level and $\mathcal{E}^{(i)}$ are arbitrary tensors of the required rank to contract all the indices at each given term. In this case we would need to act with the relevant Virasoro generators, $L_0, \dots, L_N$, to obtain the constraints that the polarization tensors $\mathcal{E}^{(i)}$ need to satisfy for the state to be physical. Thus, finding the spectrum of physical states in covariant quantization amounts to solving the Virasoro constraints at each level. As mentioned in the main text, while this is in principle doable, in practice can be a very hard task, this being the main reason why the massive string spectrum remained largely unexplored for so long.

\subsection{Light-cone quantization}

In light-cone quantization rather than imposing the Virasoro consraints as operator equations on the states, we solve the constraints at the classical level to find the physical phase space of the theory, and then we quantize it, keeping only the physical degrees of freedom by construction. As explained in section~\ref{sec:review}, the idea is to use the residual gauge freedom remaining after fixing the gauge to set the worldsheet metric to be $\eta_{\alpha \beta}$ to chose a gauge in which the constraints are simple to solve.

As mentioned previously, apart from Poincaré and reparametrization invariance, the Polyakov action is Weyl invariant, which means that the action is invariant under local rescalings of the form
\begin{equation}
h_{\alpha \beta}(\sigma) \to \Omega^2 (\sigma) h_{\alpha \beta}(\sigma),
\end{equation} 
for an arbitrary function $\Omega(\sigma)$ and where the $X^\mu(\sigma)$ are left unchanged.

Both reparametrization and Weyl invariance are gauge symmetries of the worldsheet that can be used to fix the worldsheet metric to be flat, bring the Polyakov action to that of a set of free scalar fields in 2 dimensions:
\begin{equation}
S_P = -\frac{1}{4\pi \alpha'} \int d^2 \sigma \, \partial_\alpha X \cdot \partial^{\alpha} X.
\end{equation}

However, after gauge fixing there remains a residual gauge freedom given by any transformation $\sigma \to \tilde{\sigma}$ that changes the flat metric by
\begin{equation}
\eta_{\alpha \beta} \to \Omega^2(\sigma)\eta_{\alpha \beta},
\end{equation}
as this can be compensated by a Weyl transformation. In terms of the $\sigma^{\pm}$ worlsheet coordinates, the gauge fixed metric on the worldsheet reads
\begin{equation}
ds^2=-d\sigma^+ d\sigma^-,
\end{equation}
and therefore any transformation of the form $\sigma^+ \to \tilde{\sigma}^+(\sigma^+)$, $\sigma^- \to \tilde{\sigma}^-(\sigma^-)$ is a remaining gauge freedom, as this would simply multiply the metric by a factor removable by a Weyl rescaling.

The way to best exploit this remaining freedom is to go to the light-cone gauge. For this, let us define the spacetime light-cone coordinates as
\begin{equation}
\label{eq:lightcone-coords}
X^{\pm}=\frac{X^0 \pm X^{d-1}}{\sqrt{2}},
\end{equation}
in which the spacetime metric reads
\begin{equation}
ds^2=-2dX^-dX^+ + \sum_{i=1}^{d-2}dX^i dX^i.
\end{equation}

As in the last section we focus on the open string, but the story is completely analogous in the case of the closed string. The light-cone gauge is defined by using our remaining gauge freedom to impose
\begin{equation}
\label{eq:light-cone_gauge}
X^+ = x^+ + 2 \alpha' p^+ \tau,
\end{equation}
which fixes the residual reparametrization invariance.

The virtue of this gauge choice is that it makes solving the Virasoro constraints trivial. To see this, remember that the equations of motion are solved by a sum of right and left-moving waves. For $X^-$ we have
\begin{equation}
X^-=X_L^-(\sigma^+)+X_R^-(\sigma^-).
\end{equation}

We will see that $X^-$ is completely fixed by the Virasoro constraints in this gauge. To see this, note that the constraint $(\partial_+ X)^2=0$ in light-cone coordinates reads
\begin{equation}
2 \partial_+ X^- \partial_+ X^+ = \sum_{i=1}^{d-2}\partial_+ X^i \partial_+ X^i. 
\end{equation}
Since from our gauge choice~\eqref{eq:light-cone_gauge} we have $\partial_+ X^+ =\alpha ' p^+$, this constraint becomes
\begin{equation}
\partial_+ X^-=\frac{1}{2 \alpha' p^+}\sum_{i=1}^{d-2}\partial_+ X^i \partial_+ X^i,
\end{equation}
while the constraint $(\partial_- X)^2=0$ implies
\begin{equation}
\partial_- X^-=\frac{1}{2 \alpha' p^+}\sum_{i=1}^{d-2}\partial_- X^i \partial_- X^i,
\end{equation}
which shows that $X^-(\sigma^+,\sigma^-)$ is completely fixed by the other fields $X^i$, $X^+$, up to an integration constant. In terms of the mode expansion for $X^-$
\begin{equation}
X^-=x^-+2 \alpha' p^- \tau + i \sqrt{2 \alpha '}\sum_{n \neq 0}\frac{1}{n}\alpha_n^- \cos n \sigma \, e^{- i n \tau},
\end{equation}
$x^-$ corresponds to the integration constant, and $p^-$ and $\alpha_n^-$ are fixed by the constraints to
\begin{equation}
\label{eq:alphaminus}
\alpha_n^-=\frac{1}{2\sqrt{2 \alpha '}p^+}\sum_{m=-\infty}^\infty \sum_{i=1}^{d-2}\alpha_{n-m}^i \alpha_m^i, \quad p^- = \frac{1}{\sqrt{2\alpha'}}\alpha_0^-.
\end{equation}

While as before $p^-$ will suffer from an ordering ambiguity when we quantize, the important point is that now all the excitations of the string are expressed in terms of oscillators made up of indices pointing along the $i=1,\dots,d-2,$ directions {\it only}, in contrast to the previous case where the indices run from $0$ to $d$ and excited negative norm states in the quantum theory. The remaining oscillators are the transverse modes of the string, which correspond to the physical excitations solving the Virasoro constraints. This accomplishes our first objective of restricting to the physical phase space at the classical level. Now, the next step is quantizing.

As before, the way to canonically quantize is to take the mode expansions from the classical theory and replace the Poisson brackets by commutators. For the transverse oscillators this gives
\begin{equation}
\label{eq:alphaminus}
[\alpha_n^i,\alpha_m^j]=[\tilde{\alpha}_n^i,\tilde{\alpha}_m^j]=n \delta^{ij}\delta_{n+m,0},
\end{equation}
and for the zero-modes 
\begin{equation}
[x^i,p_j]=i \delta^{ij}, \quad [x^-,p^+]=-i, \quad [x^+, p^-]=-i,
\end{equation}

The Fock space is built in the same way as in covariant quantization except that now we only act on the vacuum with the transverse oscillators carrying spatial indices only, and therefore the Hilbert space is free of negative norm states by construction and the Virasoro constraints are automatically satisfied. This permits to bypass the cumbersome problem of determining the allowed polarization tensors for physical states that we described in the previous section, as any state of the form
\begin{equation}
\label{eq:lightcone_state}
\alpha_{n_1}^{i_1} \alpha_{n_2}^{i}\dots |0;p \rangle, \quad n_1, \, n_2, \dots <0, \quad i_1,i_2,\dots=1, \dots , d-2,
\end{equation}
is physical in this setup, no matter the polarization tensor that comes with it.

As explained in the main text, to organize states created by light-cone oscillators into $SO(25)$ irreducible representations need an expression for the Lorentz generators in terms of string oscillators, which we now derive. The Lorentz generators correspond to the conserved charges associated to Lorentz Transformations, and we can express them in terms of the oscillators by integrating the time component of the corresponding conserved current along the spatial direction of the worldsheet. 

The conserved current associated to Lorentz transformations is
\begin{equation}
X^\mu P^{\nu,\alpha} - X^\nu P^{\mu,\alpha}
\end{equation}
where 
\begin{equation}
P^{\mu}_\alpha=\frac{1}{2 \pi \alpha'}\partial_\alpha X^\mu,
\end{equation}
is the conserved current associated to translations.
Integrating its $\tau$ component along the space direction of the worldsheet we get
\begin{equation}
\label{eq:Jmunu}
J^{\mu\nu} = \int_0^\pi d \sigma\,X^\mu P^{\nu,\tau} - X^\nu P^{\mu,\tau}
\end{equation}
which can be evaluated using the mode expansion~\eqref{eq:openstringmodeexp} to give
 \begin{equation}
\label{eq:J_covariant}
J^{\mu\nu} = x^\mu p^\nu - p^\nu x^\mu +\frac{i}{2}\sum_{n\neq0}\frac{\alpha^\mu_n\alpha^\nu_{-n}-\alpha^\nu_n\alpha^\mu_{-n}}{n},
\end{equation}

which is independent of $\tau$ as expected, and where the relevant part is the sum over oscillators, as this is the one that will mix the different $SO(24)$ structures, while the contribution from the center of mass variables must vanish if the transformation belongs to the little group. This point is more subtle than one would think in the light-cone gauge, as here the center of mass degrees of freedom do not commute with the $\alpha^-_n$, but it can be checked with an explicit computation that the contribution from the center of mass terms drops.

To conclude this review of the light-cone gauge formalism, let us mention some subtleties that arise in quantization. As in the case of covariant quantization, the mass-shell constraint suffers also from an ordering ambiguity in the quantization of the operator $p^-$ from its classical definition~\eqref{eq:alphaminus} that can be represented by an a priori unkwown constant $a$.
\begin{equation}
M^2=\frac{1}{\alpha'}\left(\sum_{i=1}^{d-2}\sum_{n>0} \alpha^i_n \alpha^i_n -a \right),
\end{equation}
as can be seen from inserting eq.~\eqref{eq:alphaminus} in $p^2$. 
Even more worrisome is the fact that by choosing the light-cone gauge we broke the manifest Lorentz invariance of the classical theory, and while at the classical level this is only a cosmetic change and the theory remains Lorentz invariant, upon quantization there is no guarantee that Lorentz symmetry does not become anomalous. It turns out that these two issues are related, and it can be seen that requiring that Lorentz invariance is mantained at the quantum level requires simultaneously that the normal ordering constant is $a=1$ and that the spacetime dimension is $d=26$, the critical dimension of the bosonic string~\footnote{We point the reader to the original reference~\cite{Goddard:1973qh} for details.}.

\section{Young tableaux}
\label{app:Young_Tableaux}

In this appendix we review how Young tableaux can be used to characterize the irreducible representations of $SO(N)$ groups. A more detailed treatment can be found in most standard group theory books, see for example~\cite{Fulton:2004uyc, Georgi:1999wka, Zee:2016fuk}.

\subsection{Basics}
A rank $k$ tensor $T^{i_1\dots i_k}$ transforms under an $SO(N)$ transformation as
\begin{equation}
    T^{i_1\dots i_k} \rightarrow  T'{}^{i_1 \dots i_k}=R^{i_1}{}_{j_1}\dots R^{i_k}{}_{j_k} T^{j_1 \dots j_k},
\end{equation}
where $R^i{}_j$ is such that $R R^T = \mathds{1}$ and $\det R=1$. 

It is not hard to convince oneself that if $T^{i_1\dots i_k}$ possesses some symmetry under permutations of some of its indices, then this symmetry is preserved under the action of $SO(N)$. For example, if a rank 4 tensor $T^{i j k \ell}$ is symmetric under the exchange of $i,j$ and antisymmetric under the exchange of $k \ell$, then the tensor $T'{}^{i j k \ell}=R^i{}_{i'}\dots R^\ell{}_{\ell'} T^{i'j'k'\ell'}$ obtained upon acting on $T$ with the $SO(N)$ transformation $R$ has the same symmetries under permutations of its indices. 

Therefore, if we decompose a tensor into different components carrying distinct properties under the permutations of some subset of indices, then these are not mixed under the action of $SO(N)$. This teaches us that a generic tensor is reducible under $SO(N)$, and that in order to decompose it into irreducible representations we must split it into components carrying all  possible inequivalent permutation symmetries among the indices. This provides a link between the tensor representations of $SO(N)$ and the irreducible representations of the permutation group, which are encoded by objects called {\it Young diagrams}.

A Young diagram is a collection of boxes arranged into rows of non-decreasing length. For example, the diagram
\begin{equation}
\label{eq:Ydiagexample}
 \ydiagramalign{5,3,2}
\end{equation}
consists of three rows of lengths $s_1=5$, $s_2=3$ and $s_3=2$. A standard way of denoting a generic Young diagram with $k$ rows is $\mathbf{Y}(s_1,s_2,\dots,s_k)$, where $s_i$ is the length of the $i^{\text{th}}$ row, and thus the diagram in the previous example is $\mathbf{Y}(5,3,2)$. It is also usual to denote a Young diagram by the length of its columns. If the length of the $i^{\text{th}}$ column is $h_i$, a diagram with $n$ columns can be denoted by $\mathbf{Y}[h_1,\dots,k_n]$, where $h_1\geq h_2 \geq \dots \geq h_n$. Note that when referring to the length of the columns, the notation uses square brackets. Continuing with the same example, the Young diagram~\eqref{eq:Ydiagexample} can be denoted as $\mathbf{Y}[3,3,2,1,1]$.

Young diagrams can be used to encode the symmetries of a tensor  using different conventions: the {\it antisymmetric convention} and the {\it symmetric convention}. 

In both conventions, the first step to give the symmetry of a Young diagram $\mathbf{Y}$ with $k$ boxes to a rank $k$ tensor $T$ is to associate each index in $T$ to a box in the diagram, yielding a {\it Young tableau}. Then, in the antisymmetric convention one defines $T$ to have the symmetry of the Young tableau $\mathbf{Y}$ if it is antisymmetric under permutations of indices associated to the same column of $\mathbf{Y}$, and if the antisymmetrization of all the indices in column $a$ with any single index in column $b$, with $a<b$, vanishes. Similarly, in the symmetric convention we require that $T$ be symmetric under the exchange of indices belonging to the same row of $\mathbf{Y}$, and that the symmetrization of all the indices in row $a$ with any single index in row $b$, with $a<b$, yields zero identically. {\it We choose to work in the antisymmetric convention.}

There is an algorithmic way to create a tensor with the symmetry of a Young tableau starting from a tensor without any permutation symmetry: First one symmetrizes over all groups of indices corresponding to the rows of the Young tableau, and then antisymmetrizes over the groups of indices corresponding to columns. This yields a tensor with the desired symmetry in the antisymmetric convention.\footnote{Antisymmetrizing over the columns first, and then symmetrizing over the rows yields a tensor with the symmetry of the Young tableau in the symmetric convention.}

For example, let us consider a rank 3 tensor $T^{ijk}$ without any symmetry and the Young tableau
\begin{equation}
\ytableausetup{centertableaux}
\begin{ytableau}
i & j  \\
k  
\end{ytableau}
\end{equation}
Carrying out this procedure we obtain a tensor $Y^{ijk}$ with the right symmetry in the following way:
\begin{equation}
    T^{ijk} \to %
    T^{ijk}+T^{jik}
    \to\underbrace{
    T^{ijk}+T^{jik}-T^{kji}-T^{jki}
    }_{:=Y^{ijk}}.
\end{equation}
The same algorithm can be performed to construct tensors with any desired Young symmetry. Moreover, the above construction should make it clear that the length of the columns in a Young tableau is bounded by $N$, since antisymmetrizing over a number of indices larger than the dimension of the space in which the tensors live yields identically zero.

The above construction in terms of Young tableaux allows to characterize all possible permutation symmetries carried by $SO(N)$ tensors, but is not enough to decompose an arbitrary tensor into irreducible $SO(N)$ components. The reason for this is that the Euclidean metric $\delta_{ij}$ is invariant under $SO(N)$ by definition, and therefore index contraction is preserved under $SO(N)$ transformations. This implies that we can further decompose a tensor with a given symmetry by extracting its traces with respect to all possible pairs of indices. This exhausts the $SO(N)$ invariant operations we can perform on a tensor, and thus {\it irreducible tensors under $SO(N)$ are traceless in all pairs of indices and carry a specific Young symmetry under permutations of their indices. }

As a further remark let us mention that another $SO(N)$ invariant is the $N$-dimensional Levi-Civita tensor $\epsilon^{i_1 \dots i_N}$. This means that an antisymmetric tensor of rank $k$ can be fully contracted with $\epsilon^{i_1 \dots i_N}$ to yield an equivalent representation of rank $N-k$. In turn, this allows us to work with Young tableaux with columns of length up to $N/2$, since indices corresponding to longer columns can be contracted with $\epsilon^{i_1 \dots i_N}$ to yield a smaller, equivalent representation.

\subsection{Tensor product of irreducible representations and branching rules}

We briefly discuss the tensor product of two Young tableaux and the branching rules from the representations of a group to the ones of one of its subgroups. While both can be discussed within abstract group theory as a set of rules for manipulating Young tableaux, we will find it more convenient to think of Young tableaux as a language to describe concrete operations on tensors.

The tensor product of two Young tableaux $Y_1\otimes Y_2$ encodes the decomposition of two tensors $T_1^{i_1\dots i_n},\,T_2^{j_1\dots j_m}$ with symmetries $Y_1,\,Y_2$ into irreducible representations. The set of Young tableaux associated to this decomposition (with their multiplicities) are the result of $Y_1\otimes Y_2$, and can be computed in an algorithmic way by means of operations over the boxes building up $\mathbf{Y}_1$ and $\mathbf{Y}_2$.\footnote{See for instance~\cite{frappat:hal-00376660}. } 

Lastly, in this paper we often go back and forth between $SO(24)$ and $SO(25)$ representations. The two can be connected via branching rules, that prescribe how a $SO(25)$ irreducible representation decomposes into irreducible representations of $SO(24)$. Intuitively, because the group becomes smaller, some states are not connected by a group transformation anymore. A simple concrete example is a $SO(25)$ vector, which branches into a $SO(24)$ vector and a scalar. More generally, given a $SO(25)$ tensor $T^{\mu_1\dots\mu_n}$, each index can either be put in the preserved $24$-dimensional hyperplane, or in the broken $25$-th direction. This produces the full set of $SO(24)$ tensors that can be obtained from branching $T$, and the Young tableaux branching rules describe this decomposition. For a review on branching rules as well as an extensive list of the branching rules between several groups, including the $SO(25) \to SO(24)$ case relevant for us, see~\cite{Yamatsu:2015npn}.  
\section{List of single-particle states}
\label{app:levelbylevelspectrum}
In this appendix we provide the one-particle states in the first seven levels for the open bosonic string. For each particle we provide the $SO(25)$ Young tableau and the oscillator structure exciting the maximally spinning states. For the particles of the first five levels, we also provide the branching rule $SO(25)\to SO(24)$, listing all $SO(24)$ Young tableaux one can obtain by considering different polarizations; we also explicitly write the tensor structure exciting each one. More complete results can be found in the Mathematica notebooks.

\subsection{Level up to 2}
The string spectrum begins with a scalar tachyon, a massless vector $\alpha_{-1}^i$, and a massive spin two $\ydiagram{2}:\;\alpha_{-1}^i\alpha_{-1}^j$ at level 2 that has the additional polarizations $\ydiagram{1}:\;\alpha_{-2}^i$ and $\bullet:\;\alpha_{-1}\cdot\alpha_{-1}$.

\subsection{Level 3}
There is a spin three and an antisymmetric rank two
\begin{equation}
\def\arraystretch{1.3}
\begin{array}{lcl}
        \ydiagram{3}&:&\alpha_{-1}^i\alpha_{-1}^j\alpha_{-1}^k\\
        \rule{0pt}{35pt}
        \ydiagram{1,1}&:&\alpha_{-2}^{[i}\alpha_{-1}^{j]}
\end{array}
\end{equation}
The other polarizations of the spin three are
\begin{equation}
    \ydiagram{3}_{25}\to\;\ydiagram{3}_{24}\quad\ydiagram{2}_{24}\quad\ydiagram{1}_{24}\quad\bullet_{24}
\end{equation}
and, aside from the maximally spinning one, they are excited by
\begin{equation}
    \alpha_{-2}^{(i}\alpha_{-1}^{j)},\quad
    26\alpha_{-3}^i+3\alpha_{-1}^i (\alpha_{-1}\cdot\alpha_{-1}),\quad
    \alpha_{-2}\cdot\alpha_{-1}
\end{equation}
For the antisymmetric combination, we have instead 
\begin{equation}
    \ydiagram{1,1}\to\;\ydiagram{1,1}\quad\ydiagram{1}
\end{equation}
and the extra vector is given by $2\alpha_{-3}^i-\alpha_{-1}^i (\alpha_{-1}\cdot\alpha_{-1})$

\subsection{Level 4}
There are four particles. The spin four belongs to the leading Regge trajectory, the $\ydiagram{2,1}$ belongs to the trajectory that started with $\ydiagram{1,1}$ at the previous level, and then there are two seeds of new trajectories.

\begin{equation}
\def\arraystretch{1.3}
\begin{array}{lcl}
        \ydiagram{4}&:&\alpha_{-1}^i\alpha_{-1}^j\alpha_{-1}^k\alpha_{-1}^l\\
        \rule{0pt}{35pt}
        \ydiagram{2,1}&:&\alpha_{-2}^{[i}\alpha_{-1}^{j]}\alpha_{-1}^{k}\\
        \rule{0pt}{35pt}
        \ydiagram{2}&:& 4 \alpha_{-3}^{(i}\alpha_{-1}^{j)} -7 \alpha_{-2}^i\alpha_{-2}^j - \alpha_{-1}^i\alpha_{-1}^j(\alpha_{-1}\cdot\alpha_{-1}) \\
        \rule{0pt}{35pt}
        \bullet&:& 10 \alpha_{-1}\cdot\alpha_{-3} -7 \alpha_{-2}\cdot\alpha_{-2} - (\alpha_{-1}\cdot\alpha_{-1})^2
\end{array}
\end{equation}
The spin four has polarizations
\begin{equation}
    \ydiagram{4}_{25}\to\;\ydiagram{4}_{24}\quad\ydiagram{3}_{24}\quad\ydiagram{2}_{24}\quad\ydiagram{1}_{24}\quad\bullet_{24}
\end{equation}
excited by (omitting the maximally spinning structure, listed above)
\begin{equation}
\begin{split}
    & \hspace{3cm}\alpha_{-2}^{(i}\alpha_{-1}^j\alpha_{-1}^{k)},\quad
    21\alpha_{-3}^{(i}\alpha_{-1}^{j)}+14\alpha_{-2}^i\alpha_{-2}^j+2\alpha_{-1}^i\alpha_{-1}^j(\alpha_{-1}\cdot\alpha_{-1}),\\
    &416\alpha_{-4}^i + 21 \alpha_{-2}^i (\alpha_{-1}\cdot\alpha_{-1}) + 68 \alpha_{-1}^i (\alpha_{-2}\cdot\alpha_{-1}),\quad
    104 \alpha_{-1}\cdot\alpha_{-3} +52 \alpha_{-2}\cdot\alpha_{-2} + (\alpha_{-1}\cdot\alpha_{-1})^2
\end{split}
\end{equation}
The mixed symmetry has
\begin{equation}
    \ydiagram{2,1}_{25}\to\;\ydiagram{2,1}_{24}\quad\ydiagram{2}_{24}\quad\ydiagram{1,1}_{24}\quad\ydiagram{1}_{24}
\end{equation}
and additional oscillator structures
\begin{equation}
\begin{split}
    2 \alpha_{-2}^i\alpha_{-2}^j - \alpha_{-1}^i\alpha_{-1}^j(\alpha_{-1}\cdot\alpha_{-1}),\quad
    \alpha_{-3}^{[i}\alpha_{-1}^{j]},\quad
    46\alpha_{-4}^i - 3 \alpha_{-2}^i (\alpha_{-1}\cdot\alpha_{-1}) - 20 \alpha_{-1}^i (\alpha_{-2}\cdot\alpha_{-1})
\end{split}
\end{equation}
The spin two has oscillators
\begin{equation}
    4\alpha_{-4}^i - 9 \alpha_{-2}^i (\alpha_{-1}\cdot\alpha_{-1}) + 4 \alpha_{-1}^i (\alpha_{-2}\cdot\alpha_{-1}),\quad
    20 \alpha_{-1}\cdot\alpha_{-3} -19 \alpha_{-2}\cdot\alpha_{-2} +8(\alpha_{-1}\cdot\alpha_{-1})^2
\end{equation}

\subsection{Level 5}
We have two seeds of new trajectories, $\ydiagram{2,1}$ and $\ydiagram{1,1}$, plus four particles that extend previously existing trajectories.
\begin{equation}
\def\arraystretch{1.3}
\begin{array}{lcl}
        \ydiagram{5}&:&\alpha_{-1}^{i_1}\alpha_{-1}^{i_2}\alpha_{-1}^{i_3}\alpha_{-1}^{i_4}\alpha_{-1}^{i_5}\\
        \rule{0pt}{35pt}
        \ydiagram{3,1}&:&\alpha_{-2}^{[i}\alpha_{-1}^{j]}\alpha_{-1}^{k}\alpha_{-1}^{l}\\
        \rule{0pt}{35pt}
        \ydiagram{3}&:& 26\alpha_{-3}^{(i}\alpha_{-1}^j\alpha_{-1}^{k)} - 24 \alpha_{-2}^{(i}\alpha_{-2}^j\alpha_{-1}^{k)} - 9 \alpha_{-1}^i\alpha_{-1}^j\alpha_{-1}^k(\alpha_{-1}\cdot\alpha_{-1}) \\
        \rule{0pt}{35pt}
        \ydiagram{1}&:& 188\alpha_{-5}^i-820\alpha_{-3}^i(\alpha_{-1}\cdot\alpha_{-1})+4580\alpha_{-1}^i(\alpha_{-3}\cdot\alpha_{-1})+\\
        &&+640\alpha_{-2}^i(\alpha_{-2}\cdot\alpha_{-1})-3360\alpha_{-1}^i(\alpha_{-2}\cdot\alpha_{-2})-375\alpha_{-1}^i(\alpha_{-1}\cdot\alpha_{-1})^2\\
        \rule{0pt}{35pt}
        \ydiagram{2,1}&:& 2 \alpha_{-3}^{[i}\alpha_{-1}^{j]}\alpha_{-1}^k -3 \alpha_{-2}^{[i}\alpha_{-1}^{j]}\alpha_{-2}^k \\
        \rule{0pt}{35pt}
        \ydiagram{1,1}&:& 8 \alpha_{-4}^{[i}\alpha_{-1}^{j]} -26 \alpha_{-3}^{[i}\alpha_{-2}^{j]} -5 \alpha_{-2}^{[i}\alpha_{-1}^{j]}(\alpha_{-1}\cdot\alpha_{-1})
\end{array}
\end{equation}

The spin five has polarizations
\begin{equation}
\begin{split}
    \ydiagram{5}_{25}\to\;\ydiagram{5}_{24}\quad\ydiagram{4}_{24}\quad\ydiagram{3}_{24}\\
    \rule{0pt}{20pt}
    \ydiagram{2}_{24}\quad\ydiagram{1}_{24}\quad\bullet_{24}
\end{split}
\end{equation}
excited by
\begin{equation}
\begin{split}
    \alpha_{-2}^{(i}\alpha_{-1}^j\alpha_{-1}^k\alpha_{-1}^{l)},\quad
    90\alpha_{-3}^{(i}\alpha_{-1}^j\alpha_{-1}^{k)} + 70 \alpha_{-2}^{(i}\alpha_{-2}^j\alpha_{-1}^{k)} + 7 \alpha_{-1}^i\alpha_{-1}^j\alpha_{-1}^k(\alpha_{-1}\cdot\alpha_{-1}),\\
    224\alpha_{-4}^{(i}\alpha_{-1}^{j)}+126\alpha_{-3}^{(i}\alpha_{-2}^{j)}+9(\alpha_{-1}\cdot\alpha_{-1})\alpha_{-2}^{(i}\alpha_{-1}^{j)}+32(\alpha_{-2}\cdot\alpha_{-1})\alpha_{-1}^i\alpha_{-1}^j,\\
    45500\alpha_{-5}^i+1260\alpha_{-3}^i(\alpha_{-1}\cdot\alpha_{-1})+5796\alpha_{-1}^i(\alpha_{-3}\cdot\alpha_{-1})+3136\alpha_{-2}^i(\alpha_{-2}\cdot\alpha_{-1})+\\
    +3024\alpha_{-1}^i(\alpha_{-2}\cdot\alpha_{-2})+33\alpha_{-1}^i(\alpha_{-1}\cdot\alpha_{-1})^2,\\
    2912 \alpha_{-4}\cdot\alpha_{-1} +3042 \alpha_{-3}\cdot\alpha_{-2} +55 (\alpha_{-2}\cdot\alpha_{-1})(\alpha_{-1}\cdot\alpha_{-1})
\end{split}
\end{equation}

The long L-shaped has polarizations
\begin{equation}
\begin{split}
    \ydiagram{3,1}_{25}\to\;\ydiagram{3,1}_{24}\quad\ydiagram{2,1}_{24}\quad\ydiagram{1,1}_{24}\quad\ydiagram{3}_{24}\quad\ydiagram{2}_{24}\quad\ydiagram{1}_{24}
\end{split}
\end{equation}
excited by
\begin{equation}
\begin{split}
    2 \alpha_{-3}^{[i}\alpha_{-1}^{j]}\alpha_{-1}^k + \alpha_{-2}^{[i}\alpha_{-1}^{j]}\alpha_{-1}^k,\quad
    206 \alpha_{-4}^{[i}\alpha_{-1}^{j]} +26 \alpha_{-3}^{[i}\alpha_{-2}^{j]} +5 \alpha_{-2}^{[i}\alpha_{-1}^{j]}(\alpha_{-1}\cdot\alpha_{-1}),\\
    2\alpha_{-3}^{(i}\alpha_{-1}^j\alpha_{-1}^{k)} - 4 \alpha_{-2}^{(i}\alpha_{-2}^j\alpha_{-1}^{k)} + 3 \alpha_{-1}^i\alpha_{-1}^j\alpha_{-1}^k(\alpha_{-1}\cdot\alpha_{-1}),\\
    6\alpha_{-3}^{(i}\alpha_{-2}^{j)}-(\alpha_{-1}\cdot\alpha_{-1})\alpha_{-2}^{(i}\alpha_{-1}^{j)}-4(\alpha_{-2}\cdot\alpha_{-1})\alpha_{-1}^i\alpha_{-1}^j,\\
    5980\alpha_{-5}^i-180\alpha_{-3}^i(\alpha_{-1}\cdot\alpha_{-1})-1660\alpha_{-1}^i(\alpha_{-3}\cdot\alpha_{-1})-320\alpha_{-2}^i(\alpha_{-2}\cdot\alpha_{-1})+\\
    -784\alpha_{-1}^i(\alpha_{-2}\cdot\alpha_{-2})-23\alpha_{-1}^i(\alpha_{-1}\cdot\alpha_{-1})^2
\end{split}
\end{equation}

The symmetric spin three is excited by
\begin{equation}
\begin{split}
    56\alpha_{-4}^{(i}\alpha_{-1}^{j)}-46\alpha_{-3}^{(i}\alpha_{-2}^{j)}-21(\alpha_{-1}\cdot\alpha_{-1})\alpha_{-2}^{(i}\alpha_{-1}^{j)}+8(\alpha_{-2}\cdot\alpha_{-1})\alpha_{-1}^i\alpha_{-1}^j,\\
    1300\alpha_{-5}^i-2444\alpha_{-3}^i(\alpha_{-1}\cdot\alpha_{-1})+364\alpha_{-1}^i(\alpha_{-3}\cdot\alpha_{-1})-1696\alpha_{-2}^i(\alpha_{-2}\cdot\alpha_{-1})+\\
    +1128\alpha_{-1}^i(\alpha_{-2}\cdot\alpha_{-2})-201\alpha_{-1}^i(\alpha_{-1}\cdot\alpha_{-1})^2,\\
    152 \alpha_{-4}\cdot\alpha_{-1} -118 \alpha_{-3}\cdot\alpha_{-2} +67 (\alpha_{-2}\cdot\alpha_{-1})(\alpha_{-1}\cdot\alpha_{-1})
\end{split}
\end{equation}

The vector has a scalar excitation
\begin{equation}
    848 \alpha_{-4}\cdot\alpha_{-1} -502 \alpha_{-3}\cdot\alpha_{-2} -125 (\alpha_{-2}\cdot\alpha_{-1})(\alpha_{-1}\cdot\alpha_{-1})
\end{equation}

The short L-shaped is excited by
\begin{equation}
\begin{split}
    16\alpha_{-4}^{(i}\alpha_{-1}^{j)}-14\alpha_{-3}^{(i}\alpha_{-2}^{j)}+3(\alpha_{-1}\cdot\alpha_{-1})\alpha_{-2}^{(i}\alpha_{-1}^{j)}-8(\alpha_{-2}\cdot\alpha_{-1})\alpha_{-1}^i\alpha_{-1}^j,\\
    10 \alpha_{-3}^{[i}\alpha_{-2}^{j]} -3 \alpha_{-2}^{[i}\alpha_{-1}^{j]}(\alpha_{-1}\cdot\alpha_{-1}),\\
    460\alpha_{-5}^i+28\alpha_{-3}^i(\alpha_{-1}\cdot\alpha_{-1})+340\alpha_{-1}^i(\alpha_{-3}\cdot\alpha_{-1})-640\alpha_{-2}^i(\alpha_{-2}\cdot\alpha_{-1})+\\
    -96\alpha_{-1}^i(\alpha_{-2}\cdot\alpha_{-2})+69\alpha_{-1}^i(\alpha_{-1}\cdot\alpha_{-1})^2
\end{split}
\end{equation}

Lastly, the antisymmetric spin two has a vector polarization
\begin{equation}
    4\alpha_{-5}^i+36\alpha_{-3}^i(\alpha_{-1}\cdot\alpha_{-1})-4\alpha_{-1}^i(\alpha_{-3}\cdot\alpha_{-1})-32\alpha_{-2}^i(\alpha_{-2}\cdot\alpha_{-1})+8\alpha_{-1}^i(\alpha_{-2}\cdot\alpha_{-2})-5\alpha_{-1}^i(\alpha_{-1}\cdot\alpha_{-1})^2
\end{equation}

\subsection{Level 6}
From this level onward, we only list the particles and the maximally spinning $SO(24)$ representative oscillator structure. The seeds of new trajectories are $\ydiagram{3},\,\ydiagram{2,2},\,\ydiagram{1,1,1},\,\ydiagram{2},\,\ydiagram{1},\,\bullet$. Notice that one $\ydiagram{2}$ comes from extending the Regge trajectory of the scalar at level four up two levels, but now this trajectory splits into two degenerate ones: there are two $\ydiagram{2}$ particles at level 6, indistinguishable in mass and symmetry. Because any linear combination of them has the same mass and symmetry, the distinction of which combination is the new seed and which is the extension of the existing trajectory is quite arbitrary.

\begin{equation}
\def\arraystretch{1.3}
\begin{array}{lcl}
        \ydiagram{6}&:&\alpha_{-1}^{i_1}\alpha_{-1}^{i_2}\alpha_{-1}^{i_3}\alpha_{-1}^{i_4}\alpha_{-1}^{i_5}\alpha_{-1}^{i_6}\\
        \rule{0pt}{35pt}
        \ydiagram{4,1}&:& \alpha_{-2}^{[i_1}\alpha_{-1}^{i_5]}\alpha_{-1}^{i_2}\alpha_{-1}^{i_3}\alpha_{-1}^{i_4} \\
        \rule{0pt}{35pt}
        \ydiagram{4}&:& 14\alpha_{-3}^{(i}\alpha_{-1}^j\alpha_{-1}^k\alpha_{-1}^{l)}-9\alpha_{-2}^{(i}\alpha_{-2}^j\alpha_{-1}^k\alpha_{-1}^{l)}-6\alpha_{-1}^{i}\alpha_{-1}^j\alpha_{-1}^k\alpha_{-1}^{l}(\alpha_{-1}\cdot\alpha_{-1}) \\
        \rule{0pt}{35pt}
        \ydiagram{2}&:& 6\alpha_{-5}^{(i}\alpha_{-1}^{j)}-60\alpha_{-4}^{(i}\alpha_{-2}^{j)}+104\alpha_{-3}^{(i}\alpha_{-3}^{j)}-12\alpha_{-3}^{(i}\alpha_{-1}^{j)}(\alpha_{-1}\cdot\alpha_{-1})+\\
        &&-18\alpha_{-1}^i\alpha_{-1}^j(\alpha_{-1}\cdot\alpha_{-3})+24\alpha_{-2}^i\alpha_{-2}^j(\alpha_{-1}\cdot\alpha_{-1})+\\
        &&+15\alpha_{-1}^i\alpha_{-1}^j(\alpha_{-2}\cdot\alpha_{-2})+3\alpha_{-1}^i\alpha_{-1}^j(\alpha_{-1}\cdot\alpha_{-1})^2 \\
        \rule{0pt}{35pt}
        \ydiagram{3,1}&:& 4\alpha_{-3}^{[i}\alpha_{-1}^{l]}\alpha_{-1}^j\alpha_{-1}^k -3\alpha_{-2}^{[i}\alpha_{-1}^{l]}\alpha_{-2}^{(j}\alpha_{-1}^{k)} \\
        \rule{0pt}{35pt}
        \ydiagram{2,1}&:& 18\alpha_{-4}^{[i}\alpha_{-1}^{k]}\alpha_{-1}^j-20(\alpha_{-3}^{(i}\alpha_{-2}^{j)}\alpha_{-1}^k - \alpha_{-3}^{(k}\alpha_{-2}^{j)}\alpha_{-1}^i)-34(\alpha_{-3}^{(i}\alpha_{-1}^{j)}\alpha_{-2}^k - \alpha_{-3}^{(k}\alpha_{-1}^{j)}\alpha_{-2}^i) +\\
        &&-9\alpha_{-2}^{[i}\alpha_{-1}^{k]}\alpha_{-1}^j (\alpha_{-1}\cdot\alpha_{-1}) \\
\end{array}
\end{equation}
The new seeds are
\begin{equation}
\def\arraystretch{1.3}
\begin{array}{lcl}
        \ydiagram{3}&:& 14\alpha_{-4}^{(i}\alpha_{-1}^j\alpha_{-1}^{k)}-14\alpha_{-3}^{(i}\alpha_{-2}^j\alpha_{-1}^{k)}+44\alpha_{-2}^i\alpha_{-2}^j\alpha_{-2}^k+\alpha_{-2}^{(i}\alpha_{-1}^j\alpha_{-1}^{k)}(\alpha_{-1}\cdot\alpha_{-1})+\\
        &&-4\alpha_{-1}^i\alpha_{-1}^j\alpha_{-1}^k(\alpha_{-2}\cdot\alpha_{-1}) \\
        \rule{0pt}{35pt}
        \ydiagram{2,2}&:& \alpha_{-2}^{[i}\alpha_{-1}^{k]}\alpha_{-2}^{[j}\alpha_{-1}^{l]} \\
        \rule{0pt}{35pt}
        \ydiagram{1,1,1}&:& \alpha_{-3}^{[i}\alpha_{-2}^j\alpha_{-1}^{k]} \\
        \rule{0pt}{35pt}
        \ydiagram{2}&:& 20\alpha_{-5}^{(i}\alpha_{-1}^{j)}-488\alpha_{-4}^{(i}\alpha_{-2}^{j)}+872\alpha_{-3}^{(i}\alpha_{-3}^{j)}-8\alpha_{-3}^{(i}\alpha_{-1}^{j)}(\alpha_{-1}\cdot\alpha_{-1})+\\
        &&-684\alpha_{-1}^i\alpha_{-1}^j(\alpha_{-1}\cdot\alpha_{-3})+188\alpha_{-2}^i\alpha_{-2}^j(\alpha_{-1}\cdot\alpha_{-1})-72\alpha_{-2}^{(i}\alpha_{-1}^{j)}(\alpha_{-2}\cdot\alpha_{-1})+\\
        &&+536\alpha_{-1}^i\alpha_{-1}^j(\alpha_{-2}\cdot\alpha_{-2})+61\alpha_{-1}^i\alpha_{-1}^j(\alpha_{-1}\cdot\alpha_{-1})^2 \\
        \rule{0pt}{35pt}
        \ydiagram{1}&:& 12\alpha_{-6}^i-65\alpha_{-4}^i(\alpha_{-1}\cdot\alpha_{-1})-268\alpha_{-1}^i(\alpha_{-1}\cdot\alpha_{-4})+40\alpha_{-3}^i(\alpha_{-1}\cdot\alpha_{-2})+\\
        &&+370\alpha_{-2}^i(\alpha_{-1}\cdot\alpha_{-3})+152\alpha_{-1}^i(\alpha_{-2}\cdot\alpha_{-3})-255\alpha_{-2}^i(\alpha_{-2}\cdot\alpha_{-2})+\\
        &&-30\alpha_{-2}^i(\alpha_{-1}\cdot\alpha_{-1})^2+40\alpha_{-1}^i(\alpha_{-1}\cdot\alpha_{-1})(\alpha_{-1}\cdot\alpha_{-2}) \\
        \rule{0pt}{35pt}
        \bullet&:& 24\alpha_{-5}\cdot\alpha_{-1}-336\alpha_{-4}\cdot\alpha_{-2}+280\alpha_{-3}\cdot\alpha_{-3}-84(\alpha_{-3}\cdot\alpha_{-1})(\alpha_{-1}\cdot\alpha_{-1})+\\
        &&+54(\alpha_{-2}\cdot\alpha_{-2})(\alpha_{-1}\cdot\alpha_{-1})+24(\alpha_{-2}\cdot\alpha_{-1})^2+5(\alpha_{-1}\cdot\alpha_{-1})^3
\end{array}
\end{equation}

\subsection{Level 7}
We have six new seeds, $\ydiagram{3,1},\,\ydiagram{2,1}(\times 2),\,\ydiagram{1,1}(\times 2),\,\ydiagram{1}$. As a result of $\ydiagram{2,1}$ and $\bullet$ appearing at the previous level, the degeneracy of $\ydiagram{3,1}$ and $\ydiagram{1}$ is two. Lastly, $\ydiagram{3}$ inherits the degeneracy ($\times 2$) of $\ydiagram{2}$ at level 6.

\begin{equation}
\def\arraystretch{1.3}
\begin{array}{lcl}
        \ydiagram{7}&:& \alpha_{-1}^{i_1}\alpha_{-1}^{i_2}\alpha_{-1}^{i_3}\alpha_{-1}^{i_4}\alpha_{-1}^{i_5}\alpha_{-1}^{i_6}\alpha_{-1}^{i_7} \\
        \rule{0pt}{35pt}
        \ydiagram{5,1}&:& \alpha_{-2}^{[i_1}\alpha_{-1}^{i_6]}\alpha_{-1}^{i_2}\alpha_{-1}^{i_3}\alpha_{-1}^{i_4}\alpha_{-1}^{i_5} \\
        \rule{0pt}{35pt}
        \ydiagram{5}&:& 2\alpha_{-3}^{(i_1}\alpha_{-1}^{i_2}\alpha_{-1}^{i_3}\alpha_{-1}^{i_4}\alpha_{-1}^{i_5)} -\alpha_{-2}^{(i_1}\alpha_{-2}^{i_2}\alpha_{-1}^{i_3}\alpha_{-1}^{i_4}\alpha_{-1}^{i_5)} -\alpha_{-1}^{i_1}\alpha_{-1}^{i_2}\alpha_{-1}^{i_3}\alpha_{-1}^{i_4}\alpha_{-1}^{i_5}(\alpha_{-1}\cdot\alpha_{-1}) \\
        \rule{0pt}{35pt}
        \ydiagram{3}&:& 138\alpha_{-5}^{(i}\alpha_{-1}^j\alpha_{-1}^{k)}-450\alpha_{-4}^{(i}\alpha_{-2}^j\alpha_{-1}^{k)}+640\alpha_{-3}^{(i}\alpha_{-3}^j\alpha_{-1}^{k)}+100\alpha_{-3}^{(i}\alpha_{-2}^j\alpha_{-2}^{k)}+\\
        &&-140\alpha_{-3}^{(i}\alpha_{-1}^j\alpha_{-1}^{k)}(\alpha_{-1}\cdot\alpha_{-1})-270\alpha_{-1}^{i}\alpha_{-1}^j\alpha_{-1}^{k}(\alpha_{-3}\cdot\alpha_{-1})+\\
        &&+150\alpha_{-2}^{(i}\alpha_{-2}^j\alpha_{-1}^{k)}(\alpha_{-1}\cdot\alpha_{-1})+225\alpha_{-1}^{i}\alpha_{-1}^j\alpha_{-1}^{k}(\alpha_{-2}\cdot\alpha_{-2})+\\
        &&+45\alpha_{-1}^{i}\alpha_{-1}^j\alpha_{-1}^{k}(\alpha_{-1}\cdot\alpha_{-1})^2 \\
        \rule{0pt}{35pt}
        \ydiagram{3}&:& 346\alpha_{-5}^{(i}\alpha_{-1}^j\alpha_{-1}^{k)}-1390\alpha_{-4}^{(i}\alpha_{-2}^j\alpha_{-1}^{k)}+2040\alpha_{-3}^{(i}\alpha_{-3}^j\alpha_{-1}^{k)}+300\alpha_{-3}^{(i}\alpha_{-2}^j\alpha_{-2}^{k)}+\\
        &&-240\alpha_{-3}^{(i}\alpha_{-1}^j\alpha_{-1}^{k)}(\alpha_{-1}\cdot\alpha_{-1})-2070\alpha_{-1}^{i}\alpha_{-1}^j\alpha_{-1}^{k}(\alpha_{-3}\cdot\alpha_{-1})+\\
        &&+450\alpha_{-2}^{(i}\alpha_{-2}^j\alpha_{-1}^{k)}(\alpha_{-1}\cdot\alpha_{-1})-160\alpha_{-2}^{(i}\alpha_{-1}^j\alpha_{-1}^{k)}(\alpha_{-2}\cdot\alpha_{-1})+\\
        &&+1695\alpha_{-1}^{i}\alpha_{-1}^j\alpha_{-1}^{k}(\alpha_{-2}\cdot\alpha_{-2})+210\alpha_{-1}^{i}\alpha_{-1}^j\alpha_{-1}^{k}(\alpha_{-1}\cdot\alpha_{-1})^2 \\
        \rule{0pt}{35pt}
        \ydiagram{4,1}&:& 2\alpha_{-3}^{[i_1}\alpha_{-1}^{i_5]}\alpha_{-1}^{i_2}\alpha_{-1}^{i_3}\alpha_{-1}^{i_4} -\alpha_{-2}^{[i_1}\alpha_{-1}^{i_5]}\alpha_{-2}^{(i_2}\alpha_{-1}^{i_3}\alpha_{-1}^{i_4)} \\
        \rule{0pt}{35pt}
        \ydiagram{3,1}&:& 4\alpha_{-4}^{[i}\alpha_{-1}^{l]}\alpha_{-1}^j\alpha_{-1}^k +30(\alpha_{-3}^{(j}\alpha_{-1}^k\alpha_{-1}^{l)}\alpha_{-2}^i-\alpha_{-3}^{(i}\alpha_{-1}^j\alpha_{-1}^{k)}\alpha_{-2}^l)+\\
        &&-56\alpha_{-2}^{[i}\alpha_{-1}^{l]}\alpha_{-2}^j\alpha_{-2}^k-11\alpha_{-2}^{[i}\alpha_{-1}^{l]}\alpha_{-1}^j\alpha_{-1}^k(\alpha_{-1}\cdot\alpha_{-1}) \\
        \rule{0pt}{35pt}
        \ydiagram{4}&:& 44\alpha_{-4}^{(i}\alpha_{-1}^{j}\alpha_{-1}^{k}\alpha_{-1}^{l)}-30\alpha_{-3}^{(i}\alpha_{-2}^{j}\alpha_{-1}^{k}\alpha_{-1}^{l)}+48\alpha_{-2}^{(i}\alpha_{-2}^{j}\alpha_{-2}^{k}\alpha_{-1}^{l)}+\\
        &&+3\alpha_{-2}^{(i}\alpha_{-1}^{j}\alpha_{-1}^{k}\alpha_{-1}^{l)}(\alpha_{-1}\cdot\alpha_{-1})-16\alpha_{-1}^{i}\alpha_{-1}^{j}\alpha_{-1}^{k}\alpha_{-1}^{l}(\alpha_{-2}\cdot\alpha_{-1}) \\
        \rule{0pt}{35pt}
        \ydiagram{3,2}&:& \alpha_{-2}^{[i_1}\alpha_{-1}^{i_4]}\alpha_{-2}^{[i_2}\alpha_{-1}^{i_5]}\alpha_{-1}^{i_3} \\
        \rule{0pt}{35pt}
        \ydiagram{2,1,1}&:& \alpha_{-3}^{[i}\alpha_{-2}^k\alpha_{-1}^{l]}\alpha_{-1}^j \\
    \end{array}
\end{equation}

\begin{equation}
    \def\arraystretch{1.3}
    \begin{array}{lcl}
        \ydiagram{2}&:& 212\alpha_{-5}^{(i}\alpha_{-2}^{j)}-176\alpha_{-4}^{(i}\alpha_{-3}^{j)}-104\alpha_{-4}^{(i}\alpha_{-1}^{j)}(\alpha_{-1}\cdot\alpha_{-1})-2176\alpha_{-1}^i\alpha_{-1}^j(\alpha_{-4}\cdot\alpha_{-1})+\\
        &&-228\alpha_{-3}^{(i}\alpha_{-2}^{j)}(\alpha_{-1}\cdot\alpha_{-1})+208\alpha_{-3}^{(i}\alpha_{-1}^{j)}(\alpha_{-2}\cdot\alpha_{-1})+1404\alpha_{-2}^{(i}\alpha_{-1}^{j)}(\alpha_{-3}\cdot\alpha_{-1})+\\
        &&+1296\alpha_{-1}^i\alpha_{-1}^j(\alpha_{-3}\cdot\alpha_{-2})+352\alpha_{-2}^i\alpha_{-2}^j(\alpha_{-2}\cdot\alpha_{-1})-1104\alpha_{-2}^{(i}\alpha_{-1}^{j)}(\alpha_{-2}\cdot\alpha_{-2})+\\
        &&-105\alpha_{-2}^{(i}\alpha_{-1}^{j)}(\alpha_{-1}\cdot\alpha_{-1})^2+280\alpha_{-1}^i\alpha_{-1}^j(\alpha_{-2}\cdot\alpha_{-1})(\alpha_{-1}\cdot\alpha_{-1}) \\
        \rule{0pt}{35pt}
        \ydiagram{1}&:& 600\alpha_{-7}^i+7532\alpha_{-5}^i(\alpha_{-1}\cdot\alpha_{-1})-14840\alpha_{-1}^i(\alpha_{-5}\cdot\alpha_{-1})-17920\alpha_{-4}^i(\alpha_{-2}\cdot\alpha_{-1})+\\
        &&+42560\alpha_{-2}^i(\alpha_{-4}\cdot\alpha_{-1})+38080\alpha_{-1}^i(\alpha_{-4}\cdot\alpha_{-2})-23240\alpha_{-3}^i(\alpha_{-3}\cdot\alpha_{-1})+\\
        &&-26040\alpha_{-1}^i(\alpha_{-3}\cdot\alpha_{-3})+26600\alpha_{-3}^i(\alpha_{-2}\cdot\alpha_{-2})-28000\alpha_{-2}^i(\alpha_{-3}\cdot\alpha_{-2})+\\
        &&+490\alpha_{-3}^i(\alpha_{-1}\cdot\alpha_{-1})+12740\alpha_{-1}^i(\alpha_{-3}\cdot\alpha_{-1})(\alpha_{-1}\cdot\alpha_{-1})+\\
        &&-3920\alpha_{-2}^i(\alpha_{-2}\cdot\alpha_{-1})(\alpha_{-1}\cdot\alpha_{-1})-8820\alpha_{-1}^i(\alpha_{-2}\cdot\alpha_{-2})(\alpha_{-1}\cdot\alpha_{-1})+\\
        &&-735\alpha_{-1}^i(\alpha_{-1}\cdot\alpha_{-1})^3
\end{array}
\end{equation}
The new seeds are
\begin{equation}
\def\arraystretch{1.3}
\begin{array}{lcl}
        \ydiagram{3,1}&:& 12\alpha_{-4}^{[i}\alpha_{-1}^{l]}\alpha_{-1}^j\alpha_{-1}^k -10(\alpha_{-2}^{(j}\alpha_{-1}^k\alpha_{-1}^{l)}\alpha_{-3}^i-\alpha_{-2}^{(i}\alpha_{-1}^j\alpha_{-1}^{k)}\alpha_{-3}^l)+\\
        &&+12\alpha_{-2}^{[i}\alpha_{-1}^{l]}\alpha_{-2}^j\alpha_{-2}^k-3\alpha_{-2}^{[i}\alpha_{-1}^{l]}\alpha_{-1}^j\alpha_{-1}^k(\alpha_{-1}\cdot\alpha_{-1}) \\
        \rule{0pt}{35pt}
        \ydiagram{2,1}&:& 32\alpha_{-5}^{[i}\alpha_{-1}^{k]}\alpha_{-1}^j-88(\alpha_{-4}^{(i}\alpha_{-1}^{j)}\alpha_{-2}^k-\alpha_{-4}^{(k}\alpha_{-1}^{j)}\alpha_{-2}^i)-112\alpha_{-3}^{[i}\alpha_{-1}^{k]}\alpha_{-3}^j+190\alpha_{-3}^{[i}\alpha_{-2}^{k]}\alpha_{-2}^j+\\
        &&+4\alpha_{-3}^{[i}\alpha_{-1}^{k]}\alpha_{-1}^j(\alpha_{-1}\cdot\alpha_{-1})+9\alpha_{-2}^{[i}\alpha_{-1}^{k]}\alpha_{-2}^j(\alpha_{-1}\cdot\alpha_{-1})-20\alpha_{-2}^{[i}\alpha_{-1}^{k]}\alpha_{-1}^j(\alpha_{-2}\cdot\alpha_{-1}) \\
        \rule{0pt}{35pt}
        \ydiagram{2,1}&:& 168\alpha_{-5}^{[i}\alpha_{-1}^{k]}\alpha_{-1}^j-1320(\alpha_{-4}^{(i}\alpha_{-2}^{j)}\alpha_{-1}^k-\alpha_{-4}^{(j}\alpha_{-2}^{k)}\alpha_{-1}^i)+2360\alpha_{-3}^{[i}\alpha_{-1}^{k]}\alpha_{-3}^j-350\alpha_{-3}^{[i}\alpha_{-2}^{k]}\alpha_{-2}^j+\\
        &&-320\alpha_{-3}^{[i}\alpha_{-1}^{k]}\alpha_{-1}^j(\alpha_{-1}\cdot\alpha_{-1})+435\alpha_{-2}^{[i}\alpha_{-1}^{k]}\alpha_{-2}^j(\alpha_{-1}\cdot\alpha_{-1})+60\alpha_{-2}^{[i}\alpha_{-1}^{k]}\alpha_{-1}^j(\alpha_{-2}\cdot\alpha_{-1}) \\
        \rule{0pt}{35pt}
        \ydiagram{1,1}&:& 32\alpha_{-6}^{[i}\alpha_{-1}^{j]}+100\alpha_{-5}^{[i}\alpha_{-2}^{j]}-400\alpha_{-4}^{[i}\alpha_{-3}^{j]}-56\alpha_{-4}^{[i}\alpha_{-1}^{j]}(\alpha_{-1}\cdot\alpha_{-1})-28\alpha_{-3}^{[i}\alpha_{-2}^{j]}(\alpha_{-1}\cdot\alpha_{-1})+\\
        &&+420\alpha_{-2}^{[i}\alpha_{-1}^{j]}(\alpha_{-3}\cdot\alpha_{-1})-280\alpha_{-2}^{[i}\alpha_{-1}^{j]}(\alpha_{-2}\cdot\alpha_{-2})-35\alpha_{-2}^{[i}\alpha_{-1}^{j]}(\alpha_{-1}\cdot\alpha_{-1})^2 \\
        \rule{0pt}{35pt}
        \ydiagram{1,1}&:& 240\alpha_{-6}^{[i}\alpha_{-1}^{j]}+540\alpha_{-5}^{[i}\alpha_{-2}^{j]}-2342\alpha_{-4}^{[i}\alpha_{-3}^{j]}-525\alpha_{-4}^{[i}\alpha_{-1}^{j]}(\alpha_{-1}\cdot\alpha_{-1})+56\alpha_{-3}^{[i}\alpha_{-1}^{j]}(\alpha_{-2}\cdot\alpha_{-1})+\\
        &&+3276\alpha_{-2}^{[i}\alpha_{-1}^{j]}(\alpha_{-3}\cdot\alpha_{-1})-2184\alpha_{-2}^{[i}\alpha_{-1}^{j]}(\alpha_{-2}\cdot\alpha_{-2})-252\alpha_{-2}^{[i}\alpha_{-1}^{j]}(\alpha_{-1}\cdot\alpha_{-1})^2 \\
        \rule{0pt}{35pt}
        \ydiagram{1}&:& 2360\alpha_{-7}^i+39196\alpha_{-5}^i(\alpha_{-1}\cdot\alpha_{-1})-72120\alpha_{-1}^i(\alpha_{-5}\cdot\alpha_{-1})-88000\alpha_{-4}^i(\alpha_{-2}\cdot\alpha_{-1})+\\
        &&+225280\alpha_{-2}^i(\alpha_{-4}\cdot\alpha_{-1})+150400\alpha_{-1}^i(\alpha_{-4}\cdot\alpha_{-2})-133960\alpha_{-3}^i(\alpha_{-3}\cdot\alpha_{-1})+\\
        &&-96360\alpha_{-1}^i(\alpha_{-3}\cdot\alpha_{-3})+143200\alpha_{-3}^i(\alpha_{-2}\cdot\alpha_{-2})-146720\alpha_{-2}^i(\alpha_{-3}\cdot\alpha_{-2})+\\
        &&+4310\alpha_{-3}^i(\alpha_{-1}\cdot\alpha_{-1})+56260\alpha_{-1}^i(\alpha_{-3}\cdot\alpha_{-1})(\alpha_{-1}\cdot\alpha_{-1})+\\
        &&-22000\alpha_{-2}^i(\alpha_{-2}\cdot\alpha_{-1})(\alpha_{-1}\cdot\alpha_{-1})-39600\alpha_{-1}^i(\alpha_{-2}\cdot\alpha_{-2})(\alpha_{-1}\cdot\alpha_{-1})+\\
        &&+3200\alpha_{-1}^i(\alpha_{-2}\cdot\alpha_{-1})^2-3345\alpha_{-1}^i(\alpha_{-1}\cdot\alpha_{-1})^3
\end{array}
\end{equation}

\subsection{Level 8}
At level 8, these are the particles. Their oscillator structures can be recovered from the notebooks.
\begin{center}
    \begin{tabular}{lll}
        \ydiagram{8} & \ydiagram{6}  & \ydiagram{5}\\
        \rule{0pt}{7ex}
        \ydiagram{4}($\times3$) &\ydiagram{3}($\times2$) &\ydiagram{2}($\times4$) \quad \ydiagram{1}($\times2$)\quad ${}^\bullet$($\times2$)\\
        \rule{0pt}{7ex}
        \ydiagram{6,1}& \ydiagram{5,1}& \ydiagram{4,1}($\times2$)\\
        \rule{0pt}{7ex}
        \ydiagram{3,1}($\times2$)& \ydiagram{2,1}($\times3$)& \ydiagram{1,1}\\
        \rule{0pt}{7ex}
        \ydiagram{4,2}& \ydiagram{3,2}& \ydiagram{2,2}($\times2$)\\
        \rule{0pt}{7ex}
        \ydiagram{3,1,1} & \ydiagram{2,1,1}& \ydiagram{1,1,1}
    \end{tabular}
\end{center} 
\section{List of full Regge trajectories}
\label{app:trajectories}
In this appendix we provide the maximally spinning representative for each particle in the open bosonic string spectrum at depth up to 4, where the depth of a trajectory is defined as
\begin{equation}
    d=\text{level}-(\text{number of boxes in the Young tableau})\,,
\end{equation}
constant for all particles belonging to the same trajectory. Expressions are written as functions of the level $N$, and they hold only at a level sufficiently large so that all the tensor structures involved exist. This sufficiently high level is specified for each trajectory. Trajectories could start at a lower level, but our formulas will in general involve oscillator structures that do not exist yet, so they will not apply. Dots indicate that oscillators with free Lorentz index and Virasoro index $-1$ have been omitted. This allows us to specify the tensor structure keeping the level on the Regge trajectory arbitrary.

\subsection{Depth 0}
We only have the leading Regge trajectory, that starts from level 0 with the tachyon
\begin{equation}
    \overbrace{\ydiagramalign{3}\dots\ydiagramalign{1}}^{N}
\end{equation}
and the expression for the oscillator structure is
\begin{equation}
    \alpha_{-1}^{i_1}\dots\alpha_{-1}^{i_N}
\end{equation}

\subsection{Depth 1}
We have a single $L-$shaped trajectory, that starts from level 3 with the spin 2 antisymmetric $\ydiagramalign{1,1}$
\begin{equation}    %
   \textstyle \overbrace{\ydiagramalign{3,1}}^{N-2}
\end{equation}
and the expression for the oscillator structure is
\begin{equation}
    \alpha_{-2}^{[i_1}\alpha_{-1}^{i_{2}]}\dots\alpha_{-1}^{i_{N-2}}
\end{equation}

\subsection{Depth 2}
We have three trajectories, one fully symmetric $\ydiagram{2}$ starting from level 4 (generation 2), one thin $L-$shaped $\ydiagramalign{2,1}$ starting at level 5 (generation 1), and one thick $L-$shaped $\ydiagramalign{2,2}$ that starts at level 6 (generation 0).

The fully symmetric trajectory is given by
\begin{equation}
\begin{split}
    (2 N^2+10 N-48)\alpha_{-3}^{(i_1}\dots\alpha_{-1}^{i_{N-2})} -(6 N+18)\alpha_{-2}^{(i_1}\alpha_{-2}^{i_2}\dots\alpha_{-1}^{i_{N-2})}+\\
    -(3 N^2-15 N+18)\alpha_{-1}^{i_1}\dots\alpha_{-1}^{i_{N-2}}(\alpha_{-1}\cdot\alpha_{-1})
\end{split}
\end{equation}

The thin $L-$shaped is given by
\begin{equation}
    (2N-8)\alpha_{-3}^{[i_1}\alpha_{-1}^{i_2]}\dots\alpha_{-1}^{i_{N-2}} -3\alpha_{-2}^{[i_1}\alpha_{-1}^{i_2]}\alpha_{-2}^{(i_3}\dots\alpha_{-1}^{i_{N-2})}
\end{equation}

The thick $L-$shaped is given by
\begin{equation}
    \alpha_{-2}^{[i_1}\alpha_{-1}^{i_2]}\alpha_{-2}^{[i_3}\alpha_{-1}^{i_4]}\dots\alpha_{-1}^{i_{N-4}}
\end{equation}

\subsection{Depth 3}
We have two trajectories of Weinberg states (generation 0), one $\ydiagram{1,1,1}$ starting from level 6, and one $\ydiagramalign{3,3}$ starting at level 9. In addition, there are four higher generation trajectories: one $\ydiagram{3,2}$ that starts at level 8 (generation 1), two $\ydiagram{3,1}$ (generation 2) starting at level 7, and one $\ydiagram{3}$ (generation 3) starting at level 6.

The $\ydiagram{1,1,1}$ is given by
\begin{equation}
    \alpha_{-3}^{[{i_1}}\alpha_{-2}^{i_2}\alpha_{-1}^{i_3]}\dots\alpha_{-1}^{i_{N-3}}
\end{equation}
while the $\ydiagram{3,3}$ is given by
\begin{equation}
    \alpha_{-2}^{[i_1}\alpha_{-1}^{i_2]}\alpha_{-2}^{[i_3}\alpha_{-1}^{i_4]}\alpha_{-2}^{[i_5}\alpha_{-1}^{i_6]}\dots\alpha_{-1}^{i_{N-3}}
\end{equation}

$\ydiagram{3,2}$ is given by
\begin{equation}
    (N-7)\alpha_{-3}^{[i_1}\alpha_{-1}^{i_2]}\alpha_{-2}^{[i_3}\alpha_{-1}^{i_4]}\alpha_{-1}^{i_5}\dots\alpha_{-1}^{i_{N-2}}-3\alpha_{-2}^{[i_1}\alpha_{-1}^{i_2]}\alpha_{-2}^{[i_3}\alpha_{-1}^{i_4]}\alpha_{-2}^{i_5}\dots\alpha_{-1}^{i_{N-2}}
\end{equation}

The two $\ydiagram{3,1}$ are given by
\begin{equation}
\begin{split}
    (2N^3-36N^2+198N-324)\alpha_{-4}^{[{i_1}}\alpha_{-1}^{{i_2}]}\alpha_{-1}^{i_3}\alpha_{-1}^{i_4}\dots\alpha_{-1}^{i_{N-3}}+\\
    -(2 N^3-2 N^2-108 N+288)(\alpha_{-3}^{({i_2}}\alpha_{-1}^{i_3}\alpha_{-1}^{{i_4}}\dots\alpha_{-1}^{i_{N-3})}\alpha_{-2}^{i_1}-\alpha_{-3}^{({i_1}}\alpha_{-1}^{i_3}\alpha_{-1}^{{i_4}}\dots\alpha_{-1}^{i_{N-3})}\alpha_{-2}^{i_2})+\\
    +(6 N^2-4 N-42)\alpha_{-2}^{[{i_1}}\alpha_{-1}^{{i_2}]}\alpha_{-2}^{(i_3}\alpha_{-2}^{i_4}\dots\alpha_{-1}^{i_{N-3})}+\\
    +(3 N^3-37 N^2+144 N-180)\alpha_{-2}^{[{i_1}}\alpha_{-1}^{{i_2}]}\alpha_{-1}^{i_3}\alpha_{-1}^{i_4}\dots\alpha_{-1}^{i_{N-3}}(\alpha_{-1}\cdot\alpha_{-1})
\end{split}
\end{equation}
and
\begin{equation}
\begin{split}
    (N^4-7 N^3-27 N^2+279 N-486)\alpha_{-4}^{[{i_1}}\alpha_{-1}^{{i_2}]}\alpha_{-1}^{i_3}\alpha_{-1}^{i_4}\dots\alpha_{-1}^{i_{N-3}}+\\
    -(2 N^3-2 N^2-108 N+288)(\alpha_{-2}^{({i_2}}\alpha_{-1}^{i_3}\alpha_{-1}^{{i_4}}\dots\alpha_{-1}^{i_{N-3})}\alpha_{-3}^{i_1}-\alpha_{-2}^{({i_1}}\alpha_{-1}^{i_3}\alpha_{-1}^{{i_4}}\dots\alpha_{-1}^{i_{N-3})}\alpha_{-3}^{i_2})+\\
    +(N^2+26 N-87)\alpha_{-2}^{[{i_1}}\alpha_{-1}^{{i_2}]}\alpha_{-2}^{(i_3}\alpha_{-2}^{i_4}\dots\alpha_{-1}^{i_{N-3})}+\\
    -(2 N^3-23 N^2+81 N-90)\alpha_{-2}^{[{i_1}}\alpha_{-1}^{{i_2}]}\alpha_{-1}^{i_3}\alpha_{-1}^{i_4}\dots\alpha_{-1}^{i_{N-3}}(\alpha_{-1}\cdot\alpha_{-1})
\end{split}
\end{equation}

The $\ydiagram{3}$ is given by
\begin{equation}
\begin{split}
    (2 N^3+12 N^2-230 N+600)\alpha_{-4}^{({i_1}}\alpha_{-1}^{i_2}\alpha_{-1}^{{i_3}}\dots\alpha_{-1}^{i_{N-3})}-(6 N^2+18 N-240)\alpha_{-3}^{({i_1}}\alpha_{-2}^{i_2}\alpha_{-1}^{{i_3}}\dots\alpha_{-1}^{i_{N-3})}+\\
    +(24 N+120)\alpha_{-2}^{(i_1}\alpha_{-2}^{i_2}\alpha_{-2}^{i_3}\dots\alpha_{-1}^{i_{N-3})}+(3 N^2-27 N+60)\alpha_{-2}^{({i_1}}\alpha_{-1}^{i_2}\alpha_{-1}^{{i_3}}\dots\alpha_{-1}^{i_{N-3})}(\alpha_{-1}\cdot\alpha_{-1})+\\
    -(4 N^3-48 N^2+188 N-240)\alpha_{-1}^{i_1}\alpha_{-1}^{i_2}\alpha_{-1}^{i_3}\dots\alpha_{-1}^{i_{N-3}}(\alpha_{-2}\cdot\alpha_{-1})
\end{split}
\end{equation}

\subsection{Depth 4}
We have two trajectories of Weinberg states (generation 0), one $\ydiagram{2,2,1}$ starting from level 9, and one $\ydiagramalign{4,4}$ starting at level 12. In addition, there are eleven higher generation trajectories: one $\ydiagram{4,3}$ that starts at level 11 (generation 1), one $\ydiagram{2,1,1}$ that starts at level 8 (generation 1), three $\ydiagram{4,2}$ that start at level 10 (generation 2), three $\ydiagram{4,1}$ that start at level 9 (generation 3), and three $\ydiagram{4}$ that start at level 8 (generation 4).

The $\ydiagram{2,2,1}$ is given by
\begin{equation}
    \alpha_{-3}^{[{i_1}}\alpha_{-2}^{i_2}\alpha_{-1}^{i_3]}\alpha_{-2}^{[{i_4}}\alpha_{-1}^{i_5]}\dots\alpha_{-1}^{i_{N-4}}
\end{equation}
while the $\ydiagram{4,4}$ is given by
\begin{equation}
    \alpha_{-2}^{[i_1}\alpha_{-1}^{i_2]}\alpha_{-2}^{[i_3}\alpha_{-1}^{i_4]}\alpha_{-2}^{[i_5}\alpha_{-1}^{i_6]}\alpha_{-2}^{[i_7}\alpha_{-1}^{i_8]}\dots\alpha_{-1}^{i_{N-4}}
\end{equation}

The $\ydiagram{4,3}$ is given by
\begin{equation}
    (2N-20)(\alpha_{-3}^{(i_1}\alpha_{-2}^{i_2}\alpha_{-2}^{i_3)}\alpha_{-1}^{i_4}\alpha_{-1}^{i_5}\alpha_{-1}^{i_6}-\text{antisym.})\alpha_{-1}^{i_7}\dots\alpha_{-1}^{i_{N-4}}-9
    \alpha_{-2}^{[i_1}\alpha_{-1}^{i_4]}\alpha_{-2}^{[i_2}\alpha_{-1}^{i_5]}\alpha_{-2}^{[i_3}\alpha_{-1}^{i_6]}\alpha_{-2}^{(i_{7}}\dots\alpha_{-1}^{i_{N-4})}
\end{equation}
where $\text{antisym.}$ means that one should antisymmetrize $i_1\leftrightarrow i_4,\,i_2\leftrightarrow i_5,\,i_3\leftrightarrow i_6$.

$\ydiagram{2,1,1}$ is given by
\begin{equation}
    (N-7)\alpha_{-4}^{[{i_1}}\alpha_{-2}^{i_2}\alpha_{-1}^{i_3]}\dots\alpha_{-1}^{i_{N-4}}-2\alpha_{-3}^{[{i_1}}\alpha_{-2}^{i_2}\alpha_{-1}^{i_3]}\alpha_{-2}^{(i_{4}}\dots\alpha_{-1}^{i_{N-4})}
\end{equation}

The three $\ydiagram{4,2}$ are given by
\begin{equation}
\begin{split}
    (9N-45)\bigg[\alpha_{-4}^{(i_1}\alpha_{-2}^{i_2)}\alpha_{-1}^{i_3}\alpha_{-1}^{i_4}-\{i_1\leftrightarrow i_3\}-\{i_2\leftrightarrow i_4\}+\{i_1\leftrightarrow i_3,\,i_2\leftrightarrow i_4\}\bigg]\alpha_{-1}^{i_{5}}\alpha_{-1}^{i_{6}}\dots\alpha_{-1}^{i_{N-4}}+\\
    -(20N-100)\alpha_{-3}^{[i_1}\alpha_{-1}^{i_3]}\alpha_{-3}^{[i_2}\alpha_{-1}^{i_4]}\alpha_{-1}^{i_{5}}\alpha_{-1}^{i_{6}}\dots\alpha_{-1}^{i_{N-4}}+\\
    -24\bigg[\alpha_{-3}^{(i_5}\alpha_{-1}^{i_6}\dots\alpha_{-1}^{i_{N-4})}\alpha_{-2}^{[i_1}\alpha_{-1}^{i_3]}\alpha_{-2}^{[i_2}\alpha_{-1}^{i_4]}+\alpha_{-1}^{i_5}\alpha_{-1}^{i_6}\dots\alpha_{-1}^{i_{N-4}}(\alpha_{-2}^{[i_2}\alpha_{-3}^{i_4]}\alpha_{-2}^{[i_1}\alpha_{-1}^{i_3]}+\alpha_{-2}^{[i_1}\alpha_{-3}^{i_3]}\alpha_{-2}^{[i_2}\alpha_{-1}^{i_4]})\bigg]+\\
    +12\bigg[ \alpha_{-3}^{i_1}\alpha_{-2}^{[i_2}\alpha_{-1}^{i_4]}\alpha_{-2}^{(i_3}\alpha_{-1}^{i_5}\alpha_{-1}^{i_6}\dots\alpha_{-1}^{i_{N-4})} + \{ i_1\leftrightarrow i_2,\,i_3\leftrightarrow i_4 \} +\\
    + \{ i_1\leftrightarrow i_4,\,i_2\leftrightarrow i_3 \} + \{ i_1\leftrightarrow i_3,\,i_2\leftrightarrow i_4 \} \bigg]
\end{split}
\end{equation}
and
\begin{equation}
\begin{split}
    (9N^2-135N+486)\bigg[\alpha_{-4}^{(i_1}\alpha_{-2}^{i_2)}\alpha_{-1}^{i_3}\alpha_{-1}^{i_4}-\{i_1\leftrightarrow i_3\}-\{i_2\leftrightarrow i_4\}+\\
    +\{i_1\leftrightarrow i_3,\,i_2\leftrightarrow i_4\}\bigg]\alpha_{-1}^{i_{5}}\alpha_{-1}^{i_{6}}\dots\alpha_{-1}^{i_{N-4}}+\\
    -(12N^2-164N+504)\alpha_{-3}^{[i_1}\alpha_{-1}^{i_3]}\alpha_{-3}^{[i_2}\alpha_{-1}^{i_4]}\alpha_{-1}^{i_{5}}\alpha_{-1}^{i_{6}}\dots\alpha_{-1}^{i_{N-4}}+\\
    -(24N-216)\bigg[\alpha_{-3}^{(i_5}\alpha_{-1}^{i_6}\dots\alpha_{-1}^{i_{N-4})}\alpha_{-2}^{[i_1}\alpha_{-1}^{i_3]}\alpha_{-2}^{[i_2}\alpha_{-1}^{i_4]}+\\
    +\alpha_{-1}^{i_5}\alpha_{-1}^{i_6}\dots\alpha_{-1}^{i_{N-4}}(\alpha_{-2}^{[i_2}\alpha_{-3}^{i_4]}\alpha_{-2}^{[i_1}\alpha_{-1}^{i_3]}+\alpha_{-2}^{[i_1}\alpha_{-3}^{i_3]}\alpha_{-2}^{[i_2}\alpha_{-1}^{i_4]})\bigg]+\\
    +36 \alpha_{-2}^{[i_1}\alpha_{-1}^{i_3]}\alpha_{-2}^{[i_2}\alpha_{-1}^{i_4]}\alpha_{-2}^{(i_5}\alpha_{-2}^{i_6}\dots\alpha_{-1}^{i_{N-4})}
\end{split}
\end{equation}
and
\begin{equation}
\begin{split}
    (9N-9)\bigg[\alpha_{-4}^{(i_1}\alpha_{-2}^{i_2)}\alpha_{-1}^{i_3}\alpha_{-1}^{i_4}-\{i_1\leftrightarrow i_3\}-\{i_2\leftrightarrow i_4\}+\\
    +\{i_1\leftrightarrow i_3,\,i_2\leftrightarrow i_4\}\bigg]\alpha_{-1}^{i_{5}}\alpha_{-1}^{i_{6}}\dots\alpha_{-1}^{i_{N-4}}+\\
    -(12N+20)\alpha_{-3}^{[i_1}\alpha_{-1}^{i_3]}\alpha_{-3}^{[i_2}\alpha_{-1}^{i_4]}\alpha_{-1}^{i_{5}}\alpha_{-1}^{i_{6}}\dots\alpha_{-1}^{i_{N-4}}+\\
    -12\bigg[\alpha_{-3}^{(i_5}\alpha_{-1}^{i_6}\dots\alpha_{-1}^{i_{N-4})}\alpha_{-2}^{[i_1}\alpha_{-1}^{i_3]}\alpha_{-2}^{[i_2}\alpha_{-1}^{i_4]}+\\
    +\alpha_{-1}^{i_5}\alpha_{-1}^{i_6}\dots\alpha_{-1}^{i_{N-4}}(\alpha_{-2}^{[i_2}\alpha_{-3}^{i_4]}\alpha_{-2}^{[i_1}\alpha_{-1}^{i_3]}+\alpha_{-2}^{[i_1}\alpha_{-3}^{i_3]}\alpha_{-2}^{[i_2}\alpha_{-1}^{i_4]})\bigg]+\\
    -18 \alpha_{-2}^{[i_1}\alpha_{-1}^{i_3]}\alpha_{-2}^{[i_2}\alpha_{-1}^{i_4]}(\alpha_{-1}\cdot\alpha_{-1})\alpha_{-1}^{i_5}\alpha_{-1}^{i_6}\dots\alpha_{-1}^{i_{N-4}}
\end{split}
\end{equation}
where the appearance of $\{i_a\leftrightarrow i_b,\dots\}$ after some explicit tensor structure $t$ indicates that $t$ should be added again, up to a relabeling of the indices as prescribed inside the curly brackets.

The three $\ydiagram{4,1}$ are given by
\begin{equation}
\begin{split}
    (12 N^4-300 N^3+2760 N^2-11040 N+16128)\alpha_{-5}^{[i_1}\alpha_{-1}^{i_2]}\alpha_{-1}^{i_3}\alpha_{-1}^{i_4}\alpha_{-1}^{i_5}\dots\alpha_{-1}^{i_{N-4}}+\\
    -(15 N^3-270 N^2+1515 N-2520)(\alpha_{-2}^{[i_1}\alpha_{-4}^{i_2]}\alpha_{-1}^{i_{3}}\alpha_{-1}^{i_{4}}\alpha_{-1}^{i_{5}}\dots\alpha_{-1}^{i_{N-4}} + \alpha_{-2}^{[i_1}\alpha_{-1}^{i_2]}\alpha_{-4}^{(i_{3}}\alpha_{-1}^{i_{4}}\alpha_{-1}^{i_{5}}\dots\alpha_{-1}^{i_{N-4})})+\\
    -(30 N^3-585 N^2+3705 N-7560)(\alpha_{-4}^{i_1}\alpha_{-2}^{(i_2}\alpha_{-1}^{i_3}\alpha_{-1}^{i_4}\alpha_{-1}^{i_5}\dots\alpha_{-1}^{i_{N-4})} - \{ i_1 \leftrightarrow i_2 \})+\\
    -(20 N^3-380 N^2+2320 N-4480)\alpha_{-3}^{[i_1}\alpha_{-1}^{i_2]}\alpha_{-3}^{(i_3}\alpha_{-1}^{i_4}\alpha_{-1}^{i_5}\dots\alpha_{-1}^{i_{N-4})}+\\
    +(60 N^2-690 N+1680)(\alpha_{-2}^{[i_1}\alpha_{-3}^{i_2]}\alpha_{-2}^{(i_3}\alpha_{-1}^{i_4}\alpha_{-1}^{i_5}\dots\alpha_{-1}^{i_{N-4})} + \alpha_{-2}^{[i_1}\alpha_{-1}^{i_2]}\alpha_{-3}^{(i_3}\alpha_{-2}^{i_4}\alpha_{-1}^{i_5}\dots\alpha_{-1}^{i_{N-4})} )+\\
    +(90 N^2-1140 N+3360)(\alpha_{-3}^{i_1}\alpha_{-2}^{(i_2}\alpha_{-2}^{i_3}\alpha_{-1}^{i_4}\alpha_{-1}^{i_5}\dots\alpha_{-1}^{i_{N-4})} - \{ i_1\leftrightarrow i_2 \})+\\
    -(315 N-1260)\alpha_{-2}^{[i_1}\alpha_{-1}^{i_2]}\alpha_{-2}^{(i_3}\alpha_{-2}^{i_4}\alpha_{-2}^{i_5}\dots\alpha_{-1}^{i_{N-4})}
\end{split}
\end{equation}
and
\begin{equation}
\begin{split}
    (72 N^3-1044 N^2+4968 N-7776)\alpha_{-5}^{[i_1}\alpha_{-1}^{i_2]}\alpha_{-1}^{i_3}\alpha_{-1}^{i_4}\alpha_{-1}^{i_5}\dots\alpha_{-1}^{i_{N-4}}+\\
    -(90 N^2-360 N+270)(\alpha_{-2}^{[i_1}\alpha_{-4}^{i_2]}\alpha_{-1}^{i_{3}}\alpha_{-1}^{i_{4}}\alpha_{-1}^{i_{5}}\dots\alpha_{-1}^{i_{N-4}} + \alpha_{-2}^{[i_1}\alpha_{-1}^{i_2]}\alpha_{-4}^{(i_{3}}\alpha_{-1}^{i_{4}}\alpha_{-1}^{i_{5}}\dots\alpha_{-1}^{i_{N-4})})+\\
    -(180 N^2-990 N+810)(\alpha_{-4}^{i_1}\alpha_{-2}^{(i_2}\alpha_{-1}^{i_3}\alpha_{-1}^{i_4}\alpha_{-1}^{i_5}\dots\alpha_{-1}^{i_{N-4})} - \{ i_1 \leftrightarrow i_2 \})+\\
    +(20 N^2+1160 N-4960)\alpha_{-3}^{[i_1}\alpha_{-1}^{i_2]}\alpha_{-3}^{(i_3}\alpha_{-1}^{i_4}\alpha_{-1}^{i_5}\dots\alpha_{-1}^{i_{N-4})}+\\
    +(150 N-780)(\alpha_{-2}^{[i_1}\alpha_{-3}^{i_2]}\alpha_{-2}^{(i_3}\alpha_{-1}^{i_4}\alpha_{-1}^{i_5}\dots\alpha_{-1}^{i_{N-4})} + \alpha_{-2}^{[i_1}\alpha_{-1}^{i_2]}\alpha_{-3}^{(i_3}\alpha_{-2}^{i_4}\alpha_{-1}^{i_5}\dots\alpha_{-1}^{i_{N-4})} )+\\
    +(120 N-120)(\alpha_{-3}^{i_1}\alpha_{-2}^{(i_2}\alpha_{-2}^{i_3}\alpha_{-1}^{i_4}\alpha_{-1}^{i_5}\dots\alpha_{-1}^{i_{N-4})} - \{ i_1\leftrightarrow i_2 \})+\\
    -(210 N^2-2100 N+5040)\alpha_{-3}^{[i_1}\alpha_{-1}^{i_2]}(\alpha_{-1}\cdot\alpha_{-1})\alpha_{-1}^{i_3}\alpha_{-1}^{i_4}\alpha_{-1}^{i_5}\dots\alpha_{-1}^{i_{N-4}}+\\
    +(315 N-1260)\alpha_{-2}^{[i_1}\alpha_{-1}^{i_2]}(\alpha_{-1}\cdot\alpha_{-1})\alpha_{-2}^{(i_3}\alpha_{-1}^{i_4}\alpha_{-1}^{i_5}\dots\alpha_{-1}^{i_{N-4})}
\end{split}
\end{equation}
and
\begin{equation}
\begin{split}
    (4 N^3+26 N^2-564 N+1584)\alpha_{-5}^{[i_1}\alpha_{-1}^{i_2]}\alpha_{-1}^{i_3}\alpha_{-1}^{i_4}\alpha_{-1}^{i_5}\dots\alpha_{-1}^{i_{N-4}}+\\
    +(30 N^2+300 N-2430)(\alpha_{-2}^{[i_1}\alpha_{-4}^{i_2]}\alpha_{-1}^{i_{3}}\alpha_{-1}^{i_{4}}\alpha_{-1}^{i_{5}}\dots\alpha_{-1}^{i_{N-4}} + \alpha_{-2}^{[i_1}\alpha_{-1}^{i_2]}\alpha_{-4}^{(i_{3}}\alpha_{-1}^{i_{4}}\alpha_{-1}^{i_{5}}\dots\alpha_{-1}^{i_{N-4})})+\\
    -(10 N^2+50 N-1110)(\alpha_{-4}^{i_1}\alpha_{-2}^{(i_2}\alpha_{-1}^{i_3}\alpha_{-1}^{i_4}\alpha_{-1}^{i_5}\dots\alpha_{-1}^{i_{N-4})} - \{ i_1 \leftrightarrow i_2 \})+\\
    -(30 N^2+200 N-1280)\alpha_{-3}^{[i_1}\alpha_{-1}^{i_2]}\alpha_{-3}^{(i_3}\alpha_{-1}^{i_4}\alpha_{-1}^{i_5}\dots\alpha_{-1}^{i_{N-4})}+\\
    -(50 N+300)(\alpha_{-2}^{[i_1}\alpha_{-3}^{i_2]}\alpha_{-2}^{(i_3}\alpha_{-1}^{i_4}\alpha_{-1}^{i_5}\dots\alpha_{-1}^{i_{N-4})} + \alpha_{-2}^{[i_1}\alpha_{-1}^{i_2]}\alpha_{-3}^{(i_3}\alpha_{-2}^{i_4}\alpha_{-1}^{i_5}\dots\alpha_{-1}^{i_{N-4})} )+\\
    +(100 N+600)(\alpha_{-3}^{i_1}\alpha_{-2}^{(i_2}\alpha_{-2}^{i_3}\alpha_{-1}^{i_4}\alpha_{-1}^{i_5}\dots\alpha_{-1}^{i_{N-4})} - \{ i_1\leftrightarrow i_2 \})+\\
    +(35 N^2-350 N+840)\alpha_{-3}^{[i_1}\alpha_{-1}^{i_2]}(\alpha_{-1}\cdot\alpha_{-1})\alpha_{-1}^{i_3}\alpha_{-1}^{i_4}\alpha_{-1}^{i_5}\dots\alpha_{-1}^{i_{N-4}}+\\
    -(70 N^2-700 N+1680)\alpha_{-2}^{[i_1}\alpha_{-1}^{i_2]}(\alpha_{-2}\cdot\alpha_{-1})\alpha_{-1}^{i_3}\alpha_{-1}^{i_4}\alpha_{-1}^{i_5}\dots\alpha_{-1}^{i_{N-4}}
\end{split}
\end{equation}

Lastly, the three $\ydiagram{4}$ are given by
\begin{equation}
\begin{split}
    (36 N^4-240 N^3-3492 N^2+36096 N-85680)\alpha_{-5}^{(i_1}\alpha_{-1}^{i_2}\alpha_{-1}^{i_3}\alpha_{-1}^{i_4}\dots\alpha_{-1}^{i_{N-4})}+\\
    -(120 N^3-900 N^2-3540 N+27720)\alpha_{-4}^{(i_1}\alpha_{-2}^{i_2}\alpha_{-1}^{i_3}\alpha_{-1}^{i_4}\dots\alpha_{-1}^{i_{N-4})}+\\
    -(80 N^3-160 N^2-8080 N+36960)\alpha_{-3}^{(i_1}\alpha_{-3}^{i_2}\alpha_{-1}^{i_3}\alpha_{-1}^{i_4}\dots\alpha_{-1}^{i_{N-4})}+\\
    +(450 N^2-1230 N-13440)\alpha_{-3}^{(i_1}\alpha_{-2}^{i_2}\alpha_{-2}^{i_3}\alpha_{-1}^{i_4}\dots\alpha_{-1}^{i_{N-4})}+\\
    +(60 N^3-1080 N^2+6420 N-12600)\alpha_{-3}^{(i_1}\alpha_{-1}^{i_2}\alpha_{-1}^{i_3}\alpha_{-1}^{i_4}\dots\alpha_{-1}^{i_{N-4})}(\alpha_{-1}\cdot\alpha_{-1})+\\
    -(60 N^4-1320 N^3+10740 N^2-38280 N+50400)\alpha_{-1}^{i_1}\alpha_{-1}^{i_2}\alpha_{-1}^{i_3}\alpha_{-1}^{i_4}\dots\alpha_{-1}^{i_{N-4}}(\alpha_{-3}\cdot\alpha_{-1})+\\
    -(2430 N+6210)\alpha_{-2}^{(i_1}\alpha_{-2}^{i_2}\alpha_{-2}^{i_3}\alpha_{-2}^{i_4}\dots\alpha_{-1}^{i_{N-4})}+\\
    -(135 N^2-1755 N+5670)\alpha_{-2}^{(i_1}\alpha_{-2}^{i_2}\alpha_{-1}^{i_3}\alpha_{-1}^{i_4}\dots\alpha_{-1}^{i_{N-4})}(\alpha_{-1}\cdot\alpha_{-1})+\\
    +(60 N^3-1080 N^2+6420 N-12600)\alpha_{-2}^{(i_1}\alpha_{-1}^{i_2}\alpha_{-1}^{i_3}\alpha_{-1}^{i_4}\dots\alpha_{-1}^{i_{N-4})}(\alpha_{-2}\cdot\alpha_{-1})
\end{split}
\end{equation}
and
\begin{equation}
\begin{split}
    (54 N^4-1044 N^3+7074 N^2-19044 N+15120)\alpha_{-5}^{(i_1}\alpha_{-1}^{i_2}\alpha_{-1}^{i_3}\alpha_{-1}^{i_4}\dots\alpha_{-1}^{i_{N-4})}+\\
    -(180 N^3-3060 N^2+16920 N-30240)\alpha_{-4}^{(i_1}\alpha_{-2}^{i_2}\alpha_{-1}^{i_3}\alpha_{-1}^{i_4}\dots\alpha_{-1}^{i_{N-4})}+\\
    -(80 N^3-400 N^2-4960 N+26880)\alpha_{-3}^{(i_1}\alpha_{-3}^{i_2}\alpha_{-1}^{i_3}\alpha_{-1}^{i_4}\dots\alpha_{-1}^{i_{N-4})}+\\
    +(600 N^2-5640 N+10080)\alpha_{-3}^{(i_1}\alpha_{-2}^{i_2}\alpha_{-2}^{i_3}\alpha_{-1}^{i_4}\dots\alpha_{-1}^{i_{N-4})}+\\
    +(120 N^3-2160 N^2+12840 N-25200)\alpha_{-3}^{(i_1}\alpha_{-1}^{i_2}\alpha_{-1}^{i_3}\alpha_{-1}^{i_4}\dots\alpha_{-1}^{i_{N-4})}(\alpha_{-1}\cdot\alpha_{-1})+\\
    -(90 N^4-1980 N^3+16110 N^2-57420 N+75600)\alpha_{-1}^{i_1}\alpha_{-1}^{i_2}\alpha_{-1}^{i_3}\alpha_{-1}^{i_4}\dots\alpha_{-1}^{i_{N-4}}(\alpha_{-3}\cdot\alpha_{-1})+\\
    -(3240 N-9720)\alpha_{-2}^{(i_1}\alpha_{-2}^{i_2}\alpha_{-2}^{i_3}\alpha_{-2}^{i_4}\dots\alpha_{-1}^{i_{N-4})}+\\
    -(180 N^2-2340 N+7560)\alpha_{-2}^{(i_1}\alpha_{-2}^{i_2}\alpha_{-1}^{i_3}\alpha_{-1}^{i_4}\dots\alpha_{-1}^{i_{N-4})}(\alpha_{-1}\cdot\alpha_{-1})+\\
    +(45 N^4-990 N^3+8055 N^2-28710 N+37800)\alpha_{-1}^{i_1}\alpha_{-1}^{i_2}\alpha_{-1}^{i_3}\alpha_{-1}^{i_4}\dots\alpha_{-1}^{i_{N-4}}(\alpha_{-2}\cdot\alpha_{-2})
\end{split}
\end{equation}
and
\begin{equation}
\begin{split}
    (36 N^4+216 N^3-11700 N^2+84888 N-181440)\alpha_{-5}^{(i_1}\alpha_{-1}^{i_2}\alpha_{-1}^{i_3}\alpha_{-1}^{i_4}\dots\alpha_{-1}^{i_{N-4})}+\\
    -(40 N^4+80 N^3-2440 N^2-22640 N+154560)\alpha_{-3}^{(i_1}\alpha_{-3}^{i_2}\alpha_{-1}^{i_3}\alpha_{-1}^{i_4}\dots\alpha_{-1}^{i_{N-4})}+\\
    +(120 N^3-4440 N-10080)\alpha_{-3}^{(i_1}\alpha_{-2}^{i_2}\alpha_{-2}^{i_3}\alpha_{-1}^{i_4}\dots\alpha_{-1}^{i_{N-4})}+\\
    +(60 N^4-480 N^3-4380 N^2+51600 N-126000)\alpha_{-3}^{(i_1}\alpha_{-1}^{i_2}\alpha_{-1}^{i_3}\alpha_{-1}^{i_4}\dots\alpha_{-1}^{i_{N-4})}(\alpha_{-1}\cdot\alpha_{-1})+\\
    -(180 N^4-3960 N^3+32220 N^2-114840 N+151200)\alpha_{-1}^{i_1}\alpha_{-1}^{i_2}\alpha_{-1}^{i_3}\alpha_{-1}^{i_4}\dots\alpha_{-1}^{i_{N-4}}(\alpha_{-3}\cdot\alpha_{-1})+\\
    -(1080 N^2+2160 N-3240)\alpha_{-2}^{(i_1}\alpha_{-2}^{i_2}\alpha_{-2}^{i_3}\alpha_{-2}^{i_4}\dots\alpha_{-1}^{i_{N-4})}+\\
    -(180 N^3-1800 N^2+540 N+22680)\alpha_{-2}^{(i_1}\alpha_{-2}^{i_2}\alpha_{-1}^{i_3}\alpha_{-1}^{i_4}\dots\alpha_{-1}^{i_{N-4})}(\alpha_{-1}\cdot\alpha_{-1})+\\
    -(45 N^4-990 N^3+8055 N^2-28710 N+37800)\alpha_{-1}^{i_1}\alpha_{-1}^{i_2}\alpha_{-1}^{i_3}\alpha_{-1}^{i_4}\dots\alpha_{-1}^{i_{N-4}}(\alpha_{-1}\cdot\alpha_{-1})^2
\end{split}
\end{equation} %

\clearpage
\phantomsection
\addcontentsline{toc}{section}{References}
\bibliographystyle{utphys}
{\linespread{1.075}
\bibliography{Refs}
}

\end{document}